\documentclass[useAMS,usenatbib]{mn2e}
\usepackage{footnote,graphicx,natbib,color,multirow,amsmath,url}
\usepackage{amssymb}
\usepackage{tabularx}
\usepackage{amssymb}

\def\lesssim{\mathrel{\hbox{\rlap{\hbox{\lower3pt\hbox{$\sim$}}}\hbox{\raise2pt\hbox{$<$}}}}}

\definecolor{minorcol}{rgb}{0,0,0}
\def\minor		{\color{minorcol}}

\begin{document}

\title[Quenching in the group environment]{Galaxy Zoo: The interplay of quenching mechanisms in the group environment}
\author[Smethurst et al. 2017]{R. ~J. ~Smethurst,$^{1,2}$ C. ~J. ~Lintott,$^{1}$ S.~P.~Bamford,$^{2}$ R. ~E. ~Hart,$^{2}$ \newauthor S.~J. ~Kruk,$^{1}$ K.~L.~Masters,$^{3}$  R. ~C. ~Nichol,$^{3}$ B. ~D.~Simmons$^{4}$ \footnotemark[1]
\\ $^1$ Oxford Astrophysics, Department of Physics, University of Oxford, Denys Wilkinson Building, Keble Road, Oxford, OX1 3RH, UK 
\\ $^2$ School of Physics and Astronomy, The University of Nottingham, University Park, Nottingham, NG7 2RD, UK
\\ $^3$ Institute of Cosmology and Gravitation, University of Portsmouth, Dennis Sciama Building, Barnaby Road, Portsmouth, PO1 3FX, UK 
\\ $^4$ Center for Astrophysics and Space Sciences (CASS), Department of Physics, University of California, San Diego, CA 92093, USA
\\ 
\\ Accepted 2017 April 20. Received 2017 April 20; in original form 2017 February 22}

\maketitle

\begin{abstract}
Does the environment of a galaxy directly influence the quenching history of a galaxy? Here we investigate the detailed morphological structures and star formation histories of a sample of SDSS group galaxies with both classifications from Galaxy Zoo 2 and NUV detections in GALEX. We use the optical and NUV colours to infer the quenching time and rate describing a simple exponentially declining SFH for each galaxy, along with a control sample of field galaxies. We find that the time since quenching and the rate of quenching do not correlate with the relative velocity of a satellite but are correlated with the group potential. This quenching occurs within an average quenching timescale of $\sim2.5~\rm{Gyr}$ from star forming to complete quiescence, during an average infall time (from $\sim 10R_{200}$ to $0.01R_{200}$) of $\sim 2.6~\rm{Gyr}$. Our results suggest that the environment does play a direct role in galaxy quenching through quenching mechanisms which are correlated with the group potential, such as harassment, interactions or starvation. Environmental quenching mechanisms which are correlated with satellite velocity, such as ram pressure stripping, are not the main cause of quenching in the group environment. We find that no single mechanism dominates over another, except in the most extreme environments or masses. Instead an interplay of mergers, mass \& morphological quenching and environment driven quenching mechanisms dependent on the group potential drive galaxy evolution in groups. 

\end{abstract}


\footnotetext[1]{This investigation has been made possible by the participation of over 350,000 users in the Galaxy Zoo project. Their contributions are acknowledged at \url{http://authors.galaxyzoo.org}}

\section{Introduction}\label{sec:intro}

Over half of all galaxies are found clustered together in groups \citep{zwicky38, abell58, huchra82, eke04}, sharing one large dark matter halo (groups with $\sim100$ or more galaxies are referred to as clusters; \citealt{bower04}). Conversely some galaxies are found isolated from others in less dense environments (often referred to as the field), either because they are fossil groups \citep[where all members have eventually merged;][]{ponman94, jones00, jones03} or because they have been isolated for their entire lifetimes. This environmental density is found to be correlated with morphology \citep{dressler80, smail97, poggianti99, postman05, Bamford09}, colour \citep{butcher78, pimbblet02}, quenched galaxy fraction \citep{kauffmann03, Baldry06, peng12, darvish16} and star formation rate \citep[SFR; ][]{gomez03}. Star forming disc galaxies tend to be located in low-density environments with quiescent early-type galaxies in more dense environments. This suggests that the environment may drive a galaxy's transition from star forming in the blue cloud to the quiescent red sequence through quenching of star formation. 
 
Although these correlations were originally interpreted as indicating causation, recent evidence from simulations suggests that quenching mechanisms driven by the environment may not be dominant in the galaxy lifecycle \citep{kimm09, kimm11, hirschmann14, wang14, phillips15, emerick16, fillingham16}. Perhaps, instead, the correlation of increased quenched galaxy fractions with environment density is due to a superposition of other possible quenching mechanisms each of which depend on more local factors {\minor \citep{kauffmann04, blanton06, cucciati10} than the broader environment properties \citep{balogh04, porter08, fadda08, darvish14, darvish17, alpaslan16, laigle17}.}
  
In order to isolate the cause of the density-morphology and density-SFR correlations, we need to observe how morphology and galaxy quenching timescales change in dense environments with different properties in comparison to the field. Here, we consider the group environment, as this is a more typical environment for a galaxy than the relatively rare rich cluster environment \citep{carlberg04}. We construct a sample of both group and field galaxies and use a Bayesian inference method to determine the quenching time and rate describing a simple exponentially declining SFH for a galaxy given its optical and NUV colours. From these inferred SFHs we aim to constrain the possible mechanisms at work in the group environment. We aim to determine the following: (i) How does the environment influence the detailed morphological structures of a galaxy?  (ii) Is quenching that is directly caused by the environment occurring in galaxy groups? However, dense environments are messy with many possible mechanisms at work, whose effects are difficult to disentangle. 

There are many mechanisms which have been suggested to cause quenching. They are often referred to as either internal mechanisms (caused by the galaxy's `nature') or external mechanisms (caused by the way the galaxy is `nurtured'). The properties of a galaxy and its environment are often thought to control which mechanisms will affect a galaxy throughout its lifetime and subsequently affect the morphology. {\minor In Sections~\ref{sec:intquench} \& \ref{sec:extquench} we introduce the basic principles of possible quenching mechanisms. We note, that although these mechanisms have been theorised and studied individually, they have many interdependencies which are difficult to disentangle in a population. We state how we will attempt to isolate each of the possible quenching mechanisms introduced here to study their effects on the group population at the end of each section} before we describe our data sources and samples in Section~\ref{sec:data}. We provide a more thorough discussion of quenching mechanisms in Section~\ref{sec:disc}.

\subsection{Internal Quenching Mechanisms}\label{sec:intquench}

\subsubsection{Mass quenching}\label{sec:massquench}

Mass quenching is defined by \citet{peng10, peng12} as any quenching mechanism acting independently of a galaxy's environment, but not of its mass. However, there is still much debate over the exact mechanism which is the cause of such a quench. \citet{darvish16} suggest that non-AGN driven feedback mechanisms (for example supernova feedback) are responsible for the correlation observed between the mass quenching efficiency and SFR in \citet{peng10}. However, \citet{gabor15} suggest that this is driven by ``halo quenching processes'' {\minor (which they also suggest is a driver of environmental quenching, see Section~\ref{sec:envquench})}, whereby the inflow of cool gas from the galaxy halo is either cut off or the gas is hindered from cooling at $M_{\rm{halo}} > 10^{12}~\rm{M}_{\odot}$ \citep{birnboim03, dekel06}. If this happens, a galaxy uses up the rest of its available gas for star formation as described by the Kennicutt-Schmidt law \citep{schmidt59, kennicutt98} and consequently grows in mass.

Mass quenching is thought to be a dominant mechanism for isolated galaxies in the field \citep{kormendy04}. However, it is also thought that as a galaxy infalls in to a group or cluster over long timescales, gas reservoirs can also be depleted via a mass quenching process \citep{peng12}. {\minor It is therefore difficult to separate mass quenching and environmental quenching as individual mechanisms due to their interdependence. In this work we will refer to mass quenching as encompassing any process which reduces the supply and consumption of gas in galaxies as they grow in mass, irrespective of their environment.}

We investigate this possible quenching mechanism in this study by studying the dependency of the inferred quenching parameters with stellar mass for satellite, central and field galaxies {\minor with increasing environmental density.} 
 
\subsubsection{Morphological quenching}\label{sec:morphquench}

Morphological quenching is the theorised process by which the internal structure of a galaxy can have a negative impact on its own SFR. This can happen in one of two ways, either by preventing star formation from occurring or by increasing the rate of consumption of gas for star formation. The former is thought to be caused by bulges \citep{bluck14} whereby the large gravitational potential of the bulge prevents the disc from collapsing and forming stars \citep{Fang13}. 

The latter mechanism is theorised to occur in galaxies hosting bars; the bar funnels gas to the centre of the galaxy \citep{athanassoula92a} where gas is exhausted by star formation effectively quenching the galaxy \citep{zurita04, sheth05}. {\minor Alternatively, the bar could be the consequence of another quenching mechanism entirely.}

We investigate this possible quenching mechanism in this study by studying the dependency of the bar and bulge fractions of satellite galaxies {\minor with increasing environmental density} in comparison to the field.  

\subsubsection{AGN feedback as a quenching mechanism}\label{sec:agnquench}

There are tight correlations between properties of galaxies, such as the bulge mass, total stellar mass \& stellar velocity dispersion, and the black hole mass \citep{magorrian98, marconi03, haringrix04}. This implies a co-evolution between the black hole and its host galaxy therefore suggesting that changes in the SFR and structure of a galaxy could also be tied to black hole activity. This is thought to occur via AGN feedback where the output of energetic material and radiation from the black hole is thought to either heat or expel the gas needed for SF in a galaxy, causing a quench. {\minor Alternatively, the black hole activity could be a consequence (rather than the cause) of another quenching mechanism which also gives rise to morphological changes in a galaxy (such as mass and morphological quenching mechanisms).}

AGN feedback was first suggested as a mechanism for regulating star formation due to the results of simulations wherein galaxies could grow to unrealistic stellar masses \citep{silk98, Bower06, Croton06, somerville08}. Without a prescription for the effects of AGN feedback, the shape of the galaxy luminosity function could therefore not be matched at the high luminosity end \citep{baugh98, baugh05, kauffmann99a, kauffmann99b, somerville01, kitzbichler06}. 

Indirect observational evidence has been found for both positive and negative feedback in various systems (see the comprehensive review from \citealt{fabian12}). The strongest being the indirect evidence that the largest AGN fraction is found in the green valley {\minor\citep{Martin07, cowie08, Hickox09, schawinski10a}}, suggesting a link between AGN activity and the process which moves a galaxy from the blue cloud to the red sequence. Recent statistical evidence from \cite{smethurst16} has shown the dominance of rapid, recent quenching within a population of Type 2 AGN host galaxies, suggesting that AGN feedback is indeed an important evolutionary mechanism. 

We do not directly investigate the presence of AGN in the group environment due to constraints from low number statistics, however we discuss the possible role of AGN feedback in the context of the results presented in this study in Section~\ref{sec:bigpic}.

\subsection{External Quenching Mechanisms}\label{sec:extquench}

\subsubsection{Mergers as a quenching mechanism}\label{sec:mergersquench}

Major mergers have been intrinsically linked to the formation of early-type galaxies since \citet{toomre72} showed this was possible with a simulation of the merger of two equal mass disc galaxies. The hypothesis is as follows: when two galaxies merge, the influx of cold gas funnelled by the forces in the interaction often results in energetic starbursts \citep{mihos94, mihos96, hopkins06d, hopkins08a, hopkins08b, snyder11, hayward14, sparre16}, which can exhaust the gas required for star formation, effectively quenching the post-merger remnant. This remnant galaxy will also have formed a dynamically hot bulge through the dissipation of angular momentum in the merger \citep{toomre77, walker96, kormendy04, hopkins11c, martig12}. The mass ratio of the two galaxies merging is thought to affect the size of the bulge that is formed in the remnant \citep{cox08, hopkins09c, tonini16}, with the most massive major mergers with a 1:1 mass ratio producing fully elliptical galaxies \citep{toomre72, barnes96, mihos96, kauffmann96, pontzen16}. However, recent simulations of the merger of two disc galaxies with a 1:1 mass ratio have shown that a disc remnant galaxy can be produced \citep{hopkins09c, pontzen16, sparre16}. 

Such a scenario is also intrinsically linked to the triggering of an AGN due to the influx of gas in the merger which can fuel the black hole accretion \citep{sanders88, dimatteo05, hopkins09a, treister12}. Simulations of mergers with AGN have led many to believe that a merger which triggers both a starburst and an AGN can quench a galaxy in extremely rapid timescales \citep{springel05b, bell06}. Recent simulations have also suggested that feedback from the triggered AGN (see Section \ref{sec:agnquench}) is necessary to fully remove (or heat) all the available gas, otherwise the SFR will recover back to the star formation sequence (SFS) post-merger \citep{athanassoula16, pontzen16, sparre16}. 

Mergers also have a clear environmental dependence, as they are more likely to occur in denser environments (at least until the velocity dispersion of a cluster becomes so large as to suppress mergers due to increased interaction velocities between galaxies). {\minor Here we attempt to separate their effects from those quenching mechanisms driven solely by the properties of the galaxy environment, through controlling for the environment density by comparing galaxies at the same local group density.} We investigate mergers as a possible quenching mechanism by studying the dependency of the inferred quenching parameters with the number of galaxies in a group {\minor with increasing environmental density}. 

\subsubsection{Environment driven quenching}\label{sec:envquench}

The proposed quenching mechanisms under the umbrella of environmental quenching are numerous and varied {\minor(see comprehensive review by \citealt{boselli06})}. Together with the typical gravitational galaxy-galaxy interactions \citep{moore96} which are expected to be more frequent in a dense environment, environmental quenching also includes hydrodynamic interactions occurring between the cold interstellar medium (ISM) of the in-falling galaxy and the hot intergalactic medium (IGM) of the group or cluster. Such hydrodynamic interactions include ram pressure stripping \citep{gunngott72}, viscous stripping \citep{nulsen82}, and thermal evaporation \citep[a rapid rise in temperature of the ISM due to contact with the IGM;][]{cowie77}. Another such process is starvation \citep[also called strangulation;][]{larson80} which can remove the outer galaxy halo, thus cutting off the star formation gas supply to a galaxy. Preprocessing occurs when all of the above mechanisms take place in a group of galaxies which then merges with a larger group or cluster \citep{dressler04}. 

The most likely (and therefore the most studied) candidate mechanism for the cause of the environmental density-morphology and SFR relations is ram pressure stripping \citep[RPS;][]{abadi99, poggianti99}. However, there has been mounting evidence that RPS can only strip a galaxy of $40-60\%$ of its gas supply \citep{fillingham16} and so may not be as effective a quenching mechanism as first thought \citep{emerick16}. Therefore investigations of other environmentally driven quenching mechanisms, such as strangulation \citep{peng15, hahn16, maier16, paccagnella16, roberts16, vandevoort16} and harassment \citep[high speed galaxy `fly-by' gravitational interactions][]{bialas15, smith15b} are having a recent resurgence.

We investigate the environment as a possible driver of quenching in this study by studying the dependency of the inferred quenching parameters with halo mass, the ratio of masses between satellites and their central galaxies, the relative velocity of satellites and the stellar velocity dispersions of group galaxies in comparison to the field. 
\\

This paper proceeds as follows. In Section~\ref{sec:data} we describe our data sources and group galaxy sample. We show the results of an investigation into the environmental dependence of the detailed morphological structure of group galaxies in Section~\ref{sec:morphfrac}. We describe our Bayesian inference method for determining the quenching histories of group galaxies and present the results of this method in Section~\ref{sec:starpy}. We then discuss the possible quenching mechanisms that could be responsible for our results and how they fit into the bigger picture of quenching in Section~\ref{sec:disc}. We summarise our findings in Section~\ref{sec:conc}. The zero points of all magnitudes are in the AB system. Where necessary, we adopt the WMAP Seven-Year Cosmology \citep{jarosik11} with $(\Omega_m , ~\Omega_\Lambda , ~h) = (0.26, 0.73, 0.71)$.

\section{Data and Methods}\label{sec:data}

\subsection{Data Sources}\label{sec:photo}

\begin{figure*}
\centering
\includegraphics[width=\textwidth]{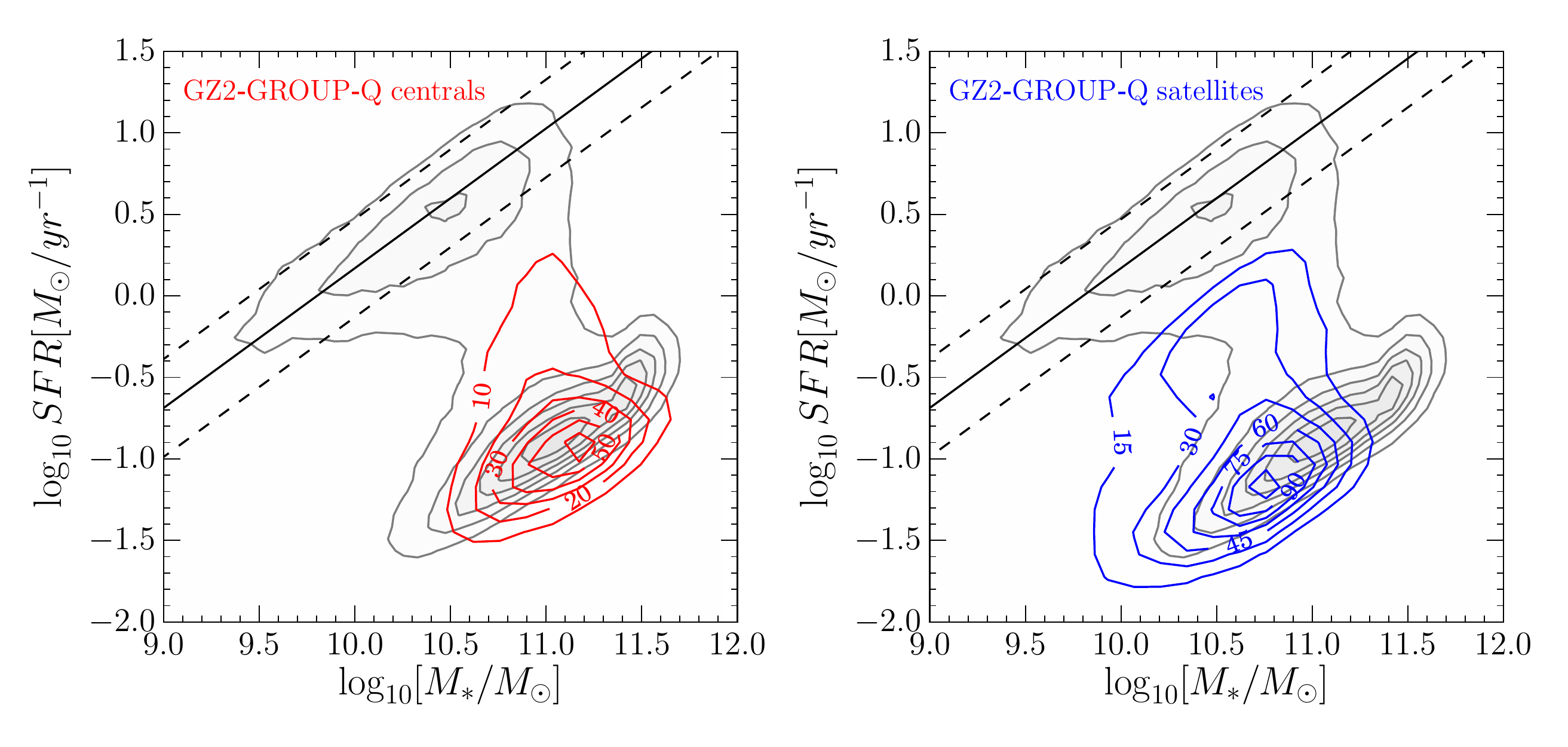}
\caption[Stellar mass-SFR plane for the centrals and satellites of the \textsc{gz2-group-q} sample]{The stellar mass-SFR plane showing central (left; red contours) and satellite (right; blue contours) galaxies in the \textsc{gz2-group-q} sample, selected to be $1\sigma$ below the SFS. In both panels the entire SDSS sample from the MPA-JHU catalogue is shown by the grey contours. The definition of the SFS from \cite{peng10} at $\overline{z} = 0.053$ (solid line, the mean redshift of the \textsc{gz2-group-q} sample) with $\pm1\sigma$ (dashed lines) is shown.}
\label{fig:sfrmass}
\end{figure*}

In this investigation we use visual classifications of galaxy morphologies from the Galaxy Zoo 2\footnote{\url{http://zoo2.galaxyzoo.org/}} (GZ2) citizen science project \citep{lintott09, GZ2}, which obtains multiple independent classifications for each optical image. The full question tree for an image is shown in Figure~1 of \citeauthor{GZ2}  The GZ2 project used $304022$ images from the Sloan Digital Sky Survey Data Release 7 (SDSS; \citealt{york00, abazajian09}) all classified by at least 17 independent volunteers, with a mean number of classifications of $\sim42$. We also utilise the Petrosian magnitude, {\tt petroMag}, values for the $u$ ($3543 \rm{\AA}$) and $r$ ($6231 \rm{\AA}$) wavebands provided by the SDSS DR7 pipeline \citep{stoughton02} for the GZ2 galaxies.

We also required NUV ($2267 \rm{\AA}$) photometry from the GALEX survey \citep{martin05}, within which $\sim42\%$ of the GZ2 sample was observed, giving a total of $126316$ galaxies with $0.01 \lesssim z \lesssim 0.25$ {\minor and $-22 < M_{u} < -15$}. This will be referred to as the \textsc{gz2-galex} sample. The completeness of this sample is shown in Figure~2 of \cite{smethurst15}, {\minor wherein typical Milky Way $L_{*}$ galaxies ($M_u \sim −20.5$) are still detected out to the highest redshift ($z \sim 0.25$); however dwarf and lower mass galaxies are only detected at the lowest redshifts ($z\lesssim0.05$).}

Magnitudes are corrected for galactic extinction \citep{Oh11} by applying the \citet{Cardelli89} law, giving a typical correction of $u-r \sim 0.05$. K-corrections are also adopted to $z=0.0$ and absolute magnitudes obtained from the NYU-VAGC \citep{Blanton05, padmanabhan08, blanton07}, giving a typical $u-r$ correction of $\sim 0.15$ mag. 


\subsection{Group Identification}\label{sec:groups}

The construction of a robust cluster or group catalogue is a challenge, with many studies attempting this across the SDSS \citep[e.g. ][]{merchan05, miller05, berlind06, yang07, tago08, tago10, tinker11, munoz12, tempel14} and other large surveys \citep{tucker00, merchan02, eke04, cucciati10, robotham11, knobel12}. The difficulties arise in removing projection effects, understanding the selection function used, covering large ranges in mass and redshift, and dealing with spectral fibre collisions (see the comprehensive review by \citealt{postman02} for an in depth discussion).

In this work we use the \citet{berlind06} catalogue, which employs a friends-of-friends algorithm to identify group galaxies in the SDSS DR4. This group catalogue was then cross matched with the \textsc{gz2-galex} sample. We limit our sample to $z < 0.1$ to ensure GALEX completeness to the red sequence, as in \citealt{wyder07} and \citealt{yesuf14}, so that we do not introduce any bias in our sample due to the necessity for NUV colours, which could otherwise be attributed to environmental effects. This results in $10423$ galaxies in groups with the number of group members, $N_{\rm{group}} \geq 3$. {\minor We chose to retain those galaxies residing in smaller groups with $N_{\rm{group}} \leq 5$ ($5201$ galaxies) in order to maximise the sample size and ensure a sample representative of the entire group population.} Centrals are identified as the brightest group galaxy in the $r$-band (as opposed to the most massive group galaxy which necessitates an assumption of the mass-to-light ratio of a galaxy), with all other galaxies in a group designated as satellites.

The projected group-centric radius, $R$, of all satellite galaxies from their central galaxies was calculated to quantify environmental density. In order to compare groups of different sizes, the virial radius is used as a normalisation factor to this projected group-centric radius. Here we use a proxy to the virial radius, $R_{200}$ \citep[see][]{navarro95}, the radius within which the group mass overdensity is 200 times the critical density, $\rho_{\rm{crit}}(z)$, as defined by \citealt{finn05}:
\begin{equation}\label{eq:overdense}
200\rho_{\rm{crit}}(z) = \frac{M_{cl}}{\frac{4}{3}\pi R_{200}^3},
\end{equation}

where $M_{cl}$ is the total mass of the group. \citeauthor{finn05} then use the redshift dependence of the critical density and the virial mass to relate the line-of-sight velocity dispersion of the group, $\sigma_x$, to the group mass so that $R_{200}$ becomes:
\begin{equation}\label{eq:r200}
R_{200}~=~1.73 \left ( \frac{\sigma_x}{1000 \rm{km}~\rm{s}^{-1}} \right) \cdot \frac{1}{\sqrt{\Omega_{\Lambda} +\Omega_o(1+z)^3}} ~ h_{100}^{-1} ~\rm{Mpc}. 
\end{equation}
$\sigma_x$ is provided in the \cite{berlind06} catalogue and is calculated as:
\begin{equation}
\sigma_x = \frac{1}{1+\overline{z}} \sqrt{\frac{1}{N_{\rm{group}}-1} \sum_{i=1}^{N_{\rm{group}}}(cz_i - c\overline{z})^2}
\end{equation}
where $\overline{z}$ is the mean redshift of the group and $N_{\rm{group}}$, the number of galaxies in a group. Since most groups in the sample have low $N_{\rm{group}}$, using the mean redshift for $z_{\rm{group}}$, rather than the central galaxy redshift is most appropriate in this case. These calculations resulted in a sample of $2,209$ centrals and $8,214$ satellites within a projected group-centric radius range of $0.01 < R/R_{200} < 10.0$ and $z < 0.084$ which shall be referred to as the \textsc{gz2-group} sample. Note that for a galaxy (central or satellite) to be included in the \textsc{gz2-group} sample, the rest of its group does not need to be included. However the properties of that group are still inherited by the included galaxy. 

We obtain SFRs, stellar masses and stellar velocity dispersions of galaxies in the \textsc{gz2-group} sample from the MPA-JHU catalogue \citep{kauffmann03, brinchmann04}. The measurements of stellar velocity dispersion, $\sigma_*$, from the MPA-JHU catalogue are limited by the SDSS instrument dispersion of $\sim69~\rm{km}~\rm{s}^{-1}$ \citep{stoughton02}. Therefore any $\sigma_*$ values derived below the instrument dispersion are assumed to be upper limits at $70~\rm{km}~\rm{s}^{-1}$. 

In this study we specifically focus on galaxies that are below the star forming sequence (SFS). We select a subsample of the \textsc{gz2-group} galaxies that are $1\sigma$ below the SFS (as defined by \citealt{peng10} for a given galaxy mass and redshift), giving $3,867$ satellite and $1,564$ central galaxies which will collectively be referred to as the \textsc{gz2-group-q} sample (with a median $N_{\rm{group}}=8$, mean $N_{\rm{group}}\sim26$ and maximum $N_{\rm{group}}=311$). Note that centrals consist of a larger proportion, $\sim40\%$, of the \textsc{gz2-group-q} sample, compared to $\sim27\%$ of the \textsc{gz2-group} sample, as expected when applying a star formation rate threshold. These galaxies are shown in Figure \ref{fig:sfrmass} and can be seen to lie below the SFS.

\begin{figure}
\centering{
\includegraphics[width=0.45\textwidth]{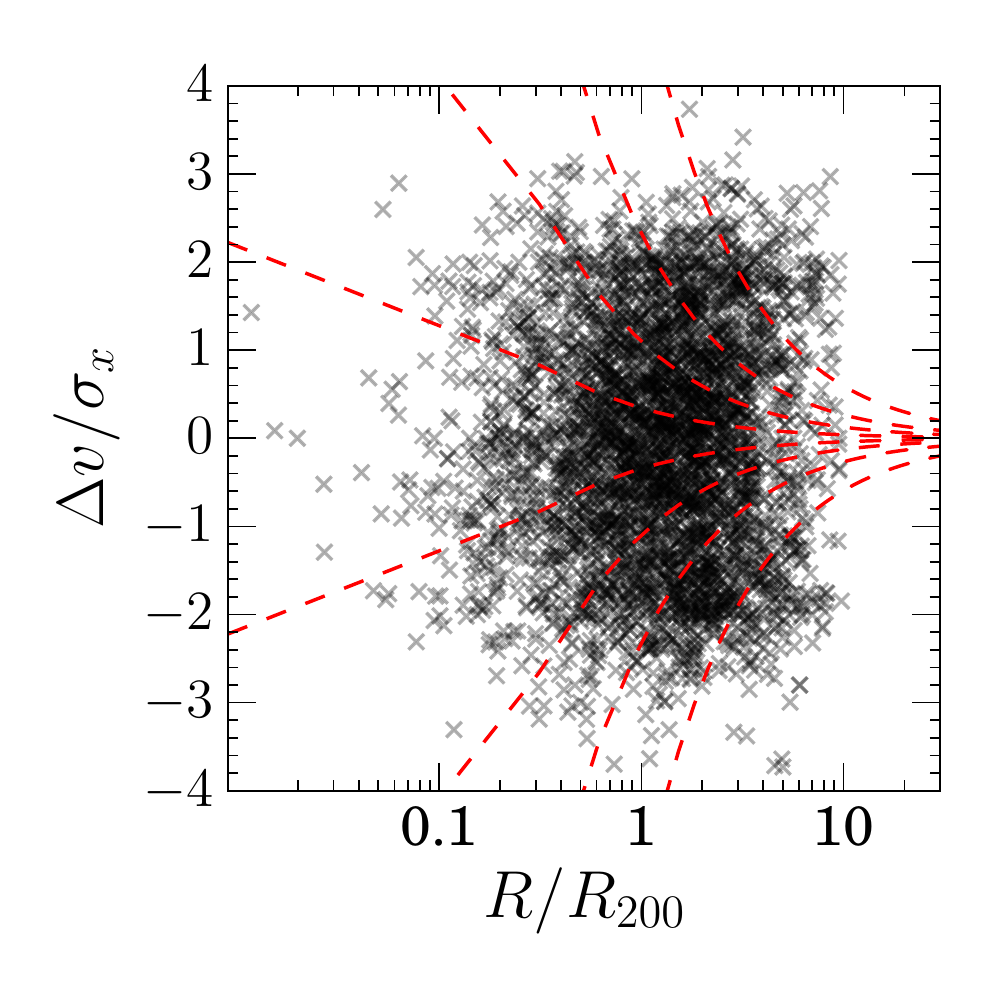}}
\caption[]{Phase space distribution of satellite galaxies in the \textsc{gz2-group-q} sample showing caustics at constant $(\Delta v/\sigma)\times(R/R_{200}) = \{0.2, 0.64, 1.35, 3\}$ as in \cite{noble16}.}
\label{fig:phasespace}
\end{figure}

We also show the \textsc{gz2-group-q} satellite galaxies on a phase space diagram, shown in Figure~\ref{fig:phasespace}, with the normalised projected group-centric radius and normalised relative velocity of the satellites to their central galaxy. We also over plot caustics at constant $(\Delta v/\sigma)\times(R/R_{200}) = \{0.2, 0.64, 1.35, 3\}$ as in \cite{noble16}. They define those galaxies with $1.35<(\Delta v/\sigma)\times(R/R_{200})<3$ as infalling satellites, $0.64<(\Delta v/\sigma)\times(R/R_{200})<1.35$ as recently accreted satellites, $0.2<(\Delta v/\sigma)\times(R/R_{200})<0.64$ as intermediate/backsplashing satellites and $0~<~(\Delta~v/\sigma)~\times~(R/R_{200})~<0.2$ as inner satellites. 

We find that $23\%$ of the \textsc{gz2-group-q} satellites lie outside of the outer caustic\footnote{{\minor$11\%$ of galaxies in groups with $N_{group} \leq 5$ lie outside the outer caustic with $(\Delta v/\sigma)\times(R/R_{200})~>~3$. Similarly, $12\%$ of galaxies with $N_{group} > 5$ lie outside the same outer caustic. Therefore this percentage is not dependent on the size of the group.}} with $(\Delta v/\sigma)\times(R/R_{200})~>~3$. Either the true velocity has been overestimated due to projection effects or the satellite has been misidentified as a member of a group. We chose to retain these galaxies within our sample since we cannot distinguish between these two effects (however we note that removing them does not change our conclusions). This is only an issue for satellites with $R/R_{200} > 1$.

We also consider the projected neighbour density, as defined by \cite{Baldry06} as $\Sigma_N = N/4\pi d_N^2$, where $d_N$ is the distance to the $N^{\rm{th}}$ nearest neighbour, for the \textsc{gz2-group} and \textsc{gz2-group-q} samples. $\Sigma$ is a more direct probe of the local density of a galaxy's environment, and although it does not allow for the identification of groups and their properties, it is still a useful probe of the local density inside a group  \cite[see][for a comparison of various environment parametrisations]{muldrew12}.

Here we use the estimates of \cite{Bamford09} who averaged $\log\Sigma_N$ for $N = 4$ and $N=5$ by the method outlined in \citet{Baldry06}, for the entirety of the GZ2 sample. We find that $90\%$ of the \textsc{gz2-group} sample have $\log\Sigma > -0.8$ (the threshold quoted by \citealt{Baldry06} to define non-field galaxies), suggesting a purity of $\sim90\%$ for the \textsc{gz2-group} sample.

\subsection{Field sample}\label{sec:field}

\begin{figure}
\centering{
\includegraphics[width=0.45\textwidth]{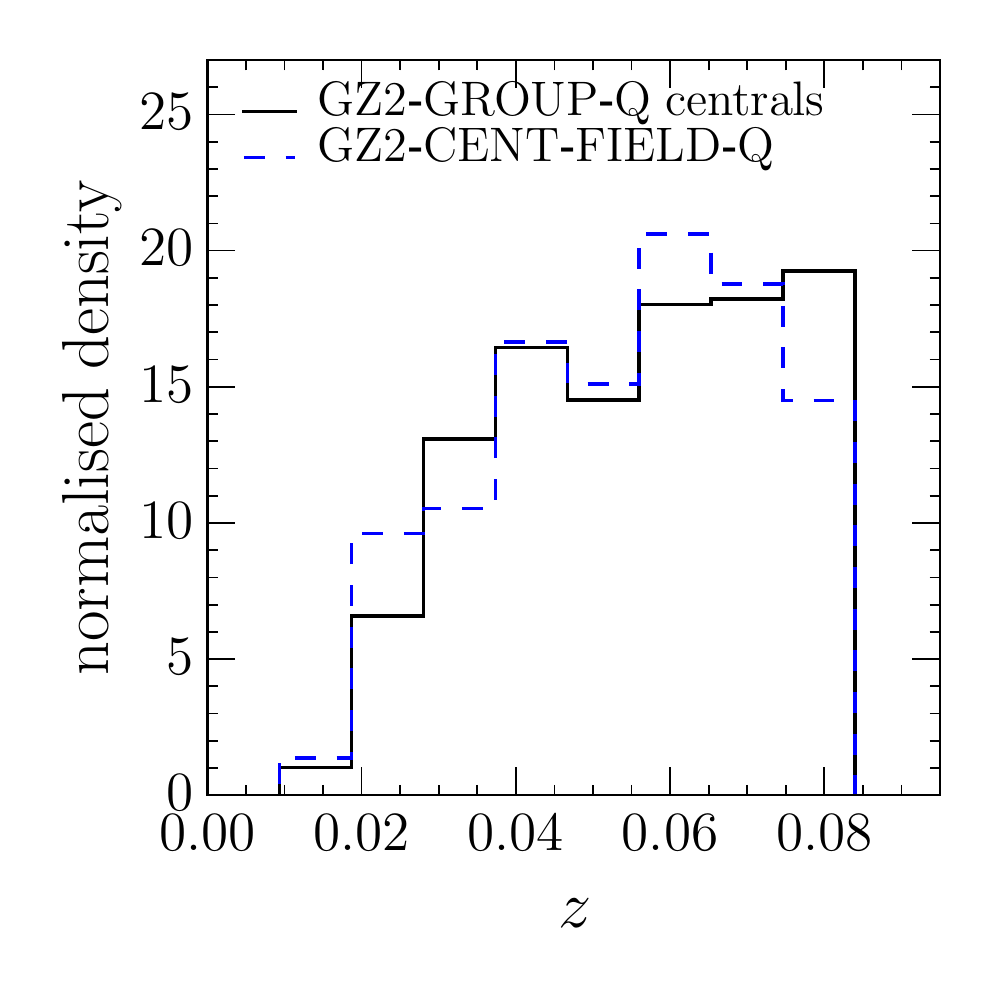}
\includegraphics[width=0.45\textwidth]{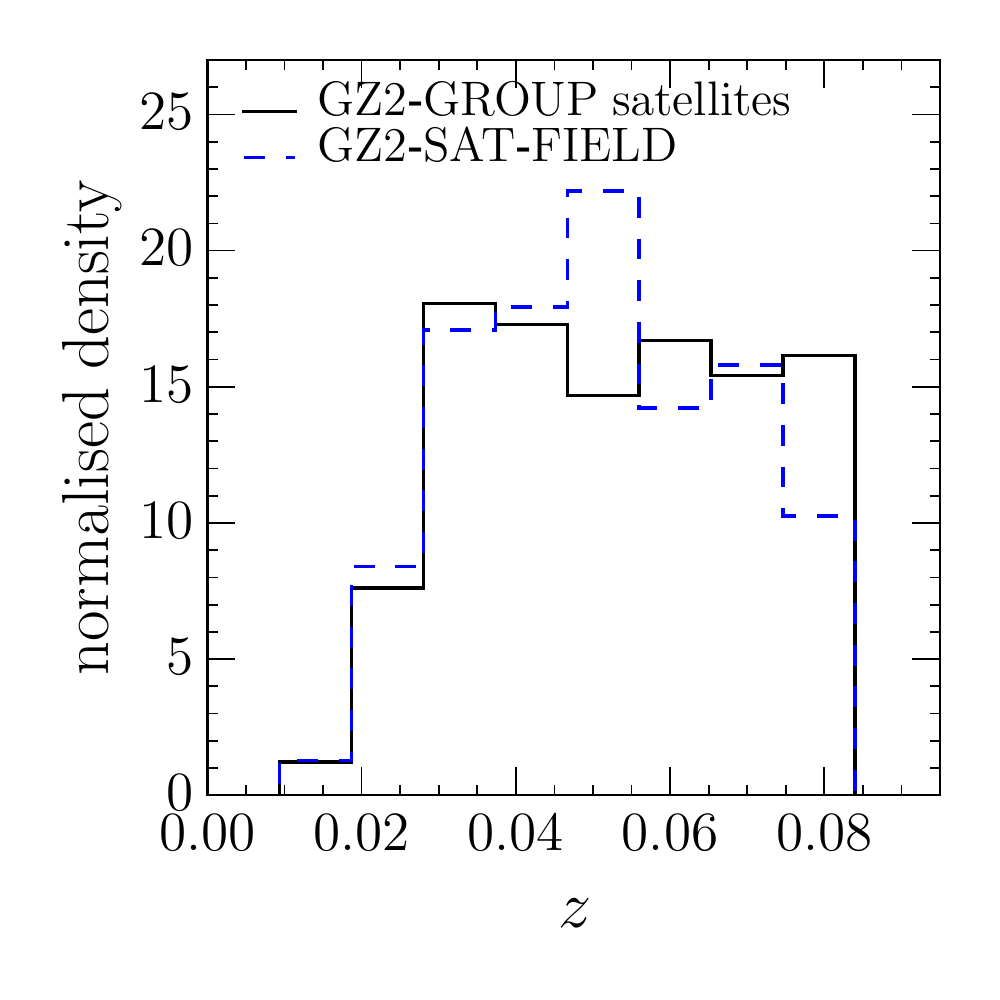}}
\caption[Redshift distribution of galaxies in the \textsc{gz2-group} and \textsc{gz2-group-q} samples]{Redshift distributions of central galaxies in the \textsc{gz2-group-q} sample (left; black solid line) and satellite galaxies in the \textsc{gz2-group} sample (right; black solid line) in comparison the redshift matched \textsc{gz2-cent-field-q} (left; blue dashed line) and \textsc{gz2-sat-field} samples (right; blue dashed line).}
\label{fig:zcompare}
\end{figure}

We constructed a sample of field galaxies for use as a control sample to the \textsc{gz2-group-q} sample. For all galaxies in the \textsc{gz2-galex} sample, we calculated the smallest projected group-centric radii, $R/R_{200}$, from each of the central galaxies in the \citet{berlind06} catalogue (regardless of whether the central was included in the \textsc{gz2-group} sample). We also use the measurement of the projected neighbour density, $\Sigma$, from \cite{Baldry06}. We select candidate field galaxies as those with (i) $R/R_{200} > 25$ and (ii) $\log\Sigma < -0.8$ \citep[the threshold on the local environment density which selects field galaxies as defined by][]{Baldry06}. We chose to use both of these environmental density measures to ensure a pure sample of candidate field galaxies.

\begin{figure}
\centering{
\includegraphics[width=0.45\textwidth]{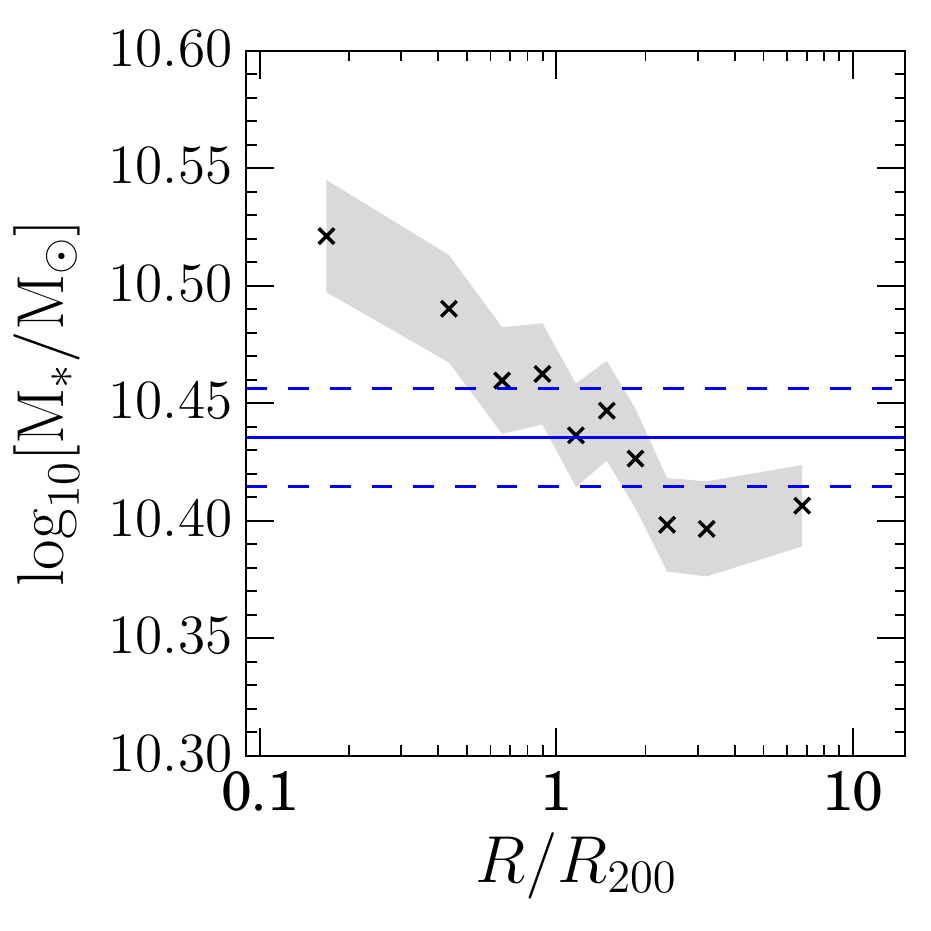}}
\caption[Average mass with group radius in the \textsc{gz2-group} sample]{The median stellar mass as a function of radius from the group centre for the \textsc{gz2-group} satellite galaxies. The shaded regions show the $\pm1\sigma$ uncertainties on the median values in each bin of $R/R_{200}$. Note the small y-axis range in comparison to the median error on the stellar mass measurements, $\sigma_{\log_{10}[M_*/M_{\odot}]}~\sim~0.09$. The average stellar mass of the \textsc{gz2-sat-field} sample is also shown (blue solid line) with the blue dashed lines showing $\pm1\sigma$ uncertainty on the median value. The points are plotted at the linear centre of each bin at $R/R_{200}~=~[0.17,  0.43,  0.66,  0.9 ,  1.17,  1.48,  1.85,  2.36,  3.22,  6.75]$.}
\label{fig:massdep}
\end{figure}

\begin{figure}
\includegraphics[width=0.45\textwidth]{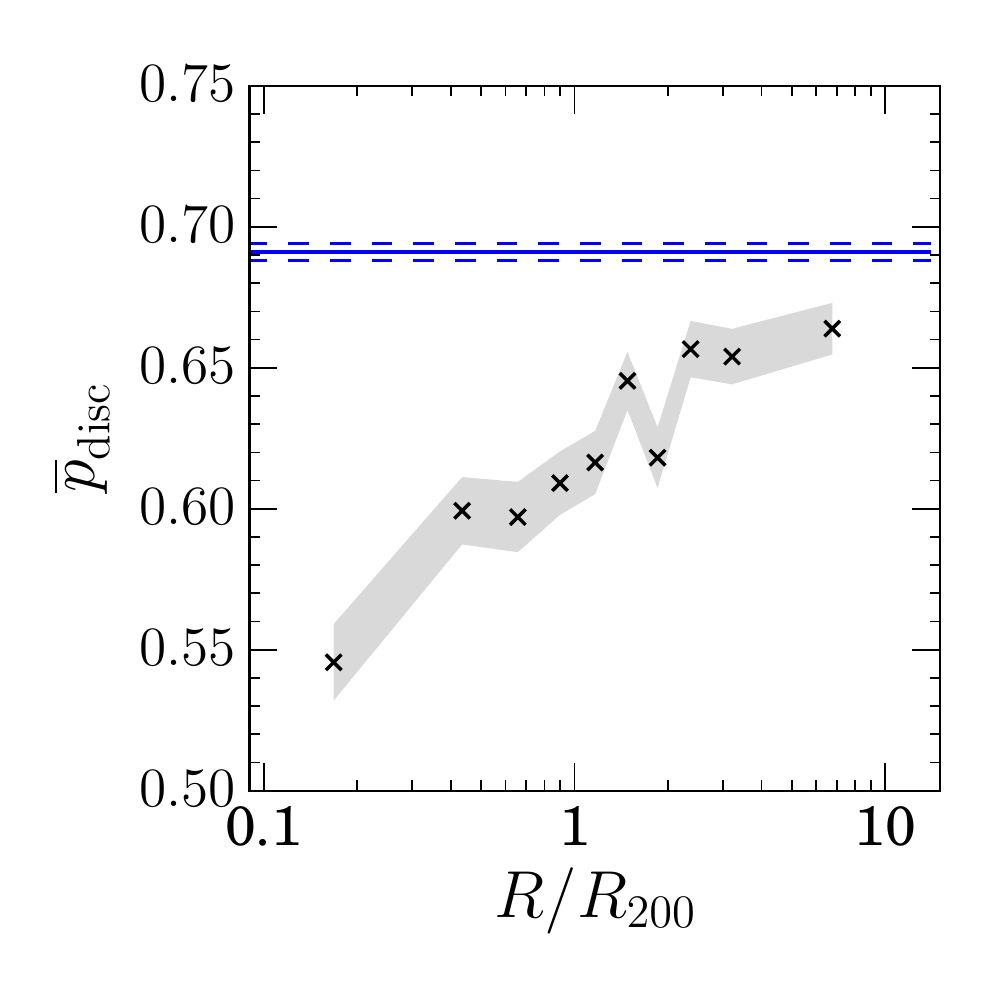}
\includegraphics[width=0.45\textwidth]{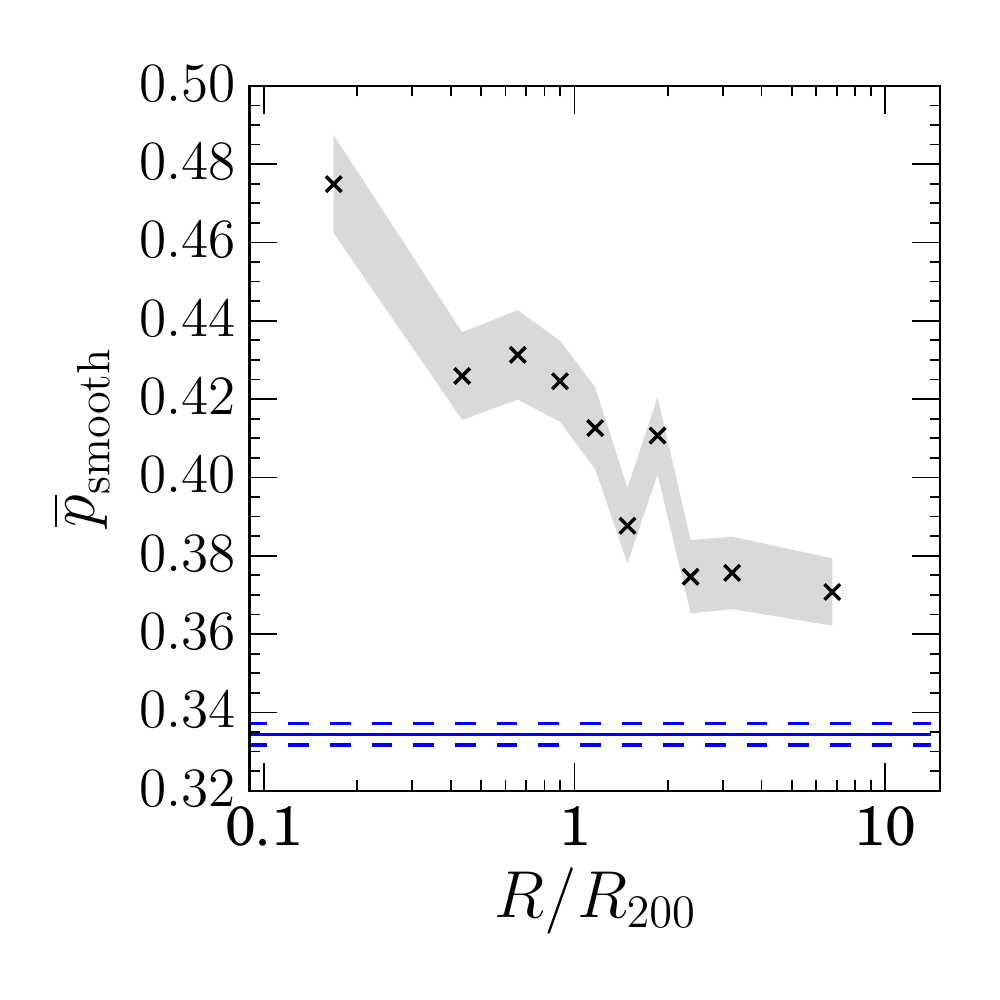}
\caption[Mean $p_d$ and $p_s$ with group radius in the \textsc{gz2-group} sample]{Mean GZ2 vote fraction for disc (top) and smooth (bottom) galaxies in the \textsc{gz2-group} sample binned by projected group-centric radius, normalised by $R_{200}$, a proxy for the virial radius of a group. The shaded region shows $\pm1\sigma$, the standard error on the mean vote fraction. The mean vote fraction of the \textsc{gz2-sat-field} sample are also shown (blue solid lines) with $\pm1\sigma$ (blue dashed lines). The points are plotted at the linear centre of each bin at $R/R_{200}~=~ [0.17,  0.43,  0.66,  0.9 ,  1.17,  1.48,  1.85,  2.36,  3.22,  6.75]$.}
\label{fig:morphradius}
\end{figure}

Firstly, each of the central galaxies of the \textsc{gz2-group-q} sample were matched to at least one candidate field galaxy in both redshift ($\pm10\%$) and stellar mass ($\pm10\%$), to give $2,309$ field galaxies with $z < 0.084$. As with the \textsc{gz2-group-q} sample in Section~\ref{sec:groups} we focus on galaxies which are either quenching or quenched and are more than $1\sigma$ below the SFS and so the same constraints must be placed on this field control sample. This encompasses $1,596$ field galaxies with $z < 0.084$ which will be referred to as the \textsc{gz2-cent-field-q} sample. It will be used as a control sample when investigating the inferred quenching parameters of both the central and satellite galaxies of the \textsc{gz2-group-q} sample. {\minor This ensures that the underlying halo mass (which is proportional to the stellar mass of the central galaxy of the group) distribution is the same across central, satellite and field samples}. The redshift distribution of the \textsc{gz2-cent-field-q} sample is shown in comparison to the distribution of central galaxies in the \textsc{gz2-group-q} sample in the top panel of Figure~\ref{fig:zcompare}. 

Secondly, each of the satellite galaxies of the \textsc{gz2-group} sample were matched to at least one candidate field galaxy in both redshift ($\pm10\%$) and stellar mass ($\pm10\%$), to give $8,444$ field galaxies with $z < 0.084$ which will be referred to as the \textsc{gz2-sat-field} sample.  Note that this sample is not restricted to being $1\sigma$ below the SFS as it will be used as a control when investigating the morphological trends of satellite galaxies in the \textsc{gz2-group} sample (i.e. those not restricted to being below the SFS) with environment. 

$237$ galaxies are present in both the \textsc{gz2-cent-field-q} and \textsc{gz2-sat-field} samples. The redshift distribution of the \textsc{gz2-sat-field} sample is shown in comparison to the distribution of satellite galaxies in the \textsc{gz2-group} sample in the bottom panel of Figure~\ref{fig:zcompare}.

We once again obtain SFRs and stellar velocity dispersions of galaxies for all of the field samples described above from the MPA-JHU catalogue \citep{kauffmann03, brinchmann04}.

\section{Effect of the group environment on detailed morphological structure}\label{sec:morphfrac}

\begin{figure}
\centering{
\includegraphics[width=0.45\textwidth]{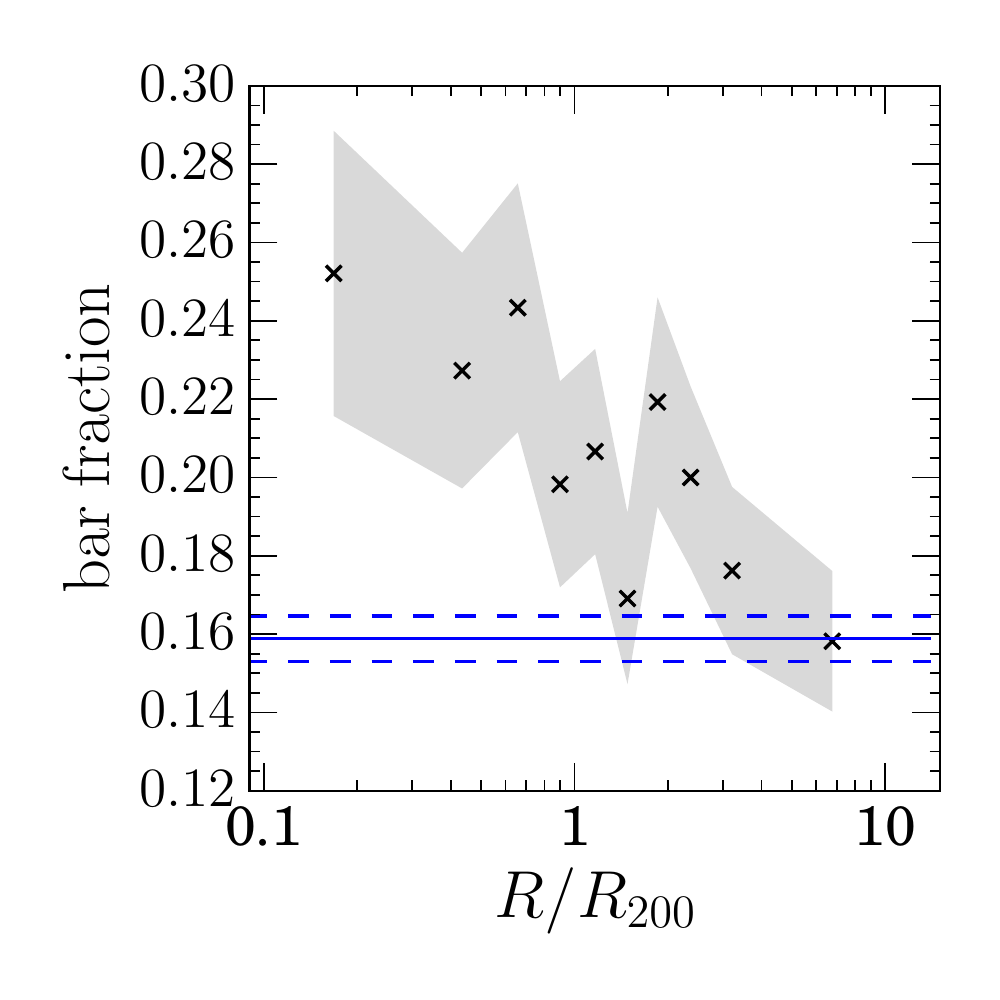}}
\caption[Bar fraction with group radius in the \textsc{gz2-group} sample]{Bar fraction (number of barred disc galaxies over number of disc galaxies) in the \textsc{gz2-group} sample binned in projected group-centric radius, normalised by $R_{200}$, a proxy for the virial radius of a group. The shaded region shows $\pm1\sigma$, the propagated counting error, on the bar fraction. The bar fraction of the \textsc{gz2-sat-field} sample is also shown (blue solid line) with $\pm1\sigma$ (blue dashed line). The points are plotted at the linear centre of each bin at $R/R_{200}~=~ [0.17,  0.43,  0.66,  0.9 ,  1.17,  1.48,  1.85,  2.36,  3.22,  6.75]$}
\label{fig:barradius}
\end{figure}

\begin{figure}
\centering{
\includegraphics[width=0.45\textwidth]{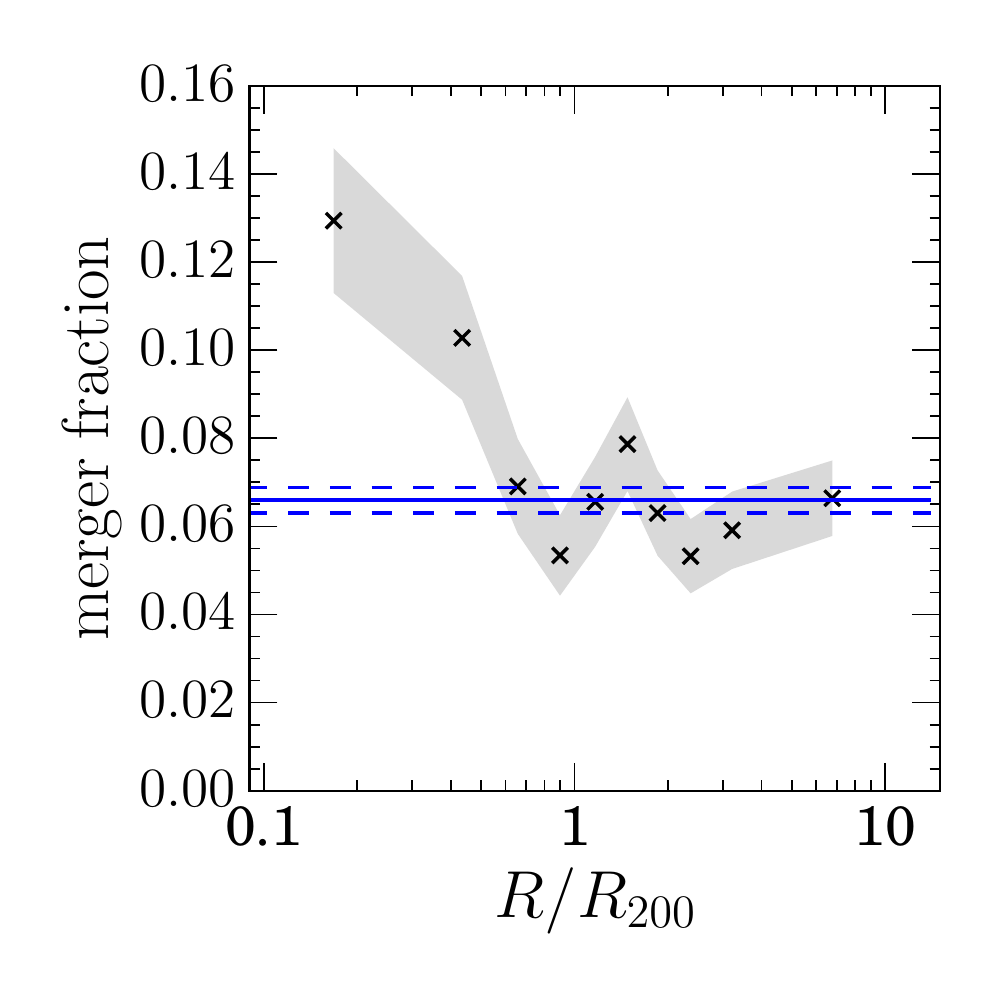}}
\caption[Merger fraction with group radius in the \textsc{gz2-group} sample]{Merger fraction in the \textsc{gz2-group} sample binned in projected group-centric radius, normalised by $R_{200}$, a proxy for the virial radius of a group. The shaded region shows $\pm1\sigma$, the propagated counting error, on the merger fraction. The merger fraction of the \textsc{gz2-sat-field} sample is also shown (blue solid line) with $\pm1\sigma$ (blue dashed line). The points are plotted at the linear centre of each bin at $R/R_{200}~=~ [0.17,  0.43,  0.66,  0.9 ,  1.17,  1.48,  1.85,  2.36,  3.22,  6.75]$.}
\label{fig:mergerradius}
\end{figure}

We utilise the GZ2 vote fractions to quantify the morphology of galaxies in the \textsc{gz2-group} sample, in order to investigate the morphological trends with group radius. We utilise $p_{\rm{disc}}$ and $p_{\rm{smooth}}$ to characterise the likelihood of galaxies being either discs or early-types. We also use vote fractions from further down the GZ2 decision tree including $p_{\rm{bar}}$, $p_{\rm{bulge}}$ and $p_{\rm{merger}}$ to calculate the bar, bulge and merger fractions in the \textsc{gz2-group} sample respectively. 

Fractions are calculated considering the number of barred (with $p_{\rm{bar}} > 0.5$; see \citealt{masters11a, Cheung13}) and bulged (with $p_{\rm{obvious}~\rm{or}~\rm{dominant}} > 0.5$ and $p_{\rm{none}~\rm{or}~\rm{noticeable}} > 0.5$) galaxies over the number of disc galaxies ($p_{\rm{disc}} > 0.43$, $p_{\rm{edge\_on, no}} > 0.715$, $N_{\rm{edge\_on, no}} > 20$; see Table 3 of \citealt{GZ2} for appropriate thresholds on the GZ2 vote fractions to select a sample of galaxies with a particular morphology) in the \textsc{gz2-group} satellite sample. The merger fraction considers the number of merging galaxies (with $p_{\rm{merger}} > 0.223$ and $N_{\rm{odd, yes}}\geq10$; see Table 3 of \citealt{GZ2}) over the number of galaxies in the \textsc{gz2-group} satellite sample. 

\begin{figure*}
\centering{
\includegraphics[width=0.95\textwidth]{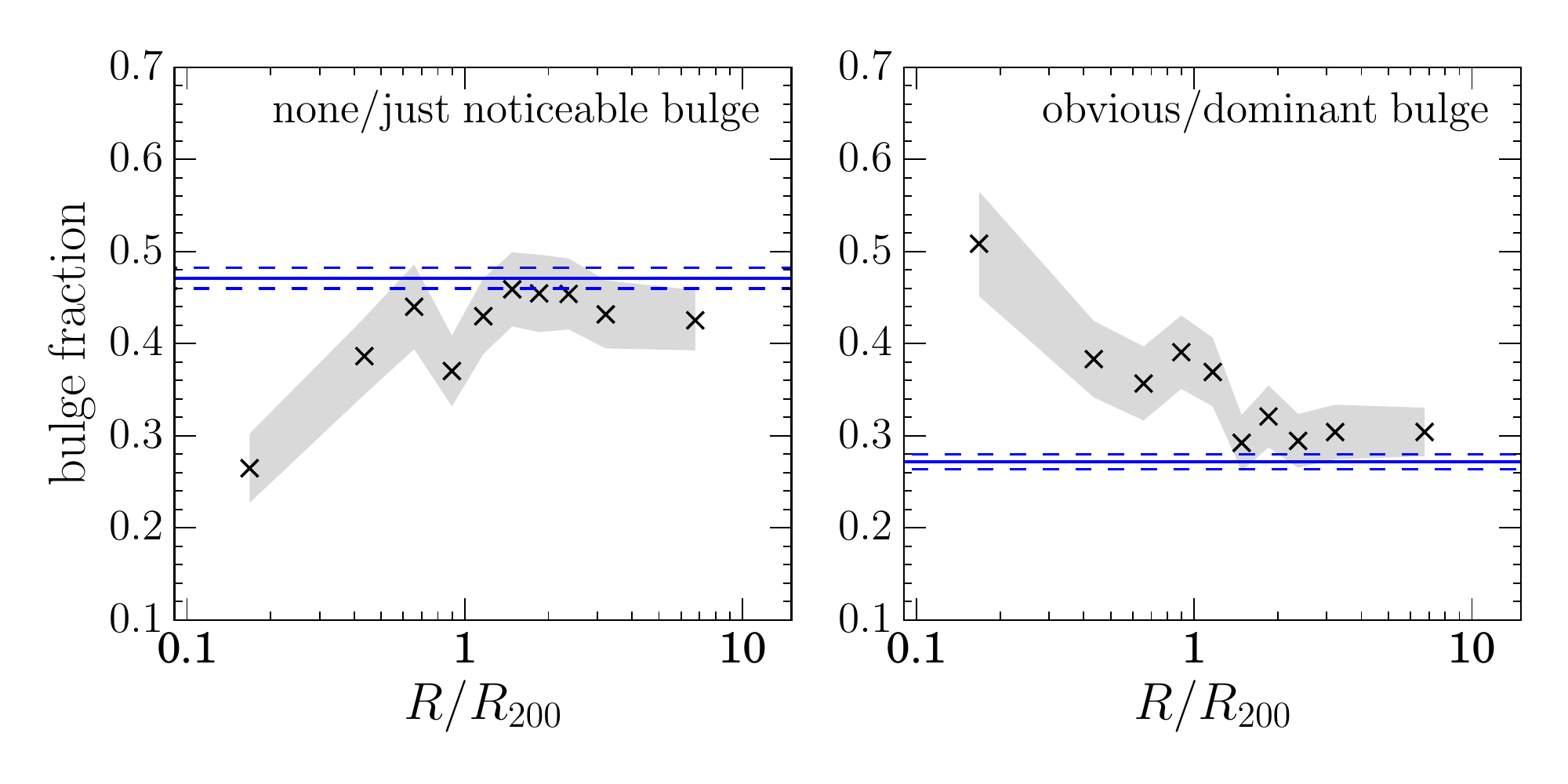}}
\caption[Bulge fraction with group radius in the \textsc{gz2-group} sample]{Fraction of galaxies with none/just noticeable bulge classifications (left) and with obvious/dominant bulge classifications (right) in the \textsc{gz2-group} sample binned in projected group-centric radius, normalised by $R_{200}$, a proxy for the virial radius of a group. The shaded regions shows $\pm1\sigma$, the propagated counting error, on the bulge fractions. The bulge fractions of the \textsc{gz2-sat-field} sample are also shown (blue solid lines) with $\pm1\sigma$ (blue dashed lines). The points are plotted at the linear centre of each bin at $R/R_{200}~=~ [0.17,  0.43,  0.66,  0.9 ,  1.17,  1.48,  1.85,  2.36,  3.22,  6.75]$. Note that in this study we use the GZ2 debiased vote fractions \citep[adjusted for classification bias, see][]{GZ2}, therefore $p_{\rm{obvious}}~+~p_{\rm{dominant}}~+~p_{\rm{just}~\rm{noticeable}}~+~p_{\rm{none}}~\neq~1$.}
\label{fig:bulgeradius}
\end{figure*}

Since morphological features have been shown to be dependent on the stellar mass of a galaxy \citep[e.g. the increase in the bar fraction with stellar mass; see][]{nair10, skibba12}, before investigating trends in the morphology with group radius in the \textsc{gz2-group} sample, the mass dependence on the group radius must be considered. This is shown in Figure~\ref{fig:massdep}. The median stellar mass is consistent with the median field value, within the uncertainties, until the two inner bins at $R/R_{200}~=~[0.17, 0.43]$. The median stellar mass increases with decreasing group-centric radius, with a difference of $\sim0.1~\rm{dex}$ between the inner and outer radius bins. However, given that this change in the median stellar mass is within the average error on the measured stellar mass values in the \textsc{gz2-group} sample, $\sigma_{\log_{10}[M_*/M_{\odot}]}\sim 0.09$, we can assume the stellar masses of the \textsc{gz2-group} satellites are independent of projected group radius. Therefore any morphological trends we observe with projected group radius are mass independent.

\subsection{Results}

We perform an initial sanity check on the \textsc{gz2-group} sample by recreating the morphology-density relation of \citet{dressler80} in Figure \ref{fig:morphradius}, which shows the mean disc and smooth vote fractions as a function of group radius. The mean disc vote fraction decreases from the mean field value (blue line) with decreasing group-centric radius. Simultaneously, the mean smooth vote fraction increases, which is in agreement with previous studies on the morphology-density relation \citep{dressler80, smail97, poggianti99, postman05, Bamford09}. The extensive morphological classifications provided by GZ2 also allow for the investigation of how more detailed morphological structure is affected by the group environment.  

Figure~\ref{fig:barradius} shows how the bar fraction (number of barred disc galaxies over the number of disc galaxies; see Section~\ref{sec:morphfrac}) increases significantly over the field fraction (blue solid line) with decreasing group-centric radius, in agreement with the findings of \cite{barazza09}, despite their alternate method for identifying bars using surface brightness profile fits. 

In Figure~\ref{fig:mergerradius} we show how the merger fraction does not significantly deviate from the field fraction (blue solid line) except for galaxies found at $< 0.5~R_{200}$. As discussed in Section~\ref{sec:mergersquench}, mergers are thought to drive bulge growth and so similarly, Figure~\ref{fig:bulgeradius} shows how the fraction of galaxies with obvious/dominant bulges increases over the field value at $< 1~R_{200}$, in the inner regions of the group (in agreement with \citealt{diaferio01}) and the fraction of those with none/just noticeable bulges decreases below the field value at $< 1~R_{200}$.

\section{Quenching Histories in the Group Environment}\label{sec:starpy}

\textsc{starpy}\footnote{Publicly available: \url{http://github.com/zooniverse/starpy/}} is a \textsc{python} code which allows the user to derive the quenching star formation history (SFH) of a single galaxy through a Bayesian Markov Chain Monte Carlo method \citep{emcee13}\footnote{\url{http://dan.iel.fm/emcee/}} with the input of the observed $u-r$ and $NUV-u$ colours\footnote{Spectral indicators of star formation are not used in this study, since the SDSS fibre is a set size and will suffer from aperture bias. Spectral SFR indicators will therefore over- or under-estimate the global average SFR of a galaxy.}, a redshift, and the use of the stellar population models of \cite{BC03}. These models are implemented using solar metallicity (varying this does not substantially affect these results; \citealt{smethurst15}) and a Chabrier IMF \citep{chabrier03} but do not model for intrinsic dust. The SFH is modelled as an exponentially declining SFR described by two parameters $[t_{q}, \tau]$, where $t_{q}$ is the time at the onset of quenching $\rm{[Gyr]}$ and $\tau$ is the exponential rate at which quenching occurs $\rm{[Gyr]}$. Under the simplifying assumption that all galaxies formed at $t=0$ $\rm{ Gyr}$ with an initial burst of star formation, the SFH can be described as: 
\begin{equation}\label{sfh}
SFR =
\begin{cases}
i_{sfr}(t_{q}) & \text{if } t < t_{q} \\
i_{sfr}(t_{q}) \times exp{\left( \frac{-(t-t_{q})}{\tau}\right)} & \text{if } t > t_{q} 
\end{cases}
\end{equation}
where $i_{sfr}$ is the constant star formation rate (SFR) defined so that at the time of quenching, $t_{q}$, the modelled galaxy resides on the SFS. We use the definition of the SFS from \cite{peng10} for a galaxy with stellar mass, $m = 10^{10.27}~M_{\odot}$ (the mean mass of the \textsc{gz2-galex} sample) at the redshift of the observed galaxy.  A smaller $\tau$ value corresponds to a rapid quench, whereas a larger $\tau$ value corresponds to a slower quench. 

We note that a galaxy undergoing a slow quench is not necessarily quiescent by the time of observation. Similarly, despite a rapid quenching rate, star formation in a galaxy may still be ongoing at very low rates, rather than being fully quenched. This SFH model has previously been shown to appropriately characterise populations of quenching or quiescent galaxies \citep{Weiner06, Martin07, Noeske07,schawinski14} which make up the \textsc{gz2-group-q} sample selected to lie $1\sigma$ below the SFS. We note also that star forming galaxies in this regime are fit by a constant SFR with a $t_{q} \simeq$ Age$(z)$, (i.e. the age of the Universe at the galaxy's observed redshift) with a very low probability.


The probabilistic fitting method used to determine the star formation history for an observed galaxy is described in full detail in Section 3.2 of \cite{smethurst15}, wherein the \textsc{starpy} code was used to characterise the SFHs of each galaxy in the \textsc{gz2-galex} sample. We assume a flat prior on all the model parameters and the difference between the observed and predicted $u-r$ and $NUV-u$ colours are modelled as independent realisations of a double Gaussian likelihood function (Equation 2 in \citealt{smethurst15}). We also make the simplifying assumption that the age of each galaxy, $t_\mathrm{age}$ corresponds to the age of the Universe at its observed redshift, $t_\mathrm{obs}$. {\minor\cite{smethurst15} tested the robustness of \textsc{starpy} with 25 synthesised galaxies, and found that the median differences in the inferred and known values of $[t_q, \tau]$ for these synthetic galaxies were $\sim[1.3, 0.5]~\rm{Gyr}$.}

An example posterior probability distribution output by \textsc{starpy} is shown for a single galaxy in Figure 5 of \cite{smethurst15}, wherein the degeneracies of the SFH model {\minor between recent, rapid quenching and earlier, slower quenching can clearly be seen}. These degeneracies are present for all galaxies run through \textsc{starpy} therefore if differences in the distributions arise when comparing two galaxies (or two populations), this is due to intrinsic differences in their SFHs and not due to the degeneracies of the model. 

We note that galaxy colours were not corrected for intrinsic dust attenuation. However,  \citet[][see Section~2.2]{smethurst16} showed that internal galactic extinction does not systematically bias the results from \textsc{starpy}; {\minor their population studies of both AGN host and inactive galaxies were consistent when comparing their sub-samples of edge-on and face-on galaxies}. 

The SFHs of all galaxies in both the \textsc{gz2-group-q} and \textsc{gz2-cent-field-q} samples were analysed using \textsc{starpy}, providing the posterior probability distribution across the two-parameter space for each individual galaxy. In \cite{smethurst15} and \cite{smethurst16} the individual SFHs of the entire \textsc{gz2-galex} sample and those hosting Type 2 AGN, respectively, were combined and weighted to give an overall distribution of the quenching parameters within a population of galaxies. 

However, in this study we adopt a more statistically rigorous method by taking the median value of an individual posterior probability distribution (i.e. the 50th percentile position of the MCMC chain, with the $\pm1\sigma$ derived from the 16th and 84th percentile positions, see Section 3.2 of \citealt{smethurst15}) to give the most likely quenching time, $t_{q}$, and quenching rate, $\tau$, for each galaxy. This allows us to investigate the trends in the quenching parameters with projected group centric radius. 

This simplifies the output from \textsc{starpy} for each galaxy from a probability distribution to just two values, with $\pm1\sigma$ uncertainties, which encompass the spread of the individual galaxy's SFH posterior probability distribution. We then calculate the time since quenching onset, $\Delta t$, for a given galaxy by calculating {\bf $\Delta t = t^\mathrm{obs} - t_{q}$} (where $t^{\rm{obs}}$ is the age of the Universe at a galaxy's observed redshift).

\begin{figure*}
\centering{
\includegraphics[width=0.43\textwidth]{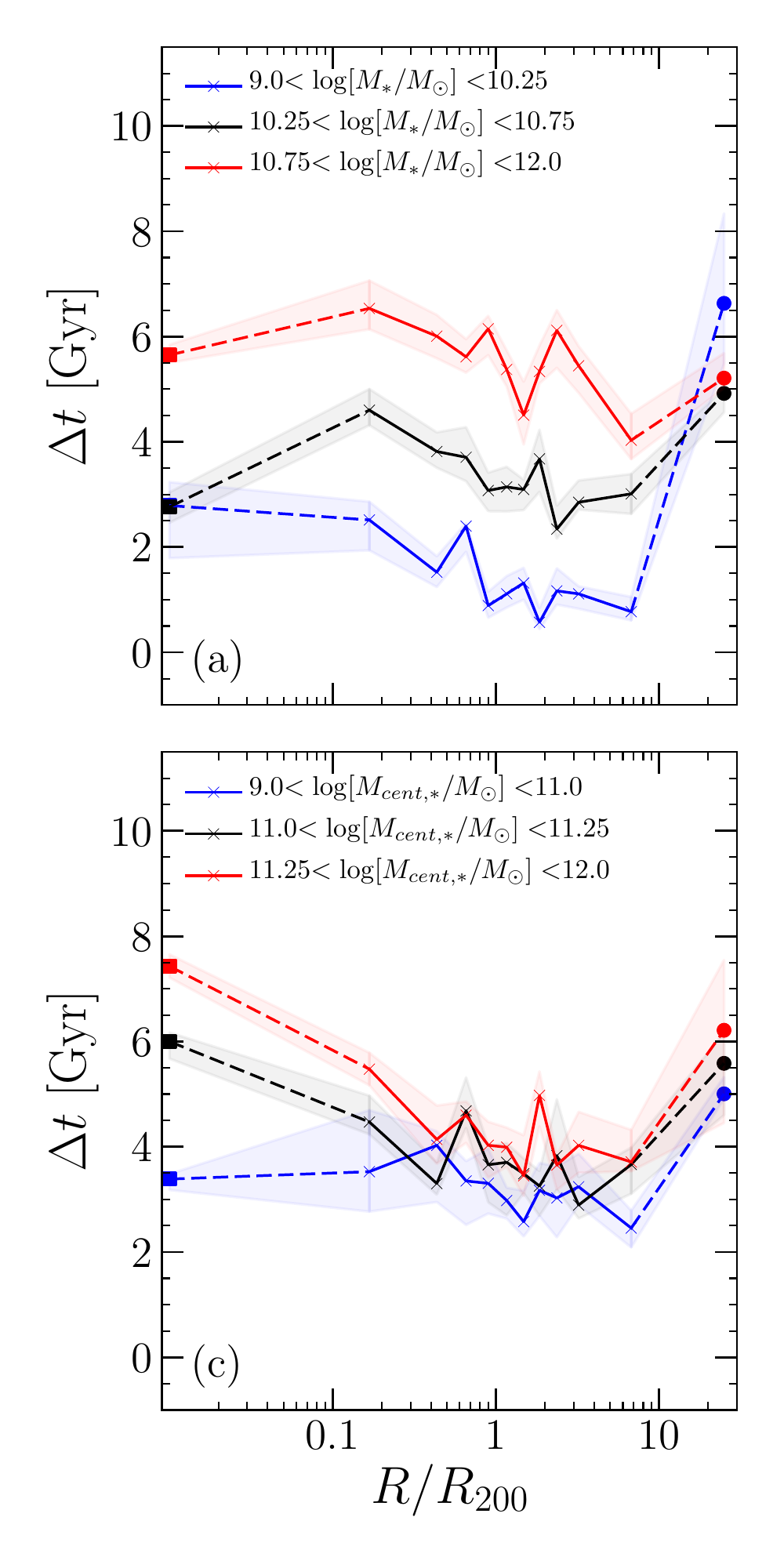}
\includegraphics[width=0.43\textwidth]{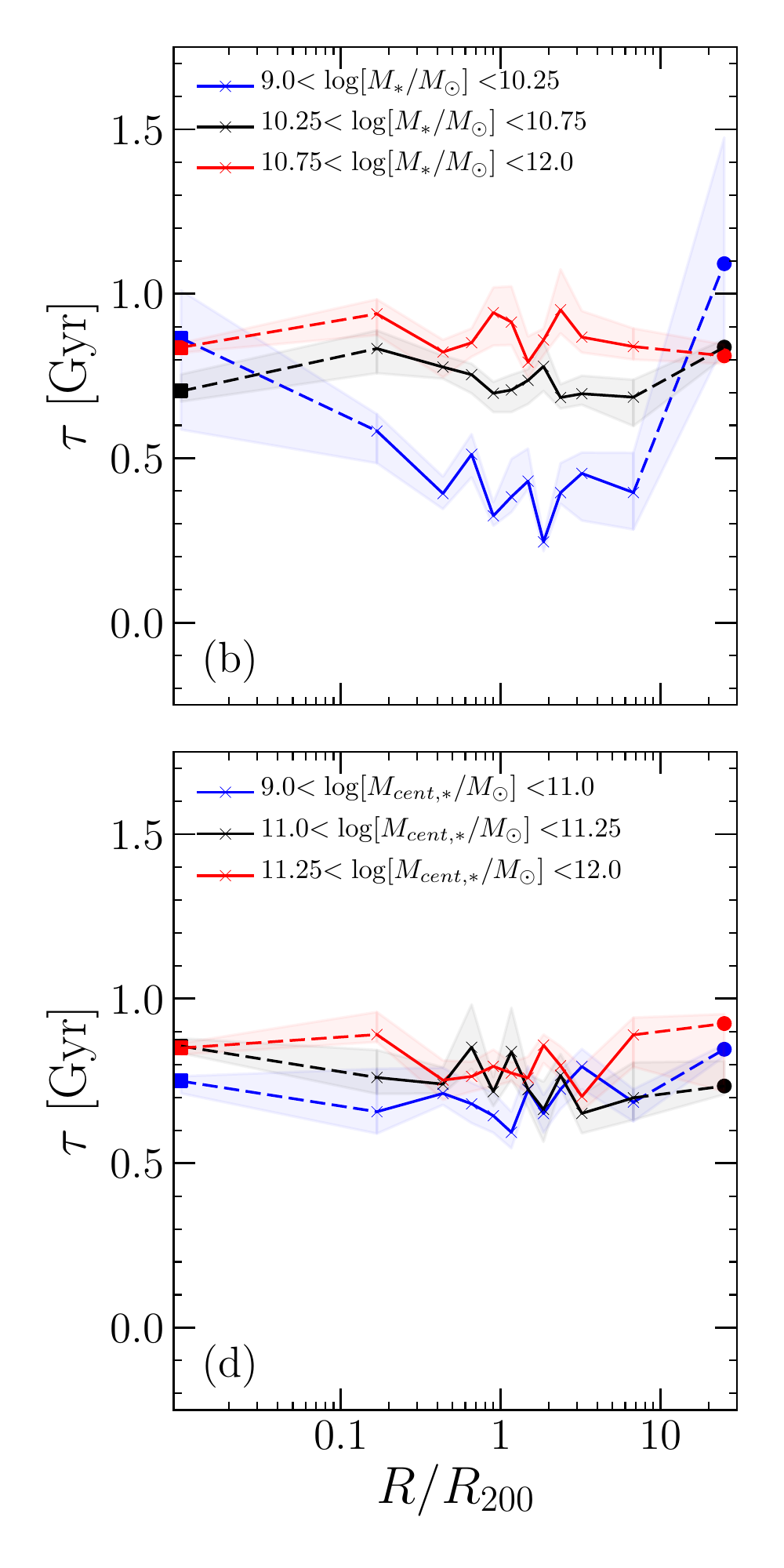}
\caption[Trend of $\Delta t$ and $\tau$ with group radius split by stellar mass and halo mass]{The bootstrapped median time since quenching onset ($\Delta t = t_{obs} - t_{q}$; left) and rate of quenching ($\tau$; right) binned in group radius, $R/R_{200}$, for \textsc{gz2-group-q} satellite galaxies (crosses) split into bins of stellar mass (top) and stellar mass of the corresponding central galaxy (bottom; a proxy for halo mass of a group). The corresponding values for central galaxies (squares, plotted at $\sim0.01 R/R_{200}$) and galaxies in the \textsc{gz2-cent-field-q} sample (circles, plotted at $25 R/R_{200}$) are shown and connected by the dashed lines to help guide the eye. The shaded regions show the $\pm1\sigma$ confidence region on the distribution of $1000$ bootstrapped median $\Delta t$ and $\tau$ values in each bin of $R/R_{200}$. Note that the median uncertainties for an individual galaxy are $\Delta t\pm_{2.6}^{2.0}$ and $\tau\pm_{0.6}^{0.5}$. The points are plotted at the linear centre of each bin at $R/R_{200}~=~ [0.17,  0.43,  0.66,  0.9 ,  1.17,  1.48,  1.85,  2.36,  3.22,  6.75]$, which were chosen to give a flat distribution for the entire \textsc{gz2-group-q} sample.}
\label{fig:timesinceradius}}
\end{figure*}

\begin{figure*}
\centering{
\includegraphics[width=0.43\textwidth]{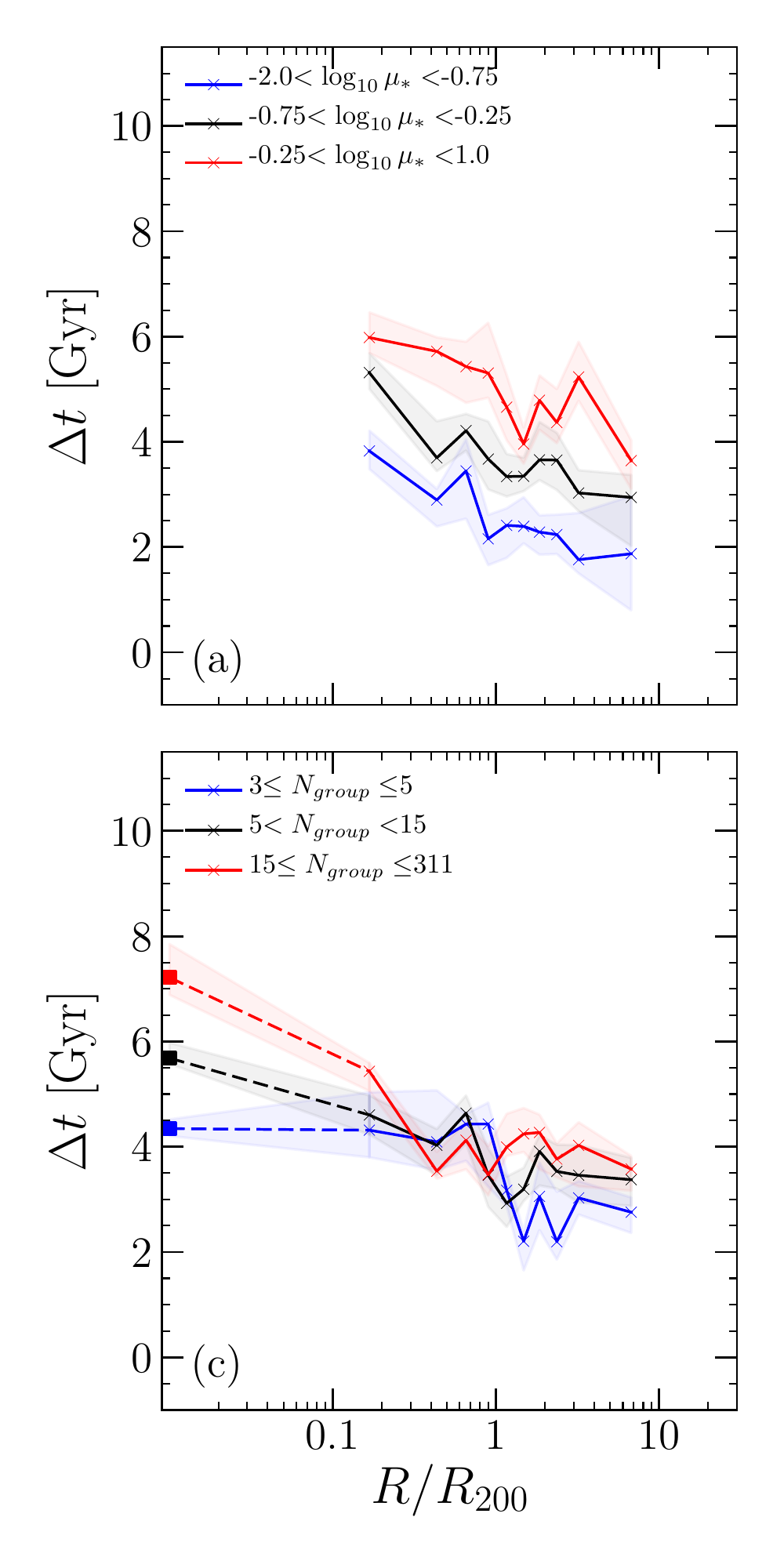}
\includegraphics[width=0.43\textwidth]{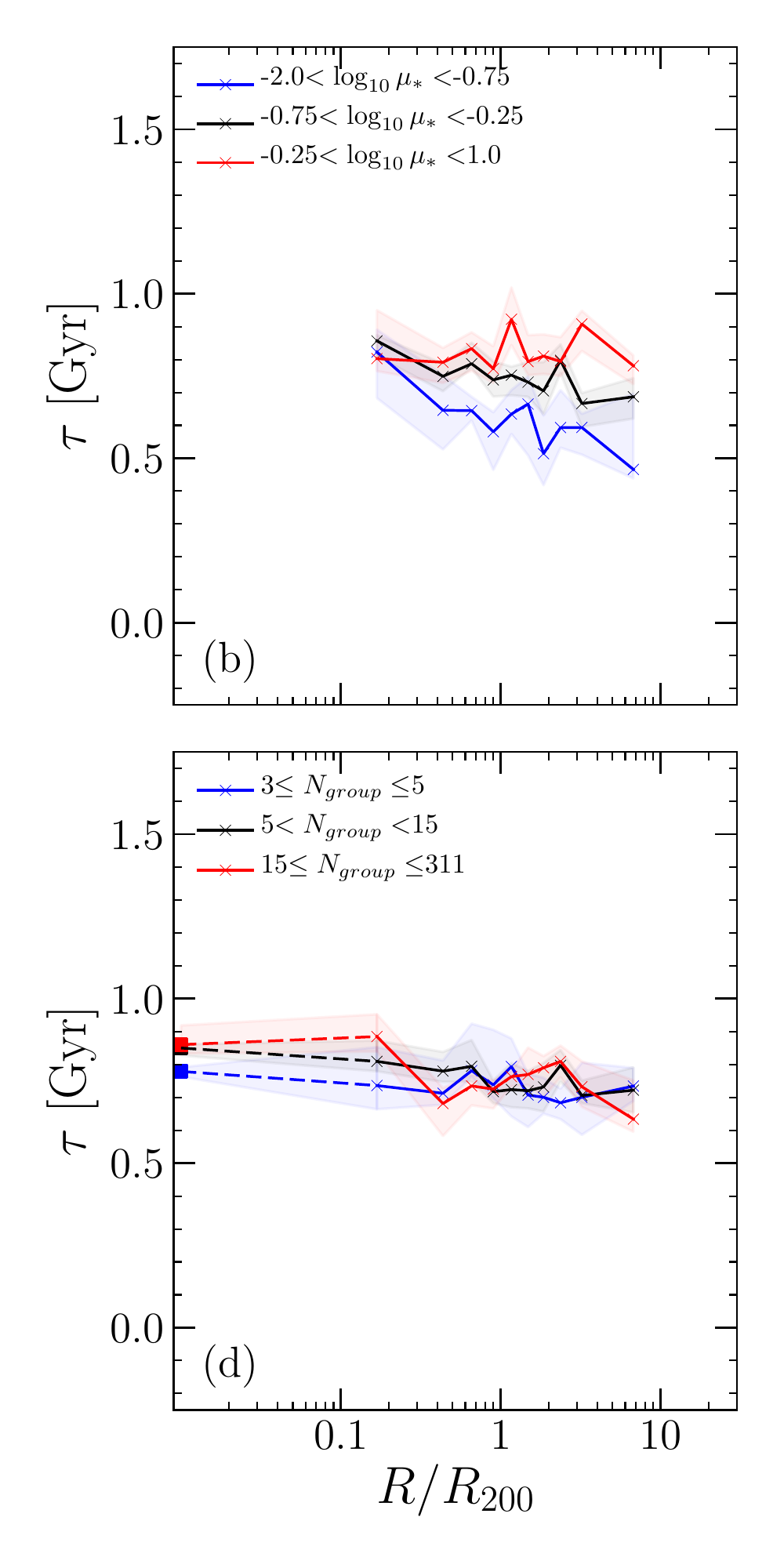}
\caption[Trend of $\Delta t$ and $\tau$ with group radius split by number in group and stellar mass ratio]{The bootstrapped median time since quenching onset ($\Delta t = t_{obs} - t_{q}$) and rate of quenching ($\tau$; right) binned in group radius, $R/R_{200}$, for \textsc{gz2-group-q} satellite galaxies (crosses) split into bins of stellar mass ratio ($\mu_* = M_*/M_{cent,*}$, top) and number of group members ($N_{group}$, bottom). The corresponding values for central galaxies (squares, plotted at $\sim0.01 R/R_{200}$) are shown, where possible, and connected by the dashed lines to help guide the eye. The shaded regions show the $\pm1\sigma$ confidence region on the distribution of $1000$ bootstrapped median $\Delta t$ and $\tau$ values in each bin of $R/R_{200}$. Note that the median uncertainties for an individual galaxy are $\Delta t\pm_{2.6}^{2.0}$ and $\tau\pm_{0.6}^{0.5}$. The points are plotted at the linear centre of each bin at $R/R_{200}~=~ [0.17,  0.43,  0.66,  0.9 ,  1.17,  1.48,  1.85,  2.36,  3.22,  6.75]$, which were chosen to give a flat distribution for the entire \textsc{gz2-group-q} sample.}
\label{fig:timesinceradiusmu}}
\end{figure*}

\begin{figure*}
\centering{
\includegraphics[width=0.43\textwidth]{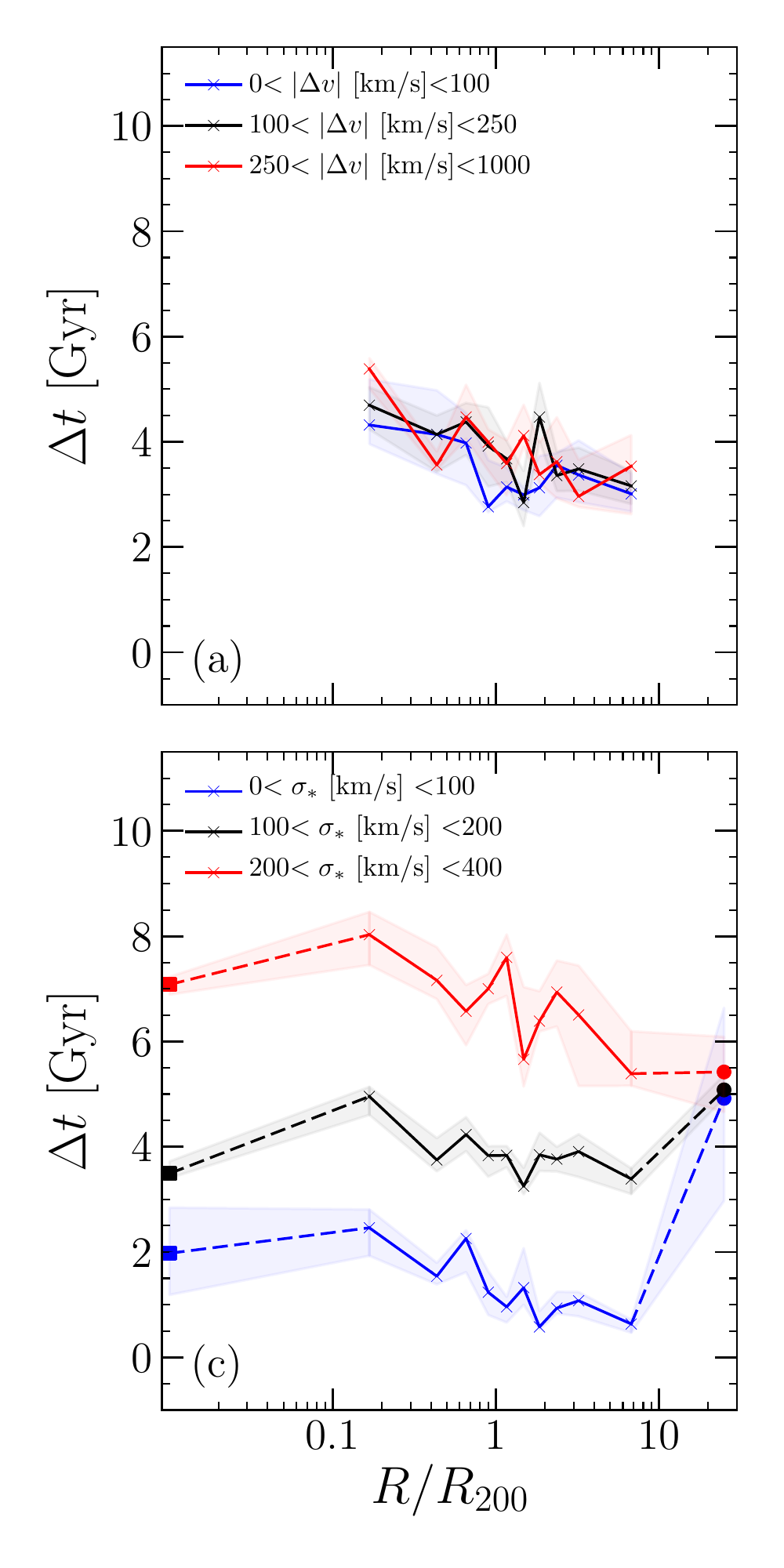}
\includegraphics[width=0.43\textwidth]{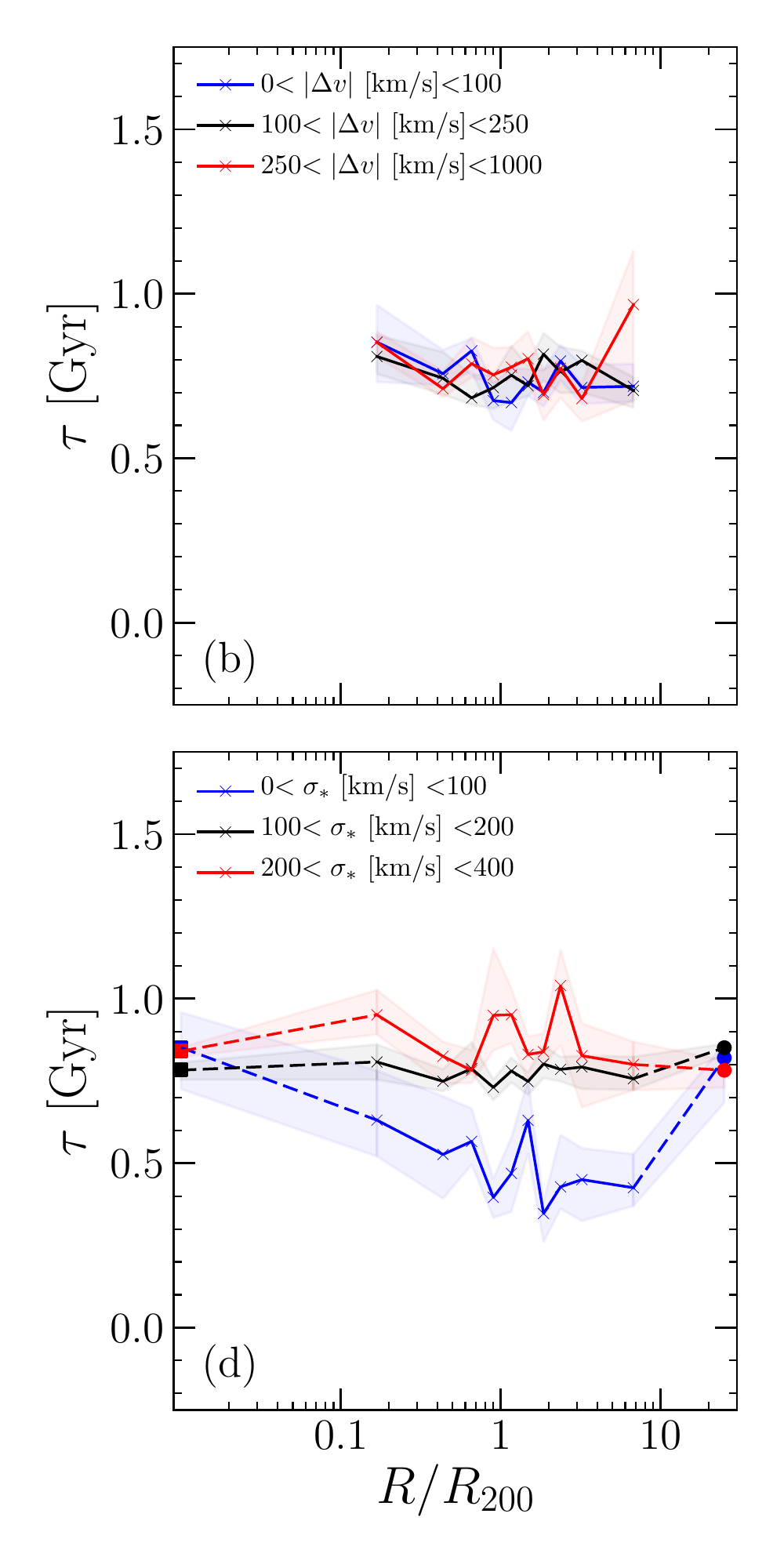}
\caption[Trend of $\Delta t$ and $\tau$ with group radius split by relative velocity and stellar velocity dispersion]{The bootstrapped median time since quenching onset ($\Delta t = t_{obs} - t_{q}$; left) and rate of quenching ($\tau$; right) binned in group radius, $R/R_{200}$, for \textsc{gz2-group-q} satellite galaxies (crosses) split by the absolute relative velocity of the satellite to its central galaxy ($|\Delta v|$, top) and stellar velocity dispersion ($\sigma_*$, bottom). The corresponding values for central galaxies (squares, plotted at $\sim0.01 R/R_{200}$) and galaxies in the \textsc{gz2-cent-field-q} sample (circles, plotted at $25 R/R_{200}$) are shown, where possible, and connected to the satellite values by the dashed lines to help guide the eye. The shaded regions show the $\pm1\sigma$ confidence region on the distribution of $1000$ bootstrapped median $\Delta t$ and $\tau$ values in each bin of $R/R_{200}$. Note that the median uncertainties for an individual galaxy are $\Delta t\pm_{2.6}^{2.0}$ and $\tau\pm_{0.6}^{0.5}$. The points are plotted at the linear centre of each bin at $R/R_{200}~=~ [0.17,  0.43,  0.66,  0.9 ,  1.17,  1.48,  1.85,  2.36,  3.22,  6.75]$, which were chosen to give a flat distribution for the entire \textsc{gz2-group-q} sample. Note that $\sigma_*$ values derived below the SDSS instrument dispersion of $70~\rm{km}~\rm{s}^{-1}$ are assumed to be upper limits (see Section~\ref{sec:groups}).}
\label{fig:timesinceradiusvel}}
\end{figure*}

\subsection{Results}\label{sec:resultssfhs}

With the output from \textsc{starpy} we can now observe the trends in the time since quenching onset, $\Delta t$, and quenching rate, $\tau$, with group radius, $R/R_{200}$, for satellite galaxies and central galaxies in the \textsc{gz2-group-q} sample, compared with galaxies in the \textsc{gz2-cent-field-q} sample. This is shown in Figures~\ref{fig:timesinceradius} - \ref{fig:timesinceradiusvel} wherein the \textsc{gz2-group-q} galaxies are binned by stellar mass (Figures~\ref{fig:timesinceradius}a-b), a proxy for halo mass (Figures~\ref{fig:timesinceradius}c-d), mass ratio (Figures~\ref{fig:timesinceradiusmu}a-b), number of group galaxies (Figures~\ref{fig:timesinceradiusmu}c-d), relative velocity (Figures~\ref{fig:timesinceradiusvel}a-b) and stellar velocity dispersion (Figures~\ref{fig:timesinceradiusvel}c-d). The bins for each of the group or galaxy parameters studied were chosen to give approximately the same number of satellite galaxies in the three bins in each case. For example, $34.7\%$ of the satellite galaxies have $3 \leq N_{group} \leq 5$, $31.7\%$ have $5 < N_{group} < 15$ and $33.6\%$ have $15 \leq N_{group} \leq 311$. 

\begin{table*}

\centering
\caption{Linear regression fits are performed on the black curves (middle bin) shown in Figures~\ref{fig:timesinceradius}-\ref{fig:timesinceradiusvel} to quantify the trends in $\Delta t$ and $\tau$ with $R/R_{200}$. The table states the median value of the posterior distribution of the inferred slope (along with $\pm1\sigma$) when fitted to both the $\Delta t$ and $\tau$ variables when the \textsc{gz2-group-q} satellite galaxies are split by the stated property (leftmost column). For clarity we state only the slopes fitted to the middle bin (shown by the black curves in Figures~\ref{fig:timesinceradius}-\ref{fig:timesinceradiusvel}), however for a given group or galaxy property the derived slopes are consistent, within the uncertainties, between the lower, middle and upper bin curves (blue, black and red curves in Figures~\ref{fig:timesinceradius}-\ref{fig:timesinceradiusvel}). {\minor We also calculate the ratio between the Gaussian likelihood, $\mathcal{L}_{\rm{flat}}$ for a flat line model and that of the linear regression model, $\mathcal{L}_{\rm{slope}}$ to quantify how likely it is that a trend is present (see Section~\ref{sec:resultssfhs}). As $\mathcal{L}_{\rm{slope}}/\mathcal{L}_{\rm{flat}}$ tends to unity, the two models become equally likely. All slope values are quoted to 2 decimal places and all likelihood ratios to 2 significant figures.}}
\label{table:resultsum}
\newcolumntype{R}{>{\raggedleft\arraybackslash}X}%
\newcolumntype{L}{>{\raggedright\arraybackslash}X}%
\newcolumntype{C}{>{\centering\arraybackslash}X}%
\setlength\extrarowheight{3pt}
\begin{tabularx}{\textwidth}{r|C|C|C|C|C}

\hline
Curve fitted & Figure & Slope in $\Delta t$ with $R/R_{200}$  & $\mathcal{L}_{\rm{slope}}/\mathcal{L}_{\rm{flat}}$ in  $\Delta t$ with $R/R_{200}$   & Slope in $\tau$ with $R/R_{200}$  & $\mathcal{L}_{\rm{slope}}/\mathcal{L}_{\rm{flat}}$ in $\tau$ with $R/R_{200}$       \\ \hline 
$10.25 < \log_{10}[M_*/M_{\odot}] < 10.75$                 & \ref{fig:timesinceradius}a,b           & $-1.18\pm_{0.38}^{0.41}$ & 49000 & $-0.09\pm_{0.05}^{0.05}$ & 35 \\
$11.0 < \log_{10}[M_{\rm{cent},*}/M_{\odot}] < 11.25$     & \ref{fig:timesinceradius}c,d           & $-0.61\pm_{0.48}^{0.48}$ & 6.2 & $-0.05\pm_{0.07}^{0.06}$ & 1.7 \\
$-0.75 < \log_{10}\mu_* < -0.25$               & \ref{fig:timesinceradiusmu}a,b         & $-1.46\pm_{0.42}^{0.44}$ & 390000 & $-0.10\pm_{0.04}^{0.04}$ & 73 \\
$5 < N_{\rm{group}}< 15$      & \ref{fig:timesinceradiusmu}c,d         & $-0.86\pm_{0.44}^{0.44}$ & 87 & $-0.06\pm_{0.05}^{0.05}$ & 4.2 \\
$100 < |\Delta v|~[\rm{km/s}] < 250$          & \ref{fig:timesinceradiusvel}a,b        & $-0.91\pm_{0.43}^{0.41}$ & 240& $-0.01\pm_{0.06}^{0.06}$ & 1.1 \\
$100 < \sigma_*~[\rm{km/s}] < 200$            & \ref{fig:timesinceradiusvel}c,d        & $-0.76\pm_{0.32}^{0.32}$ & 3300 & $-0.01\pm_{0.04}^{0.04}$ & 0.99 \\ \hline
\end{tabularx}
\end{table*}

We quantify the trends observed with $R/R_{200}$ seen across Figures ~\ref{fig:timesinceradius} - \ref{fig:timesinceradiusvel} by performing a linear regression fit to the middle bin in each figure panel (shown by the black lines), with the uncertainty on the median value in each $R/R_{200}$ bin represented by the error on the median. The fit was performed using the \textsc{linimx} module\footnote{\url{http://linmix.readthedocs.io/}}, the method for which is outlined \citealt{kelly07}. The derived slopes in $R/R_{200}$ for both the $\Delta t$ and $\tau$ variables, when the \textsc{gz2-group-q} satellite galaxies are split by each investigated galaxy or group property, are stated in Table~\ref{table:resultsum}. {\minor In order to quantify how likely it is that a trend is present when the sample is split by each property, we also calculate the ratio of Gaussian likelihoods, $\mathcal{L}$, for the linear regression model, $\mathcal{L}_{\rm{slope}}$, and for a flat line model with no slope, $\mathcal{L}_{\rm{flat}}$. Table~\ref{table:resultsum} therefore also states $\mathcal{L}_{\rm{slope}}/\mathcal{L}_{\rm{flat}}$, which provides an estimate for how likely it is that a trend is present.}

A general trend for increasing time since quenching onset, $\Delta t$ with decreasing projected group-centric radius, $R/R_{200}$, can be seen in the left panels of Figures~\ref{fig:timesinceradius} - \ref{fig:timesinceradiusvel}, supported by the fitted slopes stated in Table~\ref{table:resultsum} which range from $-1.46 < m < -0.61$. Differences from the mean field $\Delta t$ values arise at $<1~R_{200}$, similar to the results found for the environmental dependence of the morphological structure (Figures \ref{fig:morphradius}$-$\ref{fig:bulgeradius}). However, little or no trend with group radius is seen for the rate at which quenching occurs for satellites in the \textsc{gz2-group-q} sample (right panels Figures~\ref{fig:timesinceradius} - \ref{fig:timesinceradiusvel}), supported by the fitted slopes stated in Table~\ref{table:resultsum} which range from $-0.10 < m < -0.01$. This suggests that whatever mechanisms cause quenching in a group will do so at the same rate in both the dense inner and sparse outer regions. {\minor This result contradicts the results of \citet{rettura10, rettura11, ferre14} and \cite{darvish16} who find that galaxies in denser environments quench more rapidly than those in less dense environments.}

We first split the \textsc{gz2-group-q} sample by stellar mass, $M_*$, and a clear trend for increasing $\Delta t$ with increasing stellar mass for satellite, central and field galaxies can be seen (Figure~\ref{fig:timesinceradius}a). However, this trend is less apparent for the rate that quenching occurs (Figure~\ref{fig:timesinceradius}b). The central galaxies (shown by the square points) with $M_* > 10.25$ have quenched more recently than the inner satellites (plotted at $\sim0.17~R/R_{200}$) of the same mass but have done so at a similar quenching rate. The low stellar mass satellite galaxies have also quenched more recently and rapidly ($\tau\sim0.4~\rm{Gyr}$) than low stellar mass field galaxies ($\tau\sim1.1~\rm{Gyr}$), suggesting that an environmentally driven quenching mechanism could be responsible for such a rapid quench in these low stellar mass systems. 

We then split the \textsc{gz2-group-q} sample by halo mass by using the stellar mass of the corresponding central galaxy of a group, $M_{cent,*}$, as a proxy. Note that this results in a large proportion of low stellar mass satellites in the low $M_{cent,*}$ bin and a large range of satellite stellar masses in the high $M_{cent,*}$ bin, due to the definition of a central galaxy. We find a clear trend for increasing time since quenching onset with increasing halo mass for central galaxies (Figure~\ref{fig:timesinceradius}c) but this trend is less apparent within the uncertainties for satellite and field galaxies. There is also no trend for the rate of quenching with increasing halo mass for satellites (Figure~\ref{fig:timesinceradius}d) suggesting that the halo mass does not affect which quenching mechanism acts upon either central or satellite galaxies.

\begin{table*}
\centering
\caption{The results shown in Figures~\ref{fig:timesinceradius}-\ref{fig:timesinceradiusvel} are summarised by stating whether a trend with $\Delta t$ or $\tau$ is found ($\checkmark$), or not found ($\times$), for satellite galaxies for each of the galaxy or group properties investigated.}
\label{table:summary}
\newcolumntype{R}{>{\raggedleft\arraybackslash}X}%
\newcolumntype{L}{>{\raggedright\arraybackslash}X}%
\newcolumntype{C}{>{\centering\arraybackslash}X}%
\setlength\extrarowheight{3pt}
\begin{tabularx}{\textwidth}{r|C|C|C}

\hline
Property & Shown in Figure & Trend with $\Delta t$ & Trend with $\tau$        \\ \hline
$M_*$                 & \ref{fig:timesinceradius}a,b            & $\checkmark$           & $\checkmark$       \\
$M_{\rm{cent},*}$     & \ref{fig:timesinceradius}c,d            & $\times$               & $\times$           \\
$\mu_*$               & \ref{fig:timesinceradiusmu}a,b           & $\checkmark$           & $\times$           \\
$N_{\rm{group}}$      & \ref{fig:timesinceradiusmu}c,d           & $\times$               & $\times$           \\
$|\Delta v|$          & \ref{fig:timesinceradiusvel}a,b           & $\times$               & $\times$           \\
$\sigma_*$            & \ref{fig:timesinceradiusvel}c,d           & $\checkmark$           & $\checkmark$       \\ \hline

\end{tabularx}
\end{table*}

To account for the effects of conformity, whereby satellites of higher mass tend to be found in higher mass halos \citep{weinmann06, kauffmann13, hearin15, hatfield16}, we also split the satellites of the \textsc{gz2-group-q} sample by the stellar mass ratio of the satellite to its central galaxy, $\mu_* = M_*/M_{cent,*}$, again using the mass of the central as a proxy for halo mass. The time since quenching onset, $\Delta t $, increases steeply with group radius (particularly within $\sim$ one virial radius; Figure~\ref{fig:timesinceradiusmu}a) particularly for satellite galaxies with much smaller masses than their group central ($-2.0 < \log_{10}\mu_* < -0.25$, shown by the blue curve). This is confirmed by the slope derived in the linear regression fit stated in Table~\ref{table:resultsum}, $m=-1.46\pm_{0.42}^{0.44}$. Once again there is no trend for the rate that quenching occurs (Figure~\ref{fig:timesinceradiusmu}b, $m=-0.10\pm_{0.04}^{0.04}$).

Another property of the group which is expected to affect the satellite quenching histories is the number of group members, $N_{group}$, which should be roughly correlated with a satellite's local density in a  group\footnote{\minor{We cannot use the measure of local density from \cite{Bamford09}, who averaged $\log\Sigma_N$ for $N = 4$ and $N = 5$ (see Section~\ref{sec:groups}), as $\sim35\%$ of the satellites have $N_{\rm{group}} < 5$. This measure of the local environment density is therefore not appropriate for the smaller groups in the \textsc{gz2-group} sample.}}. We find that there is no trend with time since quenching onset (Figure~\ref{fig:timesinceradiusmu}c) or rate of quenching (Figure~\ref{fig:timesinceradiusmu}d) with increasing $N_{group}$ for satellite galaxies (however the general trend for increasing $\Delta t$ with $R/R_{200}$ is still apparent with a slope, $m=-0.86\pm_{0.44}^{0.44}$). The central galaxies (shown by the square points) however, do show a trend for increasing time since quenching as the number of group galaxies increases (Figure~\ref{fig:timesinceradiusmu}c), but the rate at which they quench is the same (Figure~\ref{fig:timesinceradiusmu}d) suggesting the mechanism by which this occurs is the same for all centrals regardless of halo mass. 

The \textsc{gz2-group-q} satellite galaxies are also split into bins of their relative velocity, $|\Delta v|$ to their central galaxies, i.e. the velocity at which they move through the dense group environment. There is no trend with either time since onset of quenching (Figure \ref{fig:timesinceradiusvel}a) or rate of quenching (Figure \ref{fig:timesinceradiusvel}b) with increasing relative velocity for galaxies in the \textsc{gz2-group-q} sample. In the highest relative velocity bin ($250 < |\Delta v| [\rm{km/s}] < 1000$) there is an increase in  the rate of quenching at larger projected group-centric radii (plotted at $R/R_{200}=6.75$). Although this is suggestive of the environment affecting such galaxies to a lesser extent, we  note that this feature disappears if only those galaxies inside the $(\Delta v/\sigma)\times(R/R_{200})~>~3$ caustic are used in the analysis (see Section~\ref{sec:groups}). This therefore suggests that whatever quenching mechanism is occurring in groups, it is not correlated with the velocity at which satellites move through the dense environment.

We also investigate the trend with projected group-centric radius for the \textsc{gz2-group-q} satellites when split into bins of galaxy stellar velocity dispersion, $\sigma_*$ (note that this is not the velocity dispersion of the group), which is often used as a proxy for the galaxy potential. The stellar velocity dispersion shows the largest trend in $\Delta t$ (Figure~\ref{fig:timesinceradiusvel}c) for satellite galaxies in comparison to the other properties investigated (shown in Figures~\ref{fig:timesinceradius}-\ref{fig:timesinceradiusvel}). Galaxies with low stellar velocity dispersions having quenched $\sim6~\rm{Gyr}$ more recently than those with high stellar velocity dispersion. This trend is less apparent for the rate that quenching occurs when the satellite galaxies are split by $\sigma_*$ (Figure~\ref{fig:timesinceradiusvel}d), however it is one of the largest trend seen in the rate of quenching for any of the properties investigated (shown in the right panels of Figures~\ref{fig:timesinceradius}-\ref{fig:timesinceradiusvel}), along with the stellar mass. This is not unexpected since both the stellar mass and stellar velocity dispersion will both trace the galaxy potential. Also, field galaxies (shown by the circles at $\sim 25 R/R_{200}$ in \ref{fig:timesinceradiusvel}d) with low velocity dispersions are seen to quench at much slower rates than their satellite counterparts ($\tau\sim0.9~\rm{Gyr}$ versus $\tau\sim0.4~\rm{Gyr}$). This suggests that the rapid quenching observed for the low stellar velocity dispersion satellites is directly caused by the environment. 

We summarise the results shown in Figures~\ref{fig:timesinceradius}-\ref{fig:timesinceradiusvel} in Table~\ref{table:summary}, stating whether a trend with $\Delta t$ or $\tau$ is seen for the satellites of the \textsc{gz2-group-q} sample when split by each of the group or galaxy properties investigated.

\section{Discussion}\label{sec:disc}

We shall now consider the results presented in Sections~\ref{sec:morphfrac} \& \ref{sec:starpy} in the context of the possible quenching mechanisms which could be responsible. We focus on those mechanisms first introduced in Sections~\ref{sec:intquench} \& \ref{sec:extquench}. 

\subsection{The role of mergers as quenching mechanisms in the group environment}\label{sec:rolemergerenv}

The merger classification in GZ has been shown to preferentially identify major mergers \citep{Darg10a, casteels13}; while bulge formation in disc galaxies is often associated with evolutionary histories driven by minor mergers \citep{Croton06, tonini16}.  Although we see evidence for an enhanced merger fraction in the inner regions of the group environment (Figure~\ref{fig:mergerradius}), the bulge fractions vary much more significantly from the field value than the merger fraction at $\sim1R/R_{200}$ (Figure~\ref{fig:bulgeradius}). This suggests that minor mergers may be more dominant than major mergers for satellites in the group environment, particularly at $R/R_{200} > 0.5$. 

If mergers are a dominant evolutionary mechanism for satellite galaxies (as the morphological evidence in Figures~\ref{fig:mergerradius} \& \ref{fig:bulgeradius} suggests) we would expect to see a difference in the quenching histories of satellites residing in groups with a larger number of members. However, there is no trend with time since quenching onset (Figure \ref{fig:timesinceradiusmu}c) or rate of quenching (Figure \ref{fig:timesinceradiusmu}d) with increasing $N_{group}$ for the satellite galaxies. This suggests that mergers are not the dominant quenching mechanism for satellite galaxies, but that whatever mechanism is the cause of the quenching occurs at the same rate irrespective of group size. 

Central galaxies however, do show a trend for increasing time since quenching with increasing $N_{group}$ (square points in Figure \ref{fig:timesinceradiusmu}c) occurring at a rate of $\tau \sim 0.8~\rm{Gyr}$. \cite{smethurst15} attributed these quenching rates to mergers and galaxy interactions which can transform a galaxy's morphology. Therefore, the larger the number of group members, the more likely a central galaxy has a history dominated by mergers. This is in agreement with the findings of \citet{lin10}, \citet{ellison10}, \citet{lidman13} and \citet{mcintosh08}. The latter found, by studying a sample of local groups and clusters, that half of the mergers they identified involved the central galaxy. \cite{liu09} also found that the fraction of merging centrals increases with the richness of a cluster (a measure of the number of galaxies within $1~\rm{h}^{-1}\rm{Mpc}$ of the central galaxy).

This idea is supported by the result that centrals of a given mass have quenched more recently than the inner satellites (at $\sim0.1R/R_{200}$ in Figure~\ref{fig:timesinceradius}a) of a given mass. This suggests that an episode of more recent star formation, such as a starburst, may have occurred in the central galaxies but not in the inner satellites. Mergers are thought to cause an energetic burst of star formation which can in turn quench the remnant galaxy \citep{hopkins05, treister12, pontzen16}. This result is supportive of a merger dominated history for central galaxies but not for satellite galaxies.

\subsection{The role of mass quenching in the group environment}\label{sec:rolemassenv}

A trend is seen for increasing time since quenching with increasing stellar mass and velocity dispersion (a proxy for galaxy potential) for centrals, satellites and field galaxies in Figure~\ref{fig:timesinceradius}a and Figure~\ref{fig:timesinceradiusvel}c respectively. {\minor These are the strongest trends observed across all panels of Figures~\ref{fig:timesinceradius}-\ref{fig:timesinceradiusvel}, suggesting that mass quenching is more dominant than environmentally driven quenching mechanisms in the group population.} This is suggestive of mass quenching occurring across the entire galaxy population irrespective of environmental density, supporting the work of \citet{peng10, peng12, Gabor10} and \citet{darvish16}.

\subsection{The role of morphological quenching in the group environment}\label{sec:rolemorphenv}

We find an increasing bar fraction toward the central group regions in agreement with \cite{skibba12} (Figure \ref{fig:barradius}). This increase coincides with an increase in time since quenching onset with projected group radius across the satellite galaxies of the \textsc{gz2-group-q} sample (Figures~\ref{fig:timesinceradius}-\ref{fig:timesinceradiusvel} and quantified in Table~\ref{table:resultsum}). This suggests that bars may be partly responsible for the relation between quenched fraction and environmental density. This is consistent with findings that show that bars themselves may be the cause of morphological quenching through the funnelling of gas toward the central regions of galaxies \citep{athanassoula92b, sheth05} which is then used in star formation, exhausting the available gas \citep[][and see Section~\ref{sec:morphquench}]{masters12a}.


We must therefore consider whether the environment itself may play a role in triggering the disk instabilities which can produce a bar. Indeed harassment and tidal interactions, believed to be common in the group environment, have been shown to both promote and inhibit bar formation dependent on the stellar mass \citep{noguchi88, moore96, skibba12}.  {\minor Although a bar may cause an eventual quench of a group galaxy, the bar may only be present because it was triggered by the dense group environment. It is therefore difficult to disentangle whether such a quench would be considered morphological or environmentally driven quenching.} This suggests that the polarity between internal secular processes (`nature') and external environmental processes (`nurture') may not be as extreme as first thought, in agreement with \cite{skibba12}. {\minor Similarly, some studies have suggested that internal and external processes may not be independent of each other, particularly for extreme environments and masses \citep{knobel15, darvish15, carollo16}.}

\subsection{The role of the environment in quenching}\label{sec:roleenv}

A trend for increasing time since quenching onset, $\Delta t$, with decreasing projected group-centric radius is present across the satellite population of the \textsc{gz2-group-q} sample (Figures~\ref{fig:timesinceradius}-\ref{fig:timesinceradiusvel} and quantified in Table~\ref{table:resultsum}). We interpret this as environmentally driven mechanisms causing quenching at the same rate throughout the infall time of a galaxy in a group. Galaxies which are now closer in fell into the group earlier, and as they did so they started to quench, giving rise to a larger inferred $\Delta t$.

We explain this in the context of works which consider the effects of halo mass on group galaxies. More massive halos are seen to have a greater impact on the star formation histories of their satellites than less massive halos in Figure~\ref{fig:timesinceradius}c. The halo mass is correlated with both (i) the gravitational potential of the group and (ii) the temperature of the IGM, suggesting that an environmental quenching mechanism which is correlated with one or both of these properties is responsible for this result.

Higher mass halos have hotter intra group medium (IGM) temperatures \citep{shimizu03, delpopolo05} which can have a greater impact on a galaxy through ram pressure stripping (RPS) of cold gas. \cite{gunngott72} define the ram pressure as:
\begin{equation}\label{eq:rps}
\rho_{\rm{IGM}}\cdot v^2 = 2\pi G \cdot \sigma_*(R) \cdot \sigma_g(R),
\end{equation}
where $\rho_{\rm{IGM}}$ is the density of the IGM, $\sigma_*(R)$ the star surface density, $\sigma_g(R)$ the gas surface density of the galaxy disc and $v$ the velocity of the galaxy through the IGM. Therefore if RPS is indeed a dominant environmental quenching mechanism we should see a {\minor decrease in $\tau$ (i.e. a more rapid quench) with increasing} velocity of a satellite relative to its central, $|\Delta v|$.  However we find that this is not the case (see {\minor Figure \ref{fig:timesinceradiusvel}b}). This therefore rules out RPS as the dominant environmental quenching mechanism, in support of the simulations of \citet{emerick16} and \citet{fillingham16} which showed that RPS could only remove $40-60\%$ of a satellite's gas. However, this conclusion may be due to the stellar mass range spanned by the \textsc{gz2-group-q} satellite galaxies which all have $M_* \geq 10^9 \rm{M}_{\odot}$, as simulations by \cite{fillingham16} suggest that RPS only becomes effective in lower mass satellites with $M_* \leq 10^{8-9} \rm{M}_{\odot}$, in agreement with \cite{hester06}. 

Above this mass threshold in the simulations of \cite{fillingham16}, a `starvation' (or strangulation) mode \citep{larson80, balogh00} dominates, where a galaxy's extended gaseous halo is removed causing a quench, as cold gas for use in star formation can no longer be fed from the extended halo. This idea is supported by observations by {\minor\citet{peng15}} which show that strangulation is a dominant mechanism for galaxies with $M_* < 10^{11}~M_{\odot}$ with a quenching timescale of $4~\rm{Gyr}.$ Such a mechanism will be correlated with the galaxy potential, as galaxies with a lower potential will be most easily stripped of their halos. We find that satellites with lower velocity dispersion (a proxy for the galaxy potential) are more rapidly quenched than their higher velocity dispersion counterparts and those in the field (see Figure~\ref{fig:timesinceradiusvel}d). Such a starvation mechanism is also correlated with halo mass, for which similar trends in $\Delta t$ are seen (Figure~\ref{fig:timesinceradius}c). The dominant environmental quenching mechanism occurring in the group environment must therefore be correlated with the group potential. This suggests that satellite galaxies may be most affected by gravitationally driven environmental effects, such as starvation, thermal evaporation of the galaxy halo and galaxy harassment.

We can calculate an infall timescale for the satellite galaxies in the \textsc{gz2-group-q} sample if we assume that galaxies begin their infall into a group at a radius of $\sim10\rm{R}_{200}$ and stop infalling at $\sim0.01\rm{R}_{200}$\footnote{This assumes that galaxies will then merge with their central galaxy, however it is more likely that the satellite has a close pass with the central before it `backsplashes' into the group. See, for example, \cite{pimbblet11}.}. The difference in the time since quenching onset, $\Delta t$, between these two locations in a group will provide an estimate for how long it takes a satellite to infall. This assumes (i) that the galaxy starts to quench immediately when it enters the group and (ii) that the same environmentally driven quenching process is the only quenching mechanism affecting the satellites throughout their infall. We estimate this infall time by calculating the difference in $\Delta t$ at $0.01~R_{200}$ and at $\sim10~R_{200}$ found in a given bin for each curve shown across Figures~\ref{fig:timesinceradius}-\ref{fig:timesinceradiusvel}. We define this property as $\delta \Delta t = \Delta t_{0.01R_{200}} - \Delta t_{10R_{200}}$. 

We therefore estimate a median infall time of $\delta \Delta t \sim 2.6 ~\rm{Gyr}$ for the \textsc{gz2-group-q} satellites. Similarly, the median rate of quenching of the \textsc{gz2-group-q} satellites is $\tau \sim 0.8~\rm{Gyr}$ (which is within the range of quenching rates hypothesised to result in a morphological change by \citealt{smethurst15}) and so we can also estimate the median quenching timescale (i.e. the time taken to fully quench from the SFS to $5\sigma$ below the SFS) to be $\sim 2.5~\rm{Gyr}$ for the \textsc{gz2-group-q} satellites (increasing to $\sim 3.7~\rm{Gyr}$ for those galaxies with $\tau\sim1.1\rm{Gyr}$).

This infall time and quenching timescale are in agreement with the estimates of \cite{wetzel13} who used a high resolution cosmological N-body simulation to track satellite galaxy orbits in SDSS groups and clusters and found quenching timescales of $2-6~\rm{Gyr}$. Using a similar method, \cite{oman16} derive an infall time of $\sim4~\rm{Gyr}$ and quenching timescales between $4-6~\rm{Gyr}$ for galaxies in the mass range of the \textsc{gz2-group-q} sample. {\minor However, studies such as \citet{peng10, wetzel13, hahn16, crossett17} and \citet{grootes17} have found much shorter quenching timescales of $\lesssim~1~\rm{Gyr}$ for satellite galaxies.} 

The simulations by \cite{fillingham16} and \cite{emerick16} have shown that RPS cannot remove enough gas mass to completely quench a galaxy within $\sim2~\rm{Gyr}$ but can assist in reducing the starvation timescale so that galaxies can be quenched within $\sim4~\rm{Gyr}$. This suggests that although the effects of mechanisms correlating with the group potential are detectable in the quenching parameters of the \textsc{gz2-group-q} sample, this is only made possible by the constantly present, but less dominant effects of ram pressure stripping. 

In Section~\ref{sec:rolemorphenv} we also noted that morphological quenching may only be present in the group environment due to the influence of the environment itself. Considering both this pairing of the environment and morphological quenching, and the pairing of ram pressure stripping and strangulation discussed above, suggests that all the mechanisms discussed here will affect a galaxy which is infalling through the group environment at some point in its lifetime. A single mechanism may be more dominant in the evolution of an individual galaxy but to achieve the correlations between morphology, colour and quenched galaxy fraction with density observed across the entire galaxy population, all mechanisms need to act in concert.  

\subsection{The bigger picture of quenching in galaxies}\label{sec:bigpic}

Having discussed the results presented here, we now consider the results in this paper in a broader context in conjunction with previous results found using the \textsc{starpy} method. \cite{smethurst15} infer the SFHs of the entire \textsc{gz2-galex} sample using \textsc{starpy} and investigate the morphological dependence of the derived quenching parameters for galaxy populations across the colour magnitude diagram. They find a clear difference between the quenching rates preferred by smooth and disc populations with smooth galaxies transitioning the green valley at faster rates than disc galaxies. However, intermediate quenching rates with $1 < \tau \rm{[Gyr]} < 2$, similar to the rates inferred in this study in Section~\ref{sec:starpy}, are dominant for all morphologies across the colour-magnitude diagram. Similarly, \cite{smethurst16} infer the SFHs of \textsc{gz2-galex} galaxies hosting optically selected Type 2 AGN and compare them to a control sample of currently inactive galaxies. They find evidence for rapid, recent quenching across the population of AGN host galaxies, particularly for galaxies with $M_* < 10.75 M_{\odot}$. This suggests that AGN feedback is important in the AGN host galaxy population. However, slow quenching rates are dominant for higher mass AGN host galaxies, suggesting secular evolution is also key in the evolution of galaxies currently hosting an AGN.

A parameter which is often investigated in quenching studies is the stellar mass surface density of a galaxy, which is found to correlate with SFR \citep{barro13b, whitaker16}. As a galaxy's bulge grows it is thought to be able to stabilise a disc against collapse and effectively stop it from forming stars. This is classed as a type of morphological quenching and is effective over time periods of a few $\rm{Gyr}$ \citep{Fang13} even if external gas is still fed to a galaxy. This slower quenching track of bulge dominated galaxies may help to explain the slow quenching rates observed by \cite{smethurst15} across the red and green smooth populations. They find that slow quenching with $\tau>2~\rm{Gyr}$ occurs for up to $40\%$ of the smooth green population and $24\%$ of the smooth red population. Using \textsc{starpy}, \cite{smethurst15} separated galaxies characterised by this slower quenching history, caused by processes which grow the bulge then consequently trigger morphological quenching, from those characterised by more rapid quenching histories, which are caused by processes which simultaneously quench the galaxy and grow the bulge. However, even in the latter case, morphological quenching may help in either speeding up the quenching process or in ensuring the galaxy stays quenched. This is supported by the finding of \cite{abramson16} who found that there is no threshold at which density triggered quenching occurs, but that denser systems redden faster than their less dense counterparts. This suggests that minor mergers and morphological quenching work together to fully achieve quiescence, similar to the collaboration between starvation and stripping to achieve quiescence of satellite galaxies discussed in Section~\ref{sec:roleenv}. 

This sort of partnership between two quenching mechanisms is also apparent in simulations which have shown that without AGN feedback a major merger cannot fully quench a galaxy \citep{springel05b}. In combination with a major merger however, a massive galaxy can be completely quenched by the heating or removal of gas and quiescence maintained \citep{conselice03, springel05b, hopkins08a, pontzen16}. These effects are therefore easily detectable, leading to the initial theories for the links between AGN and mergers \citep{merritt01, hopkins06b, hopkins08a, hopkins08b, peng07, jahnke11}. However, \cite{smethurst16} showed using \textsc{starpy} that galaxies hosting an AGN don't always quench at the rapid rates caused by major mergers, suggesting that a slow co-evolution of black hole and host galaxy can occur. They also showed that rapid quenching is only inferred for low mass AGN host galaxies where the AGN can have a greater impact on the galaxy SFR. 

Across the entire galaxy population we therefore have lots of examples of two quenching mechanisms working together to either quench a galaxy or ensure a galaxy stays quenched, including starvation and stripping (Section~\ref{sec:roleenv}), disc instabilities \&  environment (Section~\ref{sec:rolemorphenv}), minor mergers \& morphological quenching (see above) and mergers \& AGN \citep{smethurst15, smethurst16}. All of these mechanisms result in the same end state of galaxy quiescence (with the occasional influx of gas thwarting their progress) but no single mechanism dominates over another, except in the most extreme environments or masses. 

{\minor For example, mass and morphological quenching are dominant for galaxies in less dense environments (Figures~\ref{fig:timesinceradius}-\ref{fig:timesinceradiusvel}), but still affect galaxies in the densest environments \citep[e.g.][]{darvish16}. Similarly, mergers will dominate the evolution of galaxies in dense environments (e.g. centrals; see Section~\ref{sec:rolemergerenv}) but will drown out the more subtle effects of slower quenching mechanisms which occurred before the merger.} 

Just as the morphology of galaxies is continuous in nature from disc to bulge dominated, so too are the effects of the quenching mechanisms which can cause this change. The impact of mergers on the morphology and SFR of a galaxy depends on the mass ratio, a continuous variable from ``micro mergers'' \citep{beaton14, carlin16} through to major mergers. The strength of morphological quenching mechanisms can be measured on a continuum of stellar mass and stellar mass surface density of a galaxy; similarly the impact of environmentally driven quenching mechanisms increases with increasing halo mass. All of these processes, depending on a galaxy's environment, are likely to affect a galaxy at some point in its lifetime, acting in concert to reduce the SFR, which in turn produces the wide distribution of quenching timescales seen across the \textsc{gz2-group-q} sample. In previous works, efforts have been made to identify the dominant quenching mechanism in a galaxy sample \citep[e.g.][]{muzzin12, schawinski14, foltz15, woo15, balogh16, darvish16, huertascompany16}, yet it is clear from this study that multiple quenching mechanisms will affect galaxies across their lifetime, working in collaboration to ensure galaxies stay quenched. 

Future studies should therefore focus on disentangling the effects of these various different quenching mechanisms, rather than focussing on a single process. 

\section{Conclusions}\label{sec:conc}

We have investigated the detailed morphological structures and star formation histories (SFHs) of a sample of SDSS group galaxies \citep{berlind06}, with both classifications from Galaxy Zoo 2 and NUV detections in GALEX. SFHs were inferred using a Bayesian MCMC code, \textsc{starpy}. We have shown that although mass quenching, morphological quenching and mergers are all important mechanisms at work in quenching the galaxies in the group environment, environmentally driven quenching mechanisms do play a role in quenching galaxies as they infall into the group. 

We have discussed the possibility that no single mechanism will dominate across the group population (except in the most extreme environments or masses) with all mechanisms acting collaboratively. Our findings are summarised as follows:
\begin{enumerate}
\item The bar, obvious bulge and merger fractions are all seen to increase above the field value in the inner regions of the groups of the \textsc{gz2-group-q} sample in Figures~\ref{fig:barradius}, \ref{fig:bulgeradius} \& \ref{fig:mergerradius} respectively.  
 
\item Mergers are the dominant quenching mechanism for central galaxies but not for satellite galaxies. Satellites may undergo a minor merger in the group environment but their effects are only discernible by their indirect effect on the bulge fraction (see Figure~\ref{fig:bulgeradius}).
 
\item Mass dependent quenching is occurring across the entire \textsc{gz2-group-q} sample for both centrals and satellites irrespective of  environmental density (see Figure~\ref{fig:timesinceradius}a), {\minor the effects of which are more apparent than environmentally driven mechanisms}.
 
\item Morphological quenching is occurring for \textsc{gz2-group-q} satellite galaxies as evidenced by the heightened bar fraction in the inner group regions (see Figure~\ref{fig:barradius}). However, this may be indirectly due to environmental quenching since galaxy interactions and harassment are believed to be able to trigger bars. This suggests the polarity between `nature' vs. `nurture' may not be as extreme as previously thought, in agreement with \cite{skibba12}. 

\item The environment does cause quenching across the \textsc{gz2-group-q} sample, as evidenced by the increase in the time since quenching with decreasing group radius (Figures~\ref{fig:timesinceradius}-\ref{fig:timesinceradiusvel} and Table~\ref{table:resultsum}). Our results suggest that this is caused by a quenching mechanism correlated with the group potential, such as harassment, interactions and starvation, rather than the velocity of a satellite through the group, such as ram pressure stripping (Figures \ref{fig:timesinceradius}a \&  \ref{fig:timesinceradiusvel}c). This quenching occurs within an median quenching timescale of $\sim2.5~\rm{Gyr}$ from star forming to complete quiescence, after an average infall time of $\sim 2.6 ~\rm{Gyr}$. 
  
\end{enumerate}

It is apparent from the results presented here that many quenching mechanisms are all occurring simultaneously in the group environment; therefore a superposition of all of the effects of these mechanisms is seen in the quenching histories of the \textsc{gz2-group-q} sample, which in turn gives rise to the observed morphology-density relation. 

\section*{Acknowledgements}
{\minor The authors would like to thank the anonymous referee for their helpful comments.} We would also like to thank R. Davies and K. Pimbblet for discussion and suggestions which greatly improved the clarity of this paper. 

This publication made extensive use of the Tool for Operations on Catalogues And Tables (TOPCAT; ~\citealt{taylor05}) which can be found at \url{http://www.star.bris.ac.uk/~mbt/topcat/} and the open source Python module \emph{astroPy}\footnote{\url{http://www.astropy.org/}}; \citealt{astropy13}). This research has also made use of NASA's ADS service and Cornell's ArXiv. 

RJS would like to acknowledge funding from both the Science and Technology Facilities Council, Grant Code ST/K502236/1, and from the Ogden Trust. SJK would like to acknowledge the funding from STFC Grant Code ST/MJ0371X/1

The development of Galaxy Zoo was supported in part by the Alfred P. Sloan Foundation and by The Leverhulme Trust. 

Based on observations made with the NASA Galaxy Evolution Explorer.  GALEX is operated for NASA by the California Institute of Technology under NASA contract NAS5-98034

Funding for the SDSS and SDSS-II has been provided by the Alfred P. Sloan Foundation, the Participating Institutions, the National Science Foundation, the U.S. Department of Energy, the National Aeronautics and Space Administration, the Japanese Monbukagakusho, the Max Planck Society, and the Higher Education Funding Council for England. The SDSS Web Site is \url{http://www.sdss.org/}.
The SDSS is managed by the Astrophysical Research Consortium for the Participating Institutions. The Participating Institutions are the American Museum of Natural History, Astrophysical Institute Potsdam, University of Basel, University of Cambridge, Case Western Reserve University, University of Chicago, Drexel University, Fermilab, the Institute for Advanced Study, the Japan Participation Group, Johns Hopkins University, the Joint Institute for Nuclear Astrophysics, the Kavli Institute for Particle Astrophysics and Cosmology, the Korean Scientist Group, the Chinese Academy of Sciences (LAMOST), Los Alamos National Laboratory, the Max-Planck-Institute for Astronomy (MPIA), the Max-Planck-Institute for Astrophysics (MPA), New Mexico State University, Ohio State University, University of Pittsburgh, University of Portsmouth, Princeton University, the United States Naval Observatory, and the University of Washington.

\bibliographystyle{mn2e}
\bibliography{refs}  

\begin{thebibliography}{206}
\expandafter\ifx\csname natexlab\endcsname\relax\def\natexlab#1{#1}\fi

\bibitem[{{Abadi} {et~al}\mbox{.}(1999){Abadi}, {Moore}, \& {Bower}}]{abadi99}
{Abadi} M.~G., {Moore} B., {Bower} R.~G., 1999, \mnras, 308, 947

\bibitem[{{Abazajian} {et~al}\mbox{.}(2009){Abazajian}, {Adelman-McCarthy},
  {Ag{\"u}eros}, {Allam}, {Allende Prieto}, {An}, {Anderson}, {Anderson},
  {Annis}, {Bahcall}, \& et~al.}]{abazajian09}
{Abazajian} K.~N. {et~al.}, 2009, \apjs, 182, 543

\bibitem[{{Abell}(1958)}]{abell58}
{Abell} G.~O., 1958, \apjs, 3, 211

\bibitem[{{Abramson} {et~al}\mbox{.}(2016){Abramson}, {Gladders}, {Dressler},
  {Oemler}, {Poggianti}, \& {Vulcani}}]{abramson16}
{Abramson} L.~E., {Gladders} M.~D., {Dressler} A., {Oemler} A., {Poggianti} B.,
  {Vulcani} B., 2016, ArXiv e-prints, 1604.00016

\bibitem[{{Alpaslan} {et~al}\mbox{.}(2016){Alpaslan}, {Grootes}, {Marcum},
  {Popescu}, {Tuffs}, {Bland-Hawthorn}, {Brough}, {Brown}, {Davies}, {Driver},
  {Holwerda}, {Kelvin}, {Lara-L{\'o}pez}, {L{\'o}pez-S{\'a}nchez}, {Loveday},
  {Moffett}, {Taylor}, {Owers}, \& {Robotham}}]{alpaslan16}
{Alpaslan} M. {et~al.}, 2016, \mnras, 457, 2287

\bibitem[{{Astropy Collaboration} {et~al}\mbox{.}(2013){Astropy Collaboration},
  {Robitaille}, {Tollerud}, {Greenfield}, {Droettboom}, {Bray}, {Aldcroft},
  {Davis}, {Ginsburg}, {Price-Whelan}, {Kerzendorf}, {Conley}, {Crighton},
  {Barbary}, {Muna}, {Ferguson}, {Grollier}, {Parikh}, {Nair}, {Unther},
  {Deil}, {Woillez}, {Conseil}, {Kramer}, {Turner}, {Singer}, {Fox}, {Weaver},
  {Zabalza}, {Edwards}, {Azalee Bostroem}, {Burke}, {Casey}, {Crawford},
  {Dencheva}, {Ely}, {Jenness}, {Labrie}, {Lim}, {Pierfederici}, {Pontzen},
  {Ptak}, {Refsdal}, {Servillat}, \& {Streicher}}]{astropy13}
{Astropy Collaboration} {et~al.}, 2013, \aap, 558, A33

\bibitem[{{Athanassoula}(1992{\natexlab{a}})}]{athanassoula92a}
{Athanassoula} E., 1992{\natexlab{a}}, \mnras, 259, 328

\bibitem[{{Athanassoula}(1992{\natexlab{b}})}]{athanassoula92b}
{Athanassoula} E., 1992{\natexlab{b}}, \mnras, 259, 345

\bibitem[{{Athanassoula} {et~al}\mbox{.}(2016){Athanassoula}, {Rodionov},
  {Peschken}, \& {Lambert}}]{athanassoula16}
{Athanassoula} E., {Rodionov} S.~A., {Peschken} N., {Lambert} J.~C., 2016,
  \apj, 821, 90

\bibitem[{{Baldry} {et~al}\mbox{.}(2006){Baldry}, {Balogh}, {Bower},
  {Glazebrook}, {Nichol}, {Bamford}, \& {Budavari}}]{Baldry06}
{Baldry} I.~K., {Balogh} M.~L., {Bower} R.~G., {Glazebrook} K., {Nichol} R.~C.,
  {Bamford} S.~P., {Budavari} T., 2006, \mnras, 373, 469

\bibitem[{{Balogh} {et~al}\mbox{.}(2004){Balogh}, {Eke}, {Miller}, {Lewis},
  {Bower}, {Couch}, {Nichol}, {Bland-Hawthorn}, {Baldry}, {Baugh}, {Bridges},
  {Cannon}, {Cole}, {Colless}, {Collins}, {Cross}, {Dalton}, {de Propris},
  {Driver}, {Efstathiou}, {Ellis}, {Frenk}, {Glazebrook}, {Gomez}, {Gray},
  {Hawkins}, {Jackson}, {Lahav}, {Lumsden}, {Maddox}, {Madgwick}, {Norberg},
  {Peacock}, {Percival}, {Peterson}, {Sutherland}, \& {Taylor}}]{balogh04}
{Balogh} M. {et~al.}, 2004, \mnras, 348, 1355

\bibitem[{{Balogh} {et~al}\mbox{.}(2016){Balogh}, {McGee}, {Mok}, {Muzzin},
  {van der Burg}, {Bower}, {Finoguenov}, {Hoekstra}, {Lidman}, {Mulchaey},
  {Noble}, {Parker}, {Tanaka}, {Wilman}, {Webb}, {Wilson}, \& {Yee}}]{balogh16}
{Balogh} M.~L. {et~al.}, 2016, \mnras, 456, 4364

\bibitem[{{Balogh} {et~al}\mbox{.}(2000){Balogh}, {Navarro}, \&
  {Morris}}]{balogh00}
{Balogh} M.~L., {Navarro} J.~F., {Morris} S.~L., 2000, \apj, 540, 113

\bibitem[{{Bamford} {et~al}\mbox{.}(2009){Bamford}, {Nichol}, {Baldry}, {Land},
  {Lintott}, {Schawinski}, {Slosar}, {Szalay}, {Thomas}, {Torki}, {Andreescu},
  {Edmondson}, {Miller}, {Murray}, {Raddick}, \& {Vandenberg}}]{Bamford09}
{Bamford} S.~P. {et~al.}, 2009, \mnras, 393, 1324

\bibitem[{{Barazza} {et~al}\mbox{.}(2009){Barazza}, {Jablonka}, {Desai},
  {Jogee}, {Arag{\'o}n-Salamanca}, {De Lucia}, {Saglia}, {Halliday},
  {Poggianti}, {Dalcanton}, {Rudnick}, {Milvang-Jensen}, {Noll}, {Simard},
  {Clowe}, {Pell{\'o}}, {White}, \& {Zaritsky}}]{barazza09}
{Barazza} F.~D. {et~al.}, 2009, \aap, 497, 713

\bibitem[{{Barnes} \& {Hernquist}(1996)}]{barnes96}
{Barnes} J.~E., {Hernquist} L., 1996, \apj, 471, 115

\bibitem[{{Barro} {et~al}\mbox{.}(2013){Barro}, {Faber},
  {P{\'e}rez-Gonz{\'a}lez}, {Koo}, {Williams}, {Kocevski}, {Trump}, {Mozena},
  {McGrath}, {van der Wel}, {Wuyts}, {Bell}, {Croton}, {Ceverino}, {Dekel},
  {Ashby}, {Cheung}, {Ferguson}, {Fontana}, {Fang}, {Giavalisco}, {Grogin},
  {Guo}, {Hathi}, {Hopkins}, {Huang}, {Koekemoer}, {Kartaltepe}, {Lee},
  {Newman}, {Porter}, {Primack}, {Ryan}, {Rosario}, {Somerville}, {Salvato}, \&
  {Hsu}}]{barro13b}
{Barro} G. {et~al.}, 2013, \apj, 765, 104

\bibitem[{{Baugh} {et~al}\mbox{.}(1998){Baugh}, {Cole}, {Frenk}, \&
  {Lacey}}]{baugh98}
{Baugh} C.~M., {Cole} S., {Frenk} C.~S., {Lacey} C.~G., 1998, \apj, 498, 504

\bibitem[{{Baugh} {et~al}\mbox{.}(2005){Baugh}, {Lacey}, {Frenk}, {Granato},
  {Silva}, {Bressan}, {Benson}, \& {Cole}}]{baugh05}
{Baugh} C.~M., {Lacey} C.~G., {Frenk} C.~S., {Granato} G.~L., {Silva} L.,
  {Bressan} A., {Benson} A.~J., {Cole} S., 2005, \mnras, 356, 1191

\bibitem[{{Beaton} {et~al}\mbox{.}(2014){Beaton}, {Mart{\'{\i}}nez-Delgado},
  {Majewski}, {D'Onghia}, {Zibetti}, {Gabany}, {Johnson}, {Blanton}, \&
  {Verbiscer}}]{beaton14}
{Beaton} R.~L. {et~al.}, 2014, \apj, 790, 117

\bibitem[{{Bell} {et~al}\mbox{.}(2006){Bell}, {Phleps}, {Somerville}, {Wolf},
  {Borch}, \& {Meisenheimer}}]{bell06}
{Bell} E.~F., {Phleps} S., {Somerville} R.~S., {Wolf} C., {Borch} A.,
  {Meisenheimer} K., 2006, \apj, 652, 270

\bibitem[{{Berlind} {et~al}\mbox{.}(2006){Berlind}, {Frieman}, {Weinberg},
  {Blanton}, {Warren}, {Abazajian}, {Scranton}, {Hogg}, {Scoccimarro},
  {Bahcall}, {Brinkmann}, {Gott}, {Kleinman}, {Krzesinski}, {Lee}, {Miller},
  {Nitta}, {Schneider}, {Tucker}, {Zehavi}, \& {SDSS
  Collaboration}}]{berlind06}
{Berlind} A.~A. {et~al.}, 2006, \apjs, 167, 1

\bibitem[{{Bialas} {et~al}\mbox{.}(2015){Bialas}, {Lisker}, {Olczak},
  {Spurzem}, \& {Kotulla}}]{bialas15}
{Bialas} D., {Lisker} T., {Olczak} C., {Spurzem} R., {Kotulla} R., 2015, \aap,
  576, A103

\bibitem[{{Birnboim} \& {Dekel}(2003)}]{birnboim03}
{Birnboim} Y., {Dekel} A., 2003, \mnras, 345, 349

\bibitem[{{Blanton} {et~al}\mbox{.}(2005){Blanton}, {Eisenstein}, {Hogg},
  {Schlegel}, \& {Brinkmann}}]{Blanton05}
{Blanton} M.~R., {Eisenstein} D., {Hogg} D.~W., {Schlegel} D.~J., {Brinkmann}
  J., 2005, \apj, 629, 143

\bibitem[{{Blanton} {et~al}\mbox{.}(2006){Blanton}, {Eisenstein}, {Hogg}, \&
  {Zehavi}}]{blanton06}
{Blanton} M.~R., {Eisenstein} D., {Hogg} D.~W., {Zehavi} I., 2006, \apj, 645,
  977

\bibitem[{{Blanton} \& {Roweis}(2007)}]{blanton07}
{Blanton} M.~R., {Roweis} S., 2007, \aj, 133, 734

\bibitem[{{Bluck} {et~al}\mbox{.}(2014){Bluck}, {Mendel}, {Ellison}, {Moreno},
  {Simard}, {Patton}, \& {Starkenburg}}]{bluck14}
{Bluck} A.~F.~L., {Mendel} J.~T., {Ellison} S.~L., {Moreno} J., {Simard} L.,
  {Patton} D.~R., {Starkenburg} E., 2014, \mnras, 441, 599

\bibitem[{{Boselli} \& {Gavazzi}(2006)}]{boselli06}
{Boselli} A., {Gavazzi} G., 2006, \pasp, 118, 517

\bibitem[{{Bower} \& {Balogh}(2004)}]{bower04}
{Bower} R.~G., {Balogh} M.~L., 2004, Clusters of Galaxies: Probes of
  Cosmological Structure and Galaxy Evolution, 325

\bibitem[{{Bower} {et~al}\mbox{.}(2006){Bower}, {Benson}, {Malbon}, {Helly},
  {Frenk}, {Baugh}, {Cole}, \& {Lacey}}]{Bower06}
{Bower} R.~G., {Benson} A.~J., {Malbon} R., {Helly} J.~C., {Frenk} C.~S.,
  {Baugh} C.~M., {Cole} S., {Lacey} C.~G., 2006, \mnras, 370, 645

\bibitem[{{Brinchmann} {et~al}\mbox{.}(2004){Brinchmann}, {Charlot}, {White},
  {Tremonti}, {Kauffmann}, {Heckman}, \& {Brinkmann}}]{brinchmann04}
{Brinchmann} J., {Charlot} S., {White} S.~D.~M., {Tremonti} C., {Kauffmann} G.,
  {Heckman} T., {Brinkmann} J., 2004, \mnras, 351, 1151

\bibitem[{{Bruzual} \& {Charlot}(2003)}]{BC03}
{Bruzual} G., {Charlot} S., 2003, \mnras, 344, 1000

\bibitem[{{Butcher} \& {Oemler}(1978)}]{butcher78}
{Butcher} H., {Oemler}, Jr. A., 1978, \apj, 226, 559

\bibitem[{{Cardelli} {et~al}\mbox{.}(1989){Cardelli}, {Clayton}, \&
  {Mathis}}]{Cardelli89}
{Cardelli} J.~A., {Clayton} G.~C., {Mathis} J.~S., 1989, \apj, 345, 245

\bibitem[{{Carlberg}(2004)}]{carlberg04}
{Carlberg} R.~G., 2004, Clusters of Galaxies: Probes of Cosmological Structure
  and Galaxy Evolution, 343

\bibitem[{{Carlin} {et~al}\mbox{.}(2016){Carlin}, {Sand}, {Price}, {Willman},
  {Karunakaran}, {Spekkens}, {Bell}, {Brodie}, {Crnojevi{\'c}}, {Forbes},
  {Hargis}, {Kirby}, {Lupton}, {Peter}, {Romanowsky}, \& {Strader}}]{carlin16}
{Carlin} J.~L. {et~al.}, 2016, \apjl, 828, L5

\bibitem[{{Carollo} {et~al}\mbox{.}(2016){Carollo}, {Cibinel}, {Lilly},
  {Pipino}, {Bonoli}, {Finoguenov}, {Miniati}, {Norberg}, \&
  {Silverman}}]{carollo16}
{Carollo} C.~M. {et~al.}, 2016, \apj, 818, 180

\bibitem[{{Casteels} {et~al}\mbox{.}(2013){Casteels}, {Bamford}, {Skibba},
  {Masters}, {Lintott}, {Keel}, {Schawinski}, {Nichol}, \&
  {Smith}}]{casteels13}
{Casteels} K.~R.~V. {et~al.}, 2013, \mnras, 429, 1051

\bibitem[{{Chabrier}(2003)}]{chabrier03}
{Chabrier} G., 2003, \pasp, 115, 763

\bibitem[{{Cheung} {et~al}\mbox{.}(2013){Cheung}, {Athanassoula}, {Masters},
  {Nichol}, {Bosma}, {Bell}, {Faber}, {Koo}, {Lintott}, {Melvin}, {Schawinski},
  {Skibba}, \& {Willett}}]{Cheung13}
{Cheung} E. {et~al.}, 2013, \apj, 779, 162

\bibitem[{{Conselice}(2003)}]{conselice03}
{Conselice} C.~J., 2003, \apjs, 147, 1

\bibitem[{{Cowie} \& {Barger}(2008)}]{cowie08}
{Cowie} L.~L., {Barger} A.~J., 2008, \apj, 686, 72

\bibitem[{{Cowie} \& {Songaila}(1977)}]{cowie77}
{Cowie} L.~L., {Songaila} A., 1977, \nat, 266, 501

\bibitem[{{Cox} {et~al}\mbox{.}(2008){Cox}, {Younger}, {Hernquist}, \&
  {Hopkins}}]{cox08}
{Cox} T.~J., {Younger} J., {Hernquist} L., {Hopkins} P.~F., 2008, in IAU
  Symposium, Vol. 245, Formation and Evolution of Galaxy Bulges, {Bureau} M.,
  {Athanassoula} E., {Barbuy} B., eds., pp. 63--66

\bibitem[{{Crossett} {et~al}\mbox{.}(2017){Crossett}, {Pimbblet}, {Jones},
  {Brown}, \& {Stott}}]{crossett17}
{Crossett} J.~P., {Pimbblet} K.~A., {Jones} D.~H., {Brown} M.~J.~I., {Stott}
  J.~P., 2017, \mnras, 464, 480

\bibitem[{{Croton} {et~al}\mbox{.}(2006){Croton}, {Springel}, {White}, {De
  Lucia}, {Frenk}, {Gao}, {Jenkins}, {Kauffmann}, {Navarro}, \&
  {Yoshida}}]{Croton06}
{Croton} D.~J. {et~al.}, 2006, \mnras, 365, 11

\bibitem[{{Cucciati} {et~al}\mbox{.}(2010){Cucciati}, {Marinoni}, {Iovino},
  {Bardelli}, {Adami}, {Mazure}, {Scodeggio}, {Maccagni}, {Temporin}, {Zucca},
  {De Lucia}, {Blaizot}, {Garilli}, {Meneux}, {Zamorani}, {Le F{\`e}vre},
  {Cappi}, {Guzzo}, {Bottini}, {Le Brun}, {Tresse}, {Vettolani}, {Zanichelli},
  {Arnouts}, {Bolzonella}, {Charlot}, {Ciliegi}, {Contini}, {Foucaud},
  {Franzetti}, {Gavignaud}, {Ilbert}, {Lamareille}, {McCracken}, {Marano},
  {Merighi}, {Paltani}, {Pell{\`o}}, {Pollo}, {Pozzetti}, {Vergani}, \&
  {P{\'e}rez-Montero}}]{cucciati10}
{Cucciati} O. {et~al.}, 2010, \aap, 520, A42

\bibitem[{{Darg} {et~al}\mbox{.}(2010){Darg}, {Kaviraj}, {Lintott},
  {Schawinski}, {Sarzi}, {Bamford}, {Silk}, {Proctor}, {Andreescu}, {Murray},
  {Nichol}, {Raddick}, {Slosar}, {Szalay}, {Thomas}, \& {Vandenberg}}]{Darg10a}
{Darg} D.~W. {et~al.}, 2010, \mnras, 401, 1043

\bibitem[{{Darvish} {et~al}\mbox{.}(2017){Darvish}, {Mobasher}, {Martin},
  {Sobral}, {Scoville}, {Stroe}, {Hemmati}, \& {Kartaltepe}}]{darvish17}
{Darvish} B., {Mobasher} B., {Martin} D.~C., {Sobral} D., {Scoville} N.,
  {Stroe} A., {Hemmati} S., {Kartaltepe} J., 2017, \apj, 837, 16

\bibitem[{{Darvish} {et~al}\mbox{.}(2016){Darvish}, {Mobasher}, {Sobral},
  {Rettura}, {Scoville}, {Faisst}, \& {Capak}}]{darvish16}
{Darvish} B., {Mobasher} B., {Sobral} D., {Rettura} A., {Scoville} N., {Faisst}
  A., {Capak} P., 2016, \apj, 825, 113

\bibitem[{{Darvish} {et~al}\mbox{.}(2015){Darvish}, {Mobasher}, {Sobral},
  {Scoville}, \& {Aragon-Calvo}}]{darvish15}
{Darvish} B., {Mobasher} B., {Sobral} D., {Scoville} N., {Aragon-Calvo} M.,
  2015, \apj, 805, 121

\bibitem[{{Darvish} {et~al}\mbox{.}(2014){Darvish}, {Sobral}, {Mobasher},
  {Scoville}, {Best}, {Sales}, \& {Smail}}]{darvish14}
{Darvish} B., {Sobral} D., {Mobasher} B., {Scoville} N.~Z., {Best} P., {Sales}
  L.~V., {Smail} I., 2014, \apj, 796, 51

\bibitem[{{Dekel} \& {Birnboim}(2006)}]{dekel06}
{Dekel} A., {Birnboim} Y., 2006, \mnras, 368, 2

\bibitem[{{Del Popolo} {et~al}\mbox{.}(2005){Del Popolo}, {Hiotelis}, \&
  {Pe{\~n}arrubia}}]{delpopolo05}
{Del Popolo} A., {Hiotelis} N., {Pe{\~n}arrubia} J., 2005, \apj, 628, 76

\bibitem[{{Di Matteo} {et~al}\mbox{.}(2005){Di Matteo}, {Springel}, \&
  {Hernquist}}]{dimatteo05}
{Di Matteo} T., {Springel} V., {Hernquist} L., 2005, \nat, 433, 604

\bibitem[{{Diaferio} {et~al}\mbox{.}(2001){Diaferio}, {Kauffmann}, {Balogh},
  {White}, {Schade}, \& {Ellingson}}]{diaferio01}
{Diaferio} A., {Kauffmann} G., {Balogh} M.~L., {White} S.~D.~M., {Schade} D.,
  {Ellingson} E., 2001, \mnras, 323, 999

\bibitem[{{Dressler}(1980)}]{dressler80}
{Dressler} A., 1980, \apj, 236, 351

\bibitem[{{Dressler}(2004)}]{dressler04}
{Dressler} A., 2004, in IAU Colloq. 195: Outskirts of Galaxy Clusters: Intense
  Life in the Suburbs, {Diaferio} A., ed., pp. 341--346

\bibitem[{{Eke} {et~al}\mbox{.}(2004){Eke}, {Baugh}, {Cole}, {Frenk},
  {Norberg}, {Peacock}, {Baldry}, {Bland-Hawthorn}, {Bridges}, {Cannon},
  {Colless}, {Collins}, {Couch}, {Dalton}, {de Propris}, {Driver},
  {Efstathiou}, {Ellis}, {Glazebrook}, {Jackson}, {Lahav}, {Lewis}, {Lumsden},
  {Maddox}, {Madgwick}, {Peterson}, {Sutherland}, \& {Taylor}}]{eke04}
{Eke} V.~R. {et~al.}, 2004, \mnras, 348, 866

\bibitem[{{Ellison} {et~al}\mbox{.}(2010){Ellison}, {Patton}, {Simard},
  {McConnachie}, {Baldry}, \& {Mendel}}]{ellison10}
{Ellison} S.~L., {Patton} D.~R., {Simard} L., {McConnachie} A.~W., {Baldry}
  I.~K., {Mendel} J.~T., 2010, \mnras, 407, 1514

\bibitem[{{Emerick} {et~al}\mbox{.}(2016){Emerick}, {Mac Low}, {Grcevich}, \&
  {Gatto}}]{emerick16}
{Emerick} A., {Mac Low} M.-M., {Grcevich} J., {Gatto} A., 2016, \apj, 826, 148

\bibitem[{{Fabian}(2012)}]{fabian12}
{Fabian} A.~C., 2012, \araa, 50, 455

\bibitem[{{Fadda} {et~al}\mbox{.}(2008){Fadda}, {Biviano}, {Marleau},
  {Storrie-Lombardi}, \& {Durret}}]{fadda08}
{Fadda} D., {Biviano} A., {Marleau} F.~R., {Storrie-Lombardi} L.~J., {Durret}
  F., 2008, \apjl, 672, L9

\bibitem[{{Fang} {et~al}\mbox{.}(2013){Fang}, {Faber}, {Koo}, \&
  {Dekel}}]{Fang13}
{Fang} J.~J., {Faber} S.~M., {Koo} D.~C., {Dekel} A., 2013, \apj, 776, 63

\bibitem[{{Ferr{\'e}-Mateu} {et~al}\mbox{.}(2014){Ferr{\'e}-Mateu},
  {S{\'a}nchez-Bl{\'a}zquez}, {Vazdekis}, \& {de la Rosa}}]{ferre14}
{Ferr{\'e}-Mateu} A., {S{\'a}nchez-Bl{\'a}zquez} P., {Vazdekis} A., {de la
  Rosa} I.~G., 2014, \apj, 797, 136

\bibitem[{{Fillingham} {et~al}\mbox{.}(2016){Fillingham}, {Cooper}, {Pace},
  {Boylan-Kolchin}, {Bullock}, {Garrison-Kimmel}, \& {Wheeler}}]{fillingham16}
{Fillingham} S.~P., {Cooper} M.~C., {Pace} A.~B., {Boylan-Kolchin} M.,
  {Bullock} J.~S., {Garrison-Kimmel} S., {Wheeler} C., 2016, \mnras

\bibitem[{{Finn} {et~al}\mbox{.}(2005){Finn}, {Zaritsky}, {McCarthy},
  {Poggianti}, {Rudnick}, {Halliday}, {Milvang-Jensen}, {Pell{\'o}}, \&
  {Simard}}]{finn05}
{Finn} R.~A. {et~al.}, 2005, \apj, 630, 206

\bibitem[{{Foltz} {et~al}\mbox{.}(2015){Foltz}, {Rettura}, {Wilson}, {van der
  Burg}, {Muzzin}, {Lidman}, {Demarco}, {Nantais}, {DeGroot}, \&
  {Yee}}]{foltz15}
{Foltz} R. {et~al.}, 2015, \apj, 812, 138

\bibitem[{{Foreman-Mackey} {et~al}\mbox{.}(2013){Foreman-Mackey}, {Hogg},
  {Lang}, \& {Goodman}}]{emcee13}
{Foreman-Mackey} D., {Hogg} D.~W., {Lang} D., {Goodman} J., 2013, \pasp, 125,
  306

\bibitem[{{Gabor} \& {Dav{\'e}}(2015)}]{gabor15}
{Gabor} J.~M., {Dav{\'e}} R., 2015, \mnras, 447, 374

\bibitem[{{Gabor} {et~al}\mbox{.}(2010){Gabor}, {Dav{\'e}}, {Finlator}, \&
  {Oppenheimer}}]{Gabor10}
{Gabor} J.~M., {Dav{\'e}} R., {Finlator} K., {Oppenheimer} B.~D., 2010, \mnras,
  407, 749

\bibitem[{{G{\'o}mez} {et~al}\mbox{.}(2003){G{\'o}mez}, {Nichol}, {Miller},
  {Balogh}, {Goto}, {Zabludoff}, {Romer}, {Bernardi}, {Sheth}, {Hopkins},
  {Castander}, {Connolly}, {Schneider}, {Brinkmann}, {Lamb}, {SubbaRao}, \&
  {York}}]{gomez03}
{G{\'o}mez} P.~L. {et~al.}, 2003, \apj, 584, 210

\bibitem[{{Grootes} {et~al}\mbox{.}(2017){Grootes}, {Tuffs}, {Popescu},
  {Norberg}, {Robotham}, {Liske}, {Andrae}, {Baldry}, {Gunawardhana}, {Kelvin},
  {Madore}, {Seibert}, {Taylor}, {Alpaslan}, {Brown}, {Cluver}, {Driver},
  {Bland-Hawthorn}, {Holwerda}, {Hopkins}, {Lopez-Sanchez}, {Loveday}, \&
  {Rushton}}]{grootes17}
{Grootes} M.~W. {et~al.}, 2017, \aj, 153, 111

\bibitem[{{Gunn} \& {Gott}(1972)}]{gunngott72}
{Gunn} J.~E., {Gott}, III J.~R., 1972, \apj, 176, 1

\bibitem[{{Hahn} {et~al}\mbox{.}(2016){Hahn}, {Tinker}, \& {Wetzel}}]{hahn16}
{Hahn} C., {Tinker} J.~L., {Wetzel} A.~R., 2016, ArXiv e-prints, 1609.04398

\bibitem[{{H{\"a}ring} \& {Rix}(2004)}]{haringrix04}
{H{\"a}ring} N., {Rix} H.-W., 2004, \apjl, 604, L89

\bibitem[{{Hatfield} \& {Jarvis}(2016)}]{hatfield16}
{Hatfield} P.~W., {Jarvis} M.~J., 2016, ArXiv e-prints, 1606.08989

\bibitem[{{Hayward} {et~al}\mbox{.}(2014){Hayward}, {Torrey}, {Springel},
  {Hernquist}, \& {Vogelsberger}}]{hayward14}
{Hayward} C.~C., {Torrey} P., {Springel} V., {Hernquist} L., {Vogelsberger} M.,
  2014, \mnras, 442, 1992

\bibitem[{{Hearin} {et~al}\mbox{.}(2015){Hearin}, {Watson}, \& {van den
  Bosch}}]{hearin15}
{Hearin} A.~P., {Watson} D.~F., {van den Bosch} F.~C., 2015, \mnras, 452, 1958

\bibitem[{{Hester}(2006)}]{hester06}
{Hester} J.~A., 2006, \apj, 647, 910

\bibitem[{{Hickox} {et~al}\mbox{.}(2009){Hickox}, {Jones}, {Forman}, {Murray},
  {Kochanek}, {Eisenstein}, {Jannuzi}, {Dey}, {Brown}, {Stern}, {Eisenhardt},
  {Gorjian}, {Brodwin}, {Narayan}, {Cool}, {Kenter}, {Caldwell}, \&
  {Anderson}}]{Hickox09}
{Hickox} R.~C. {et~al.}, 2009, \apj, 696, 891

\bibitem[{{Hirschmann} {et~al}\mbox{.}(2014){Hirschmann}, {De Lucia}, {Wilman},
  {Weinmann}, {Iovino}, {Cucciati}, {Zibetti}, \& {Villalobos}}]{hirschmann14}
{Hirschmann} M., {De Lucia} G., {Wilman} D., {Weinmann} S., {Iovino} A.,
  {Cucciati} O., {Zibetti} S., {Villalobos} {\'A}., 2014, \mnras, 444, 2938

\bibitem[{{Hopkins} {et~al}\mbox{.}(2008{\natexlab{a}}){Hopkins}, {Cox},
  {Kere{\v s}}, \& {Hernquist}}]{hopkins08b}
{Hopkins} P.~F., {Cox} T.~J., {Kere{\v s}} D., {Hernquist} L.,
  2008{\natexlab{a}}, \apjs, 175, 390

\bibitem[{{Hopkins} {et~al}\mbox{.}(2009){Hopkins}, {Cox}, {Younger}, \&
  {Hernquist}}]{hopkins09c}
{Hopkins} P.~F., {Cox} T.~J., {Younger} J.~D., {Hernquist} L., 2009, \apj, 691,
  1168

\bibitem[{{Hopkins} \& {Hernquist}(2009)}]{hopkins09a}
{Hopkins} P.~F., {Hernquist} L., 2009, \apj, 694, 599

\bibitem[{{Hopkins} {et~al}\mbox{.}(2006{\natexlab{a}}){Hopkins}, {Hernquist},
  {Cox}, {Di Matteo}, {Robertson}, \& {Springel}}]{hopkins06d}
{Hopkins} P.~F., {Hernquist} L., {Cox} T.~J., {Di Matteo} T., {Robertson} B.,
  {Springel} V., 2006{\natexlab{a}}, \apjs, 163, 1

\bibitem[{{Hopkins} {et~al}\mbox{.}(2008{\natexlab{b}}){Hopkins}, {Hernquist},
  {Cox}, \& {Kere{\v s}}}]{hopkins08a}
{Hopkins} P.~F., {Hernquist} L., {Cox} T.~J., {Kere{\v s}} D.,
  2008{\natexlab{b}}, \apjs, 175, 356

\bibitem[{{Hopkins} {et~al}\mbox{.}(2005){Hopkins}, {Hernquist}, {Cox},
  {Robertson}, {Di Matteo}, {Springel}, {Martini}, {Somerville}, \&
  {Li}}]{hopkins05}
{Hopkins} P.~F. {et~al.}, 2005, in Bulletin of the American Astronomical
  Society, Vol.~37, American Astronomical Society Meeting Abstracts, p. 1354

\bibitem[{{Hopkins} {et~al}\mbox{.}(2012){Hopkins}, {Kere{\v s}}, {Murray},
  {Quataert}, \& {Hernquist}}]{hopkins11c}
{Hopkins} P.~F., {Kere{\v s}} D., {Murray} N., {Quataert} E., {Hernquist} L.,
  2012, \mnras, 427, 968

\bibitem[{{Hopkins} {et~al}\mbox{.}(2006{\natexlab{b}}){Hopkins}, {Somerville},
  {Hernquist}, {Cox}, {Robertson}, \& {Li}}]{hopkins06b}
{Hopkins} P.~F., {Somerville} R.~S., {Hernquist} L., {Cox} T.~J., {Robertson}
  B., {Li} Y., 2006{\natexlab{b}}, \apj, 652, 864

\bibitem[{{Huchra} \& {Geller}(1982)}]{huchra82}
{Huchra} J.~P., {Geller} M.~J., 1982, \apj, 257, 423

\bibitem[{{Huertas-Company} {et~al}\mbox{.}(2016){Huertas-Company}, {Bernardi},
  {P{\'e}rez-Gonz{\'a}lez}, {Ashby}, {Barro}, {Conselice}, {Daddi}, {Dekel},
  {Dimauro}, {Faber}, {Grogin}, {Kartaltepe}, {Kocevski}, {Koekemoer}, {Koo},
  {Mei}, \& {Shankar}}]{huertascompany16}
{Huertas-Company} M. {et~al.}, 2016, \mnras, 462, 4495

\bibitem[{{Jahnke} \& {Macci{\`o}}(2011)}]{jahnke11}
{Jahnke} K., {Macci{\`o}} A.~V., 2011, \apj, 734, 92

\bibitem[{{Jarosik} {et~al}\mbox{.}(2011){Jarosik}, {Bennett}, {Dunkley},
  {Gold}, {Greason}, {Halpern}, {Hill}, {Hinshaw}, {Kogut}, {Komatsu},
  {Larson}, {Limon}, {Meyer}, {Nolta}, {Odegard}, {Page}, {Smith}, {Spergel},
  {Tucker}, {Weiland}, {Wollack}, \& {Wright}}]{jarosik11}
{Jarosik} N. {et~al.}, 2011, \apjs, 192, 14

\bibitem[{{Jones} {et~al}\mbox{.}(2000){Jones}, {Ponman}, \&
  {Forbes}}]{jones00}
{Jones} L.~R., {Ponman} T.~J., {Forbes} D.~A., 2000, \mnras, 312, 139

\bibitem[{{Jones} {et~al}\mbox{.}(2003){Jones}, {Ponman}, {Horton}, {Babul},
  {Ebeling}, \& {Burke}}]{jones03}
{Jones} L.~R., {Ponman} T.~J., {Horton} A., {Babul} A., {Ebeling} H., {Burke}
  D.~J., 2003, \mnras, 343, 627

\bibitem[{{Kauffmann}(1996)}]{kauffmann96}
{Kauffmann} G., 1996, \mnras, 281, 487

\bibitem[{{Kauffmann} {et~al}\mbox{.}(1999{\natexlab{a}}){Kauffmann},
  {Colberg}, {Diaferio}, \& {White}}]{kauffmann99a}
{Kauffmann} G., {Colberg} J.~M., {Diaferio} A., {White} S.~D.~M.,
  1999{\natexlab{a}}, \mnras, 303, 188

\bibitem[{{Kauffmann} {et~al}\mbox{.}(1999{\natexlab{b}}){Kauffmann},
  {Colberg}, {Diaferio}, \& {White}}]{kauffmann99b}
{Kauffmann} G., {Colberg} J.~M., {Diaferio} A., {White} S.~D.~M.,
  1999{\natexlab{b}}, \mnras, 307, 529

\bibitem[{{Kauffmann} {et~al}\mbox{.}(2003){Kauffmann}, {Heckman}, {White},
  {Charlot}, {Tremonti}, {Brinchmann}, {Bruzual}, {Peng}, {Seibert},
  {Bernardi}, {Blanton}, {Brinkmann}, {Castander}, {Cs{\'a}bai}, {Fukugita},
  {Ivezic}, {Munn}, {Nichol}, {Padmanabhan}, {Thakar}, {Weinberg}, \&
  {York}}]{kauffmann03}
{Kauffmann} G. {et~al.}, 2003, \mnras, 341, 33

\bibitem[{{Kauffmann} {et~al}\mbox{.}(2013){Kauffmann}, {Li}, {Zhang}, \&
  {Weinmann}}]{kauffmann13}
{Kauffmann} G., {Li} C., {Zhang} W., {Weinmann} S., 2013, \mnras, 430, 1447

\bibitem[{{Kauffmann} {et~al}\mbox{.}(2004){Kauffmann}, {White}, {Heckman},
  {M{\'e}nard}, {Brinchmann}, {Charlot}, {Tremonti}, \&
  {Brinkmann}}]{kauffmann04}
{Kauffmann} G., {White} S.~D.~M., {Heckman} T.~M., {M{\'e}nard} B.,
  {Brinchmann} J., {Charlot} S., {Tremonti} C., {Brinkmann} J., 2004, \mnras,
  353, 713

\bibitem[{{Kelly}(2007)}]{kelly07}
{Kelly} B.~C., 2007, \apj, 665, 1489

\bibitem[{{Kennicutt}(1998)}]{kennicutt98}
{Kennicutt}, Jr. R.~C., 1998, \apj, 498, 541

\bibitem[{{Kimm} {et~al}\mbox{.}(2009){Kimm}, {Somerville}, {Yi}, {van den
  Bosch}, {Salim}, {Fontanot}, {Monaco}, {Mo}, {Pasquali}, {Rich}, \&
  {Yang}}]{kimm09}
{Kimm} T. {et~al.}, 2009, \mnras, 394, 1131

\bibitem[{{Kimm} {et~al}\mbox{.}(2011){Kimm}, {Yi}, \& {Khochfar}}]{kimm11}
{Kimm} T., {Yi} S.~K., {Khochfar} S., 2011, \apj, 729, 11

\bibitem[{{Kitzbichler} \& {White}(2006)}]{kitzbichler06}
{Kitzbichler} M.~G., {White} S.~D.~M., 2006, \mnras, 366, 858

\bibitem[{{Knobel} {et~al}\mbox{.}(2012){Knobel}, {Lilly}, {Iovino}, {Kova{\v
  c}}, {Bschorr}, {Presotto}, {Oesch}, {Kampczyk}, {Carollo}, {Contini},
  {Kneib}, {Le Fevre}, {Mainieri}, {Renzini}, {Scodeggio}, {Zamorani},
  {Bardelli}, {Bolzonella}, {Bongiorno}, {Caputi}, {Cucciati}, {de la Torre},
  {de Ravel}, {Franzetti}, {Garilli}, {Lamareille}, {Le Borgne}, {Le Brun},
  {Maier}, {Mignoli}, {Pello}, {Peng}, {Perez Montero}, {Silverman}, {Tanaka},
  {Tasca}, {Tresse}, {Vergani}, {Zucca}, {Barnes}, {Bordoloi}, {Cappi},
  {Cimatti}, {Coppa}, {Koekemoer}, {L{\'o}pez-Sanjuan}, {McCracken}, {Moresco},
  {Nair}, {Pozzetti}, \& {Welikala}}]{knobel12}
{Knobel} C. {et~al.}, 2012, \apj, 753, 121

\bibitem[{{Knobel} {et~al}\mbox{.}(2015){Knobel}, {Lilly}, {Woo}, \& {Kova{\v
  c}}}]{knobel15}
{Knobel} C., {Lilly} S.~J., {Woo} J., {Kova{\v c}} K., 2015, \apj, 800, 24

\bibitem[{{Kormendy} \& {Kennicutt}(2004)}]{kormendy04}
{Kormendy} J., {Kennicutt}, Jr. R.~C., 2004, \araa, 42, 603

\bibitem[{{Laigle} {et~al}\mbox{.}(2017){Laigle}, {Pichon}, {Arnouts},
  {McCracken}, {Dubois}, {Devriendt}, {Slyz}, {Le Borgne}, {Benoit-Levy},
  {Hwang}, {Ilbert}, {Kraljic}, {Malavasi}, {Park}, \& {Vibert}}]{laigle17}
{Laigle} C. {et~al.}, 2017, ArXiv e-prints, 1702.08810

\bibitem[{{Larson} {et~al}\mbox{.}(1980){Larson}, {Tinsley}, \&
  {Caldwell}}]{larson80}
{Larson} R.~B., {Tinsley} B.~M., {Caldwell} C.~N., 1980, \apj, 237, 692

\bibitem[{{Lidman} {et~al}\mbox{.}(2013){Lidman}, {Iacobuta}, {Bauer},
  {Barrientos}, {Cerulo}, {Couch}, {Delaye}, {Demarco}, {Ellingson}, {Faloon},
  {Gilbank}, {Huertas-Company}, {Mei}, {Meyers}, {Muzzin}, {Noble}, {Nantais},
  {Rettura}, {Rosati}, {S{\'a}nchez-Janssen}, {Strazzullo}, {Webb}, {Wilson},
  {Yan}, \& {Yee}}]{lidman13}
{Lidman} C. {et~al.}, 2013, \mnras, 433, 825

\bibitem[{{Lin} {et~al}\mbox{.}(2010){Lin}, {Cooper}, {Jian}, {Koo}, {Patton},
  {Yan}, {Willmer}, {Coil}, {Chiueh}, {Croton}, {Gerke}, {Lotz},
  {Guhathakurta}, \& {Newman}}]{lin10}
{Lin} L. {et~al.}, 2010, \apj, 718, 1158

\bibitem[{{Lintott} {et~al}\mbox{.}(2009){Lintott}, {Schawinski}, {Keel}, {van
  Arkel}, {Bennert}, {Edmondson}, {Thomas}, {Smith}, {Herbert}, {Jarvis},
  {Virani}, {Andreescu}, {Bamford}, {Land}, {Murray}, {Nichol}, {Raddick},
  {Slosar}, {Szalay}, \& {Vandenberg}}]{lintott09}
{Lintott} C.~J. {et~al.}, 2009, \mnras, 399, 129

\bibitem[{{Liu} {et~al}\mbox{.}(2009){Liu}, {Mao}, {Deng}, {Xia}, \&
  {Wen}}]{liu09}
{Liu} F.~S., {Mao} S., {Deng} Z.~G., {Xia} X.~Y., {Wen} Z.~L., 2009, \mnras,
  396, 2003

\bibitem[{{Magorrian} {et~al}\mbox{.}(1998){Magorrian}, {Tremaine},
  {Richstone}, {Bender}, {Bower}, {Dressler}, {Faber}, {Gebhardt}, {Green},
  {Grillmair}, {Kormendy}, \& {Lauer}}]{magorrian98}
{Magorrian} J. {et~al.}, 1998, \aj, 115, 2285

\bibitem[{{Maier} {et~al}\mbox{.}(2016){Maier}, {Kuchner}, {Ziegler},
  {Verdugo}, {Balestra}, {Girardi}, {Mercurio}, {Rosati}, {Fritz}, {Grillo},
  {Nonino}, \& {Sartoris}}]{maier16}
{Maier} C. {et~al.}, 2016, \aap, 590, A108

\bibitem[{{Marconi} \& {Hunt}(2003)}]{marconi03}
{Marconi} A., {Hunt} L.~K., 2003, \apjl, 589, L21

\bibitem[{{Martig} {et~al}\mbox{.}(2012){Martig}, {Bournaud}, {Croton},
  {Dekel}, \& {Teyssier}}]{martig12}
{Martig} M., {Bournaud} F., {Croton} D.~J., {Dekel} A., {Teyssier} R., 2012,
  \apj, 756, 26

\bibitem[{{Martin} {et~al}\mbox{.}(2005){Martin}, {Fanson}, {Schiminovich},
  {Morrissey}, {Friedman}, {Barlow}, {Conrow}, {Grange}, {Jelinsky},
  {Milliard}, {Siegmund}, {Bianchi}, {Byun}, {Donas}, {Forster}, {Heckman},
  {Lee}, {Madore}, {Malina}, {Neff}, {Rich}, {Small}, {Surber}, {Szalay},
  {Welsh}, \& {Wyder}}]{martin05}
{Martin} D.~C. {et~al.}, 2005, \apjl, 619, L1

\bibitem[{{Martin} {et~al}\mbox{.}(2007){Martin}, {Wyder}, {Schiminovich},
  {Barlow}, {Forster}, {Friedman}, {Morrissey}, {Neff}, {Seibert}, {Small},
  {Welsh}, {Bianchi}, {Donas}, {Heckman}, {Lee}, {Madore}, {Milliard}, {Rich},
  {Szalay}, \& {Yi}}]{Martin07}
{Martin} D.~C. {et~al.}, 2007, \apjs, 173, 342

\bibitem[{{Masters} {et~al}\mbox{.}(2012){Masters}, {Nichol}, {Haynes}, {Keel},
  {Lintott}, {Simmons}, {Skibba}, {Bamford}, {Giovanelli}, \&
  {Schawinski}}]{masters12a}
{Masters} K.~L. {et~al.}, 2012, \mnras, 424, 2180

\bibitem[{{Masters} {et~al}\mbox{.}(2011){Masters}, {Nichol}, {Hoyle},
  {Lintott}, {Bamford}, {Edmondson}, {Fortson}, {Keel}, {Schawinski}, {Smith},
  \& {Thomas}}]{masters11a}
{Masters} K.~L. {et~al.}, 2011, \mnras, 411, 2026

\bibitem[{{McIntosh} {et~al}\mbox{.}(2008){McIntosh}, {Guo}, {Hertzberg},
  {Katz}, {Mo}, {van den Bosch}, \& {Yang}}]{mcintosh08}
{McIntosh} D.~H., {Guo} Y., {Hertzberg} J., {Katz} N., {Mo} H.~J., {van den
  Bosch} F.~C., {Yang} X., 2008, \mnras, 388, 1537

\bibitem[{{Merch{\'a}n} \& {Zandivarez}(2002)}]{merchan02}
{Merch{\'a}n} M., {Zandivarez} A., 2002, \mnras, 335, 216

\bibitem[{{Merch{\'a}n} \& {Zandivarez}(2005)}]{merchan05}
{Merch{\'a}n} M.~E., {Zandivarez} A., 2005, \apj, 630, 759

\bibitem[{{Merritt} \& {Ferrarese}(2001)}]{merritt01}
{Merritt} D., {Ferrarese} L., 2001, \mnras, 320, L30

\bibitem[{{Mihos} \& {Hernquist}(1994)}]{mihos94}
{Mihos} J.~C., {Hernquist} L., 1994, \apjl, 431, L9

\bibitem[{{Mihos} \& {Hernquist}(1996)}]{mihos96}
{Mihos} J.~C., {Hernquist} L., 1996, \apj, 464, 641

\bibitem[{{Miller} {et~al}\mbox{.}(2005){Miller}, {Nichol}, {Reichart},
  {Wechsler}, {Evrard}, {Annis}, {McKay}, {Bahcall}, {Bernardi}, {Boehringer},
  {Connolly}, {Goto}, {Kniazev}, {Lamb}, {Postman}, {Schneider}, {Sheth}, \&
  {Voges}}]{miller05}
{Miller} C.~J. {et~al.}, 2005, \aj, 130, 968

\bibitem[{{Moore} {et~al}\mbox{.}(1996){Moore}, {Katz}, {Lake}, {Dressler}, \&
  {Oemler}}]{moore96}
{Moore} B., {Katz} N., {Lake} G., {Dressler} A., {Oemler} A., 1996, Nature,
  379, 613–616

\bibitem[{{Mu{\~n}oz-Cuartas} \& {M{\"u}ller}(2012)}]{munoz12}
{Mu{\~n}oz-Cuartas} J.~C., {M{\"u}ller} V., 2012, \mnras, 423, 1583

\bibitem[{{Muldrew} {et~al}\mbox{.}(2012){Muldrew}, {Croton}, {Skibba},
  {Pearce}, {Ann}, {Baldry}, {Brough}, {Choi}, {Conselice}, {Cowan},
  {Gallazzi}, {Gray}, {Gr{\"u}tzbauch}, {Li}, {Park}, {Pilipenko}, {Podgorzec},
  {Robotham}, {Wilman}, {Yang}, {Zhang}, \& {Zibetti}}]{muldrew12}
{Muldrew} S.~I. {et~al.}, 2012, \mnras, 419, 2670

\bibitem[{{Muzzin} {et~al}\mbox{.}(2012){Muzzin}, {Wilson}, {Yee}, {Gilbank},
  {Hoekstra}, {Demarco}, {Balogh}, {van Dokkum}, {Franx}, {Ellingson}, {Hicks},
  {Nantais}, {Noble}, {Lacy}, {Lidman}, {Rettura}, {Surace}, \&
  {Webb}}]{muzzin12}
{Muzzin} A. {et~al.}, 2012, \apj, 746, 188

\bibitem[{{Nair} \& {Abraham}(2010)}]{nair10}
{Nair} P.~B., {Abraham} R.~G., 2010, ApJL, 714, L260Ã¢â‚¬â€œL264

\bibitem[{{Navarro} {et~al}\mbox{.}(1995){Navarro}, {Frenk}, \&
  {White}}]{navarro95}
{Navarro} J.~F., {Frenk} C.~S., {White} S.~D.~M., 1995, \mnras, 275, 56

\bibitem[{{Noble} {et~al}\mbox{.}(2016){Noble}, {Webb}, {Yee}, {Muzzin},
  {Wilson}, {van der Burg}, {Balogh}, \& {Shupe}}]{noble16}
{Noble} A.~G., {Webb} T.~M.~A., {Yee} H.~K.~C., {Muzzin} A., {Wilson} G., {van
  der Burg} R.~F.~J., {Balogh} M.~L., {Shupe} D.~L., 2016, \apj, 816, 48

\bibitem[{{Noeske} {et~al}\mbox{.}(2007){Noeske}, {Weiner}, {Faber},
  {Papovich}, {Koo}, {Somerville}, {Bundy}, {Conselice}, {Newman},
  {Schiminovich}, {Le Floc'h}, {Coil}, {Rieke}, {Lotz}, {Primack}, {Barmby},
  {Cooper}, {Davis}, {Ellis}, {Fazio}, {Guhathakurta}, {Huang}, {Kassin},
  {Martin}, {Phillips}, {Rich}, {Small}, {Willmer}, \& {Wilson}}]{Noeske07}
{Noeske} K.~G. {et~al.}, 2007, \apjl, 660, L43

\bibitem[{{Noguchi}(1988)}]{noguchi88}
{Noguchi} M., 1988, \aap, 203, 259

\bibitem[{{Nulsen}(1982)}]{nulsen82}
{Nulsen} P.~E.~J., 1982, \mnras, 198, 1007

\bibitem[{{Oh} {et~al}\mbox{.}(2011){Oh}, {Sarzi}, {Schawinski}, \&
  {Yi}}]{Oh11}
{Oh} K., {Sarzi} M., {Schawinski} K., {Yi} S.~K., 2011, \apjs, 195, 13

\bibitem[{{Oman} \& {Hudson}(2016)}]{oman16}
{Oman} K.~A., {Hudson} M.~J., 2016, \mnras, 463, 3083

\bibitem[{{Paccagnella} {et~al}\mbox{.}(2016){Paccagnella}, {Vulcani},
  {Poggianti}, {Moretti}, {Fritz}, {Gullieuszik}, {Couch}, {Bettoni}, {Cava},
  {D'Onofrio}, \& {Fasano}}]{paccagnella16}
{Paccagnella} A. {et~al.}, 2016, \apjl, 816, L25

\bibitem[{{Padmanabhan} {et~al}\mbox{.}(2008){Padmanabhan}, {Schlegel},
  {Finkbeiner}, {Barentine}, {Blanton}, {Brewington}, {Gunn}, {Harvanek},
  {Hogg}, {Ivezi{\'c}}, {Johnston}, {Kent}, {Kleinman}, {Knapp}, {Krzesinski},
  {Long}, {Neilsen}, {Nitta}, {Loomis}, {Lupton}, {Roweis}, {Snedden},
  {Strauss}, \& {Tucker}}]{padmanabhan08}
{Padmanabhan} N. {et~al.}, 2008, \apj, 674, 1217

\bibitem[{{Peng}(2007)}]{peng07}
{Peng} C.~Y., 2007, \apj, 671, 1098

\bibitem[{{Peng} {et~al}\mbox{.}(2015){Peng}, {Maiolino}, \&
  {Cochrane}}]{peng15}
{Peng} Y., {Maiolino} R., {Cochrane} R., 2015, \nat, 521, 192

\bibitem[{{Peng} {et~al}\mbox{.}(2010){Peng}, {Lilly}, {Kova{\v c}},
  {Bolzonella}, {Pozzetti}, {Renzini}, {Zamorani}, {Ilbert}, {Knobel},
  {Iovino}, {Maier}, {Cucciati}, {Tasca}, {Carollo}, {Silverman}, {Kampczyk},
  {de Ravel}, {Sanders}, {Scoville}, {Contini}, {Mainieri}, {Scodeggio},
  {Kneib}, {Le F{\`e}vre}, {Bardelli}, {Bongiorno}, {Caputi}, {Coppa}, {de la
  Torre}, {Franzetti}, {Garilli}, {Lamareille}, {Le Borgne}, {Le Brun},
  {Mignoli}, {Perez Montero}, {Pello}, {Ricciardelli}, {Tanaka}, {Tresse},
  {Vergani}, {Welikala}, {Zucca}, {Oesch}, {Abbas}, {Barnes}, {Bordoloi},
  {Bottini}, {Cappi}, {Cassata}, {Cimatti}, {Fumana}, {Hasinger}, {Koekemoer},
  {Leauthaud}, {Maccagni}, {Marinoni}, {McCracken}, {Memeo}, {Meneux}, {Nair},
  {Porciani}, {Presotto}, \& {Scaramella}}]{peng10}
{Peng} Y.-j. {et~al.}, 2010, \apj, 721, 193

\bibitem[{{Peng} {et~al}\mbox{.}(2012){Peng}, {Lilly}, {Renzini}, \&
  {Carollo}}]{peng12}
{Peng} Y.-j., {Lilly} S.~J., {Renzini} A., {Carollo} M., 2012, \apj, 757, 4

\bibitem[{{Phillips} {et~al}\mbox{.}(2015){Phillips}, {Wheeler}, {Cooper},
  {Boylan-Kolchin}, {Bullock}, \& {Tollerud}}]{phillips15}
{Phillips} J.~I., {Wheeler} C., {Cooper} M.~C., {Boylan-Kolchin} M., {Bullock}
  J.~S., {Tollerud} E., 2015, \mnras, 447, 698

\bibitem[{{Pimbblet}(2011)}]{pimbblet11}
{Pimbblet} K.~A., 2011, \mnras, 411, 2637

\bibitem[{{Pimbblet} {et~al}\mbox{.}(2002){Pimbblet}, {Smail}, {Kodama},
  {Couch}, {Edge}, {Zabludoff}, \& {O'Hely}}]{pimbblet02}
{Pimbblet} K.~A., {Smail} I., {Kodama} T., {Couch} W.~J., {Edge} A.~C.,
  {Zabludoff} A.~I., {O'Hely} E., 2002, \mnras, 331, 333

\bibitem[{{Poggianti} {et~al}\mbox{.}(1999){Poggianti}, {Smail}, {Dressler},
  {Couch}, {Barger}, {Butcher}, {Ellis}, \& {Oemler}}]{poggianti99}
{Poggianti} B.~M., {Smail} I., {Dressler} A., {Couch} W.~J., {Barger} A.~J.,
  {Butcher} H., {Ellis} R.~S., {Oemler}, Jr. A., 1999, \apj, 518, 576

\bibitem[{{Ponman} {et~al}\mbox{.}(1994){Ponman}, {Allan}, {Jones},
  {Merrifield}, {McHardy}, {Lehto}, \& {Luppino}}]{ponman94}
{Ponman} T.~J., {Allan} D.~J., {Jones} L.~R., {Merrifield} M., {McHardy} I.~M.,
  {Lehto} H.~J., {Luppino} G.~A., 1994, \nat, 369, 462

\bibitem[{{Pontzen} {et~al}\mbox{.}(2016){Pontzen}, {Tremmel}, {Roth},
  {Peiris}, {Saintonge}, {Volonteri}, {Quinn}, \& {Governato}}]{pontzen16}
{Pontzen} A., {Tremmel} M., {Roth} N., {Peiris} H.~V., {Saintonge} A.,
  {Volonteri} M., {Quinn} T., {Governato} F., 2016, ArXiv e-prints, 1607.02507

\bibitem[{{Porter} {et~al}\mbox{.}(2008){Porter}, {Raychaudhury}, {Pimbblet},
  \& {Drinkwater}}]{porter08}
{Porter} S.~C., {Raychaudhury} S., {Pimbblet} K.~A., {Drinkwater} M.~J., 2008,
  \mnras, 388, 1152

\bibitem[{{Postman}(2002)}]{postman02}
{Postman} M., 2002, in Astronomical Society of the Pacific Conference Series,
  Vol. 268, Tracing Cosmic Evolution with Galaxy Clusters, {Borgani} S.,
  {Mezzetti} M., {Valdarnini} R., eds., p.~3

\bibitem[{{Postman} {et~al}\mbox{.}(2005){Postman}, {Franx}, {Cross}, {Holden},
  {Ford}, {Illingworth}, {Goto}, {Demarco}, {Rosati}, {Blakeslee}, {Tran},
  {Ben{\'{\i}}tez}, {Clampin}, {Hartig}, {Homeier}, {Ardila}, {Bartko},
  {Bouwens}, {Bradley}, {Broadhurst}, {Brown}, {Burrows}, {Cheng}, {Feldman},
  {Golimowski}, {Gronwall}, {Infante}, {Kimble}, {Krist}, {Lesser}, {Martel},
  {Mei}, {Menanteau}, {Meurer}, {Miley}, {Motta}, {Sirianni}, {Sparks}, {Tran},
  {Tsvetanov}, {White}, \& {Zheng}}]{postman05}
{Postman} M. {et~al.}, 2005, \apj, 623, 721

\bibitem[{{Rettura} {et~al}\mbox{.}(2011){Rettura}, {Mei}, {Stanford},
  {Raichoor}, {Moran}, {Holden}, {Rosati}, {Ellis}, {Nakata}, {Nonino}, {Treu},
  {Blakeslee}, {Demarco}, {Eisenhardt}, {Ford}, {Fosbury}, {Illingworth},
  {Huertas-Company}, {Jee}, {Kodama}, {Postman}, {Tanaka}, \&
  {White}}]{rettura11}
{Rettura} A. {et~al.}, 2011, \apj, 732, 94

\bibitem[{{Rettura} {et~al}\mbox{.}(2010){Rettura}, {Rosati}, {Nonino},
  {Fosbury}, {Gobat}, {Menci}, {Strazzullo}, {Mei}, {Demarco}, \&
  {Ford}}]{rettura10}
{Rettura} A. {et~al.}, 2010, \apj, 709, 512

\bibitem[{{Roberts} {et~al}\mbox{.}(2016){Roberts}, {Parker}, \&
  {Karunakaran}}]{roberts16}
{Roberts} I.~D., {Parker} L.~C., {Karunakaran} A., 2016, \mnras, 455, 3628

\bibitem[{{Robotham} {et~al}\mbox{.}(2011){Robotham}, {Norberg}, {Driver},
  {Baldry}, {Bamford}, {Hopkins}, {Liske}, {Loveday}, {Merson}, {Peacock},
  {Brough}, {Cameron}, {Conselice}, {Croom}, {Frenk}, {Gunawardhana}, {Hill},
  {Jones}, {Kelvin}, {Kuijken}, {Nichol}, {Parkinson}, {Pimbblet}, {Phillipps},
  {Popescu}, {Prescott}, {Sharp}, {Sutherland}, {Taylor}, {Thomas}, {Tuffs},
  {van Kampen}, \& {Wijesinghe}}]{robotham11}
{Robotham} A.~S.~G. {et~al.}, 2011, \mnras, 416, 2640

\bibitem[{{Sanders} {et~al}\mbox{.}(1988){Sanders}, {Soifer}, {Elias},
  {Madore}, {Matthews}, {Neugebauer}, \& {Scoville}}]{sanders88}
{Sanders} D.~B., {Soifer} B.~T., {Elias} J.~H., {Madore} B.~F., {Matthews} K.,
  {Neugebauer} G., {Scoville} N.~Z., 1988, \apj, 325, 74

\bibitem[{{Schawinski} {et~al}\mbox{.}(2014){Schawinski}, {Urry}, {Simmons},
  {Fortson}, {Kaviraj}, {Keel}, {Lintott}, {Masters}, {Nichol}, {Sarzi},
  {Skibba}, {Treister}, {Willett}, {Wong}, \& {Yi}}]{schawinski14}
{Schawinski} K. {et~al.}, 2014, \mnras, 440, 889

\bibitem[{{Schawinski} {et~al}\mbox{.}(2010){Schawinski}, {Urry}, {Virani},
  {Coppi}, {Bamford}, {Treister}, {Lintott}, {Sarzi}, {Keel}, {Kaviraj},
  {Cardamone}, {Masters}, {Ross}, {Andreescu}, {Murray}, {Nichol}, {Raddick},
  {Slosar}, {Szalay}, {Thomas}, \& {Vandenberg}}]{schawinski10a}
{Schawinski} K. {et~al.}, 2010, \apj, 711, 284

\bibitem[{{Schmidt}(1959)}]{schmidt59}
{Schmidt} M., 1959, \apj, 129, 243

\bibitem[{{Sheth} {et~al}\mbox{.}(2005){Sheth}, {Vogel}, {Regan}, {Thornley},
  \& {Teuben}}]{sheth05}
{Sheth} K., {Vogel} S.~N., {Regan} M.~W., {Thornley} M.~D., {Teuben} P.~J.,
  2005, \apj, 632, 217

\bibitem[{{Shimizu} {et~al}\mbox{.}(2003){Shimizu}, {Kitayama}, {Sasaki}, \&
  {Suto}}]{shimizu03}
{Shimizu} M., {Kitayama} T., {Sasaki} S., {Suto} Y., 2003, \apj, 590, 197

\bibitem[{{Silk} \& {Rees}(1998)}]{silk98}
{Silk} J., {Rees} M.~J., 1998, \aap, 331, L1

\bibitem[{{Skibba} {et~al}\mbox{.}(2012){Skibba}, {Masters}, {Nichol},
  {Zehavi}, {Hoyle}, {Edmondson}, {Bamford}, {Cardamone}, {Keel}, {Lintott}, \&
  {Schawinski}}]{skibba12}
{Skibba} R.~A. {et~al.}, 2012, \mnras, 423, 1485

\bibitem[{{Smail} {et~al}\mbox{.}(1997){Smail}, {Dressler}, {Couch}, {Ellis},
  {Oemler}, {Butcher}, \& {Sharples}}]{smail97}
{Smail} I., {Dressler} A., {Couch} W.~J., {Ellis} R.~S., {Oemler}, Jr. A.,
  {Butcher} H., {Sharples} R.~M., 1997, \apjs, 110, 213

\bibitem[{{Smethurst} {et~al}\mbox{.}(2016){Smethurst}, {Lintott}, {Simmons},
  {Schawinski}, {Bamford}, {Cardamone}, {Kruk}, {Masters}, {Urry}, {Willett},
  \& {Wong}}]{smethurst16}
{Smethurst} R.~J. {et~al.}, 2016, \mnras, 463, 2986

\bibitem[{{Smethurst} {et~al}\mbox{.}(2015){Smethurst}, {Lintott}, {Simmons},
  {Schawinski}, {Marshall}, {Bamford}, {Fortson}, {Kaviraj}, {Masters},
  {Melvin}, {Nichol}, {Skibba}, \& {Willett}}]{smethurst15}
{Smethurst} R.~J. {et~al.}, 2015, \mnras, 450, 435

\bibitem[{{Smith} {et~al}\mbox{.}(2015){Smith}, {S{\'a}nchez-Janssen},
  {Beasley}, {Candlish}, {Gibson}, {Puzia}, {Janz}, {Knebe}, {Aguerri},
  {Lisker}, {Hensler}, {Fellhauer}, {Ferrarese}, \& {Yi}}]{smith15b}
{Smith} R. {et~al.}, 2015, \mnras, 454, 2502

\bibitem[{{Snyder} {et~al}\mbox{.}(2011){Snyder}, {Cox}, {Hayward},
  {Hernquist}, \& {Jonsson}}]{snyder11}
{Snyder} G.~F., {Cox} T.~J., {Hayward} C.~C., {Hernquist} L., {Jonsson} P.,
  2011, \apj, 741, 77

\bibitem[{{Somerville} {et~al}\mbox{.}(2008){Somerville}, {Hopkins}, {Cox},
  {Robertson}, \& {Hernquist}}]{somerville08}
{Somerville} R.~S., {Hopkins} P.~F., {Cox} T.~J., {Robertson} B.~E.,
  {Hernquist} L., 2008, \mnras, 391, 481

\bibitem[{{Somerville} {et~al}\mbox{.}(2001){Somerville}, {Primack}, \&
  {Faber}}]{somerville01}
{Somerville} R.~S., {Primack} J.~R., {Faber} S.~M., 2001, \mnras, 320, 504

\bibitem[{{Sparre} \& {Springel}(2016)}]{sparre16}
{Sparre} M., {Springel} V., 2016, ArXiv e-prints, 1610.03850

\bibitem[{{Springel} {et~al}\mbox{.}(2005){Springel}, {Di Matteo}, \&
  {Hernquist}}]{springel05b}
{Springel} V., {Di Matteo} T., {Hernquist} L., 2005, \apjl, 620, L79

\bibitem[{{Stoughton} {et~al}\mbox{.}(2002){Stoughton}, {Lupton}, {Bernardi},
  {Blanton}, {Burles}, {Castander}, {Connolly}, {Eisenstein}, {Frieman},
  {Hennessy}, {Hindsley}, {Ivezi{\'c}}, {Kent}, {Kunszt}, {Lee}, {Meiksin},
  {Munn}, {Newberg}, {Nichol}, {Nicinski}, {Pier}, {Richards}, {Richmond},
  {Schlegel}, {Smith}, {Strauss}, {SubbaRao}, {Szalay}, {Thakar}, {Tucker},
  {Vanden Berk}, {Yanny}, {Adelman}, {Anderson}, {Anderson}, {Annis},
  {Bahcall}, {Bakken}, {Bartelmann}, {Bastian}, {Bauer}, {Berman},
  {B{\"o}hringer}, {Boroski}, {Bracker}, {Briegel}, {Briggs}, {Brinkmann},
  {Brunner}, {Carey}, {Carr}, {Chen}, {Christian}, {Colestock}, {Crocker},
  {Csabai}, {Czarapata}, {Dalcanton}, {Davidsen}, {Davis}, {Dehnen},
  {Dodelson}, {Doi}, {Dombeck}, {Donahue}, {Ellman}, {Elms}, {Evans}, {Eyer},
  {Fan}, {Federwitz}, {Friedman}, {Fukugita}, {Gal}, {Gillespie}, {Glazebrook},
  {Gray}, {Grebel}, {Greenawalt}, {Greene}, {Gunn}, {de Haas}, {Haiman},
  {Haldeman}, {Hall}, {Hamabe}, {Hansen}, {Harris}, {Harris}, {Harvanek},
  {Hawley}, {Hayes}, {Heckman}, {Helmi}, {Henden}, {Hogan}, {Hogg}, {Holmgren},
  {Holtzman}, {Huang}, {Hull}, {Ichikawa}, {Ichikawa}, {Johnston}, {Kauffmann},
  {Kim}, {Kimball}, {Kinney}, {Klaene}, {Kleinman}, {Klypin}, {Knapp},
  {Korienek}, {Krolik}, {Kron}, {Krzesi{\'n}ski}, {Lamb}, {Leger},
  {Limmongkol}, {Lindenmeyer}, {Long}, {Loomis}, {Loveday}, {MacKinnon},
  {Mannery}, {Mantsch}, {Margon}, {McGehee}, {McKay}, {McLean}, {Menou},
  {Merelli}, {Mo}, {Monet}, {Nakamura}, {Narayanan}, {Nash}, {Neilsen},
  {Newman}, {Nitta}, {Odenkirchen}, {Okada}, {Okamura}, {Ostriker}, {Owen},
  {Pauls}, {Peoples}, {Peterson}, {Petravick}, {Pope}, {Pordes}, {Postman},
  {Prosapio}, {Quinn}, {Rechenmacher}, {Rivetta}, {Rix}, {Rockosi}, {Rosner},
  {Ruthmansdorfer}, {Sandford}, {Schneider}, {Scranton}, {Sekiguchi}, {Sergey},
  {Sheth}, {Shimasaku}, {Smee}, {Snedden}, {Stebbins}, {Stubbs}, {Szapudi},
  {Szkody}, {Szokoly}, {Tabachnik}, {Tsvetanov}, {Uomoto}, {Vogeley}, {Voges},
  {Waddell}, {Walterbos}, {Wang}, {Watanabe}, {Weinberg}, {White}, {White},
  {Wilhite}, {Wolfe}, {Yasuda}, {York}, {Zehavi}, \& {Zheng}}]{stoughton02}
{Stoughton} C. {et~al.}, 2002, \aj, 123, 485

\bibitem[{{Tago} {et~al}\mbox{.}(2008){Tago}, {Einasto}, {Saar}, {Tempel},
  {Einasto}, {Vennik}, \& {M{\"u}ller}}]{tago08}
{Tago} E., {Einasto} J., {Saar} E., {Tempel} E., {Einasto} M., {Vennik} J.,
  {M{\"u}ller} V., 2008, \aap, 479, 927

\bibitem[{{Tago} {et~al}\mbox{.}(2010){Tago}, {Saar}, {Tempel}, {Einasto},
  {Einasto}, {Nurmi}, \& {Hein{\"a}m{\"a}ki}}]{tago10}
{Tago} E., {Saar} E., {Tempel} E., {Einasto} J., {Einasto} M., {Nurmi} P.,
  {Hein{\"a}m{\"a}ki} P., 2010, \aap, 514, A102

\bibitem[{{Taylor}(2005)}]{taylor05}
{Taylor} M.~B., 2005, in Astronomical Society of the Pacific Conference Series,
  Vol. 347, Astronomical Data Analysis Software and Systems XIV, {Shopbell} P.,
  {Britton} M., {Ebert} R., eds., p.~29

\bibitem[{{Tempel} {et~al}\mbox{.}(2014){Tempel}, {Tamm}, {Gramann},
  {Tuvikene}, {Liivam{\"a}gi}, {Suhhonenko}, {Kipper}, {Einasto}, \&
  {Saar}}]{tempel14}
{Tempel} E. {et~al.}, 2014, \aap, 566, A1

\bibitem[{{Tinker} {et~al}\mbox{.}(2011){Tinker}, {Wetzel}, \&
  {Conroy}}]{tinker11}
{Tinker} J., {Wetzel} A., {Conroy} C., 2011, ArXiv e-prints, 1107.5046

\bibitem[{{Tonini} {et~al}\mbox{.}(2016){Tonini}, {Mutch}, {Croton}, \&
  {Wyithe}}]{tonini16}
{Tonini} C., {Mutch} S.~J., {Croton} D.~J., {Wyithe} J.~S.~B., 2016, \mnras,
  459, 4109

\bibitem[{{Toomre}(1977)}]{toomre77}
{Toomre} A., 1977, in Evolution of Galaxies and Stellar Populations,
  {B.~M.~Tinsley \& R.~B.~G.~Larson D.~Campbell}, ed., p. 401

\bibitem[{{Toomre} \& {Toomre}(1972)}]{toomre72}
{Toomre} A., {Toomre} J., 1972, \apj, 178, 623

\bibitem[{{Treister} {et~al}\mbox{.}(2012){Treister}, {Schawinski}, {Urry}, \&
  {Simmons}}]{treister12}
{Treister} E., {Schawinski} K., {Urry} C.~M., {Simmons} B.~D., 2012, \apjl,
  758, L39

\bibitem[{{Tucker} {et~al}\mbox{.}(2000){Tucker}, {Oemler}, {Hashimoto},
  {Shectman}, {Kirshner}, {Lin}, {Landy}, {Schechter}, \& {Allam}}]{tucker00}
{Tucker} D.~L. {et~al.}, 2000, \apjs, 130, 237

\bibitem[{{van de Voort} {et~al}\mbox{.}(2016){van de Voort}, {Bah{\'e}},
  {Bower}, {Correa}, {Crain}, {Schaye}, \& {Theuns}}]{vandevoort16}
{van de Voort} F., {Bah{\'e}} Y.~M., {Bower} R.~G., {Correa} C.~A., {Crain}
  R.~A., {Schaye} J., {Theuns} T., 2016, ArXiv e-prints, 1611.03870

\bibitem[{{Walker} {et~al}\mbox{.}(1996){Walker}, {Mihos}, \&
  {Hernquist}}]{walker96}
{Walker} I.~R., {Mihos} J.~C., {Hernquist} L., 1996, \apj, 460, 121

\bibitem[{{Wang} {et~al}\mbox{.}(2014){Wang}, {Sales}, {Henriques}, \&
  {White}}]{wang14}
{Wang} W., {Sales} L.~V., {Henriques} B.~M.~B., {White} S.~D.~M., 2014, \mnras,
  442, 1363

\bibitem[{{Weiner} {et~al}\mbox{.}(2006){Weiner}, {Willmer}, {Faber}, {Harker},
  {Kassin}, {Phillips}, {Melbourne}, {Metevier}, {Vogt}, \& {Koo}}]{Weiner06}
{Weiner} B.~J. {et~al.}, 2006, \apj, 653, 1049

\bibitem[{{Weinmann} {et~al}\mbox{.}(2006){Weinmann}, {van den Bosch}, {Yang},
  \& {Mo}}]{weinmann06}
{Weinmann} S.~M., {van den Bosch} F.~C., {Yang} X., {Mo} H.~J., 2006, \mnras,
  366, 2

\bibitem[{{Wetzel} {et~al}\mbox{.}(2013){Wetzel}, {Tinker}, {Conroy}, \& {van
  den Bosch}}]{wetzel13}
{Wetzel} A.~R., {Tinker} J.~L., {Conroy} C., {van den Bosch} F.~C., 2013,
  \mnras, 432, 336

\bibitem[{{Whitaker} {et~al}\mbox{.}(2016){Whitaker}, {Bezanson}, {van Dokkum},
  {Franx}, {van der Wel}, {Brammer}, {Forster-Schreiber}, {Giavalisco},
  {Labbe}, {Momcheva}, {Nelson}, \& {Skelton}}]{whitaker16}
{Whitaker} K.~E. {et~al.}, 2016, ArXiv e-prints, 1607.03107

\bibitem[{{Willett} {et~al}\mbox{.}(2013){Willett}, {Lintott}, {Bamford},
  {Masters}, {Simmons}, {Casteels}, {Edmondson}, {Fortson}, {Kaviraj}, {Keel},
  {Melvin}, {Nichol}, {Raddick}, {Schawinski}, {Simpson}, {Skibba}, {Smith}, \&
  {Thomas}}]{GZ2}
{Willett} K.~W. {et~al.}, 2013, \mnras, 435, 2835

\bibitem[{{Woo} {et~al}\mbox{.}(2015){Woo}, {Dekel}, {Faber}, \& {Koo}}]{woo15}
{Woo} J., {Dekel} A., {Faber} S.~M., {Koo} D.~C., 2015, \mnras, 448, 237

\bibitem[{{Wyder} {et~al}\mbox{.}(2007){Wyder}, {Martin}, {Schiminovich},
  {Seibert}, {Budav{\'a}ri}, {Treyer}, {Barlow}, {Forster}, {Friedman},
  {Morrissey}, {Neff}, {Small}, {Bianchi}, {Donas}, {Heckman}, {Lee}, {Madore},
  {Milliard}, {Rich}, {Szalay}, {Welsh}, \& {Yi}}]{wyder07}
{Wyder} T.~K. {et~al.}, 2007, \apjs, 173, 293

\bibitem[{{Yang} {et~al}\mbox{.}(2007){Yang}, {Mo}, {van den Bosch},
  {Pasquali}, {Li}, \& {Barden}}]{yang07}
{Yang} X., {Mo} H.~J., {van den Bosch} F.~C., {Pasquali} A., {Li} C., {Barden}
  M., 2007, \apj, 671, 153

\bibitem[{{Yesuf} {et~al}\mbox{.}(2014){Yesuf}, {Faber}, {Trump}, {Koo},
  {Fang}, {Liu}, {Wild}, \& {Hayward}}]{yesuf14}
{Yesuf} H.~M., {Faber} S.~M., {Trump} J.~R., {Koo} D.~C., {Fang} J.~J., {Liu}
  F.~S., {Wild} V., {Hayward} C.~C., 2014, \apj, 792, 84

\bibitem[{{York} {et~al}\mbox{.}(2000){York}, {Adelman}, {Anderson},
  {Anderson}, {Annis}, {Bahcall}, {Bakken}, {Barkhouser}, {Bastian}, {Berman},
  {Boroski}, {Bracker}, {Briegel}, {Briggs}, {Brinkmann}, {Brunner}, {Burles},
  {Carey}, {Carr}, {Castander}, {Chen}, {Colestock}, {Connolly}, {Crocker},
  {Csabai}, {Czarapata}, {Davis}, {Doi}, {Dombeck}, {Eisenstein}, {Ellman},
  {Elms}, {Evans}, {Fan}, {Federwitz}, {Fiscelli}, {Friedman}, {Frieman},
  {Fukugita}, {Gillespie}, {Gunn}, {Gurbani}, {de Haas}, {Haldeman}, {Harris},
  {Hayes}, {Heckman}, {Hennessy}, {Hindsley}, {Holm}, {Holmgren}, {Huang},
  {Hull}, {Husby}, {Ichikawa}, {Ichikawa}, {Ivezi{\'c}}, {Kent}, {Kim},
  {Kinney}, {Klaene}, {Kleinman}, {Kleinman}, {Knapp}, {Korienek}, {Kron},
  {Kunszt}, {Lamb}, {Lee}, {Leger}, {Limmongkol}, {Lindenmeyer}, {Long},
  {Loomis}, {Loveday}, {Lucinio}, {Lupton}, {MacKinnon}, {Mannery}, {Mantsch},
  {Margon}, {McGehee}, {McKay}, {Meiksin}, {Merelli}, {Monet}, {Munn},
  {Narayanan}, {Nash}, {Neilsen}, {Neswold}, {Newberg}, {Nichol}, {Nicinski},
  {Nonino}, {Okada}, {Okamura}, {Ostriker}, {Owen}, {Pauls}, {Peoples},
  {Peterson}, {Petravick}, {Pier}, {Pope}, {Pordes}, {Prosapio},
  {Rechenmacher}, {Quinn}, {Richards}, {Richmond}, {Rivetta}, {Rockosi},
  {Ruthmansdorfer}, {Sandford}, {Schlegel}, {Schneider}, {Sekiguchi}, {Sergey},
  {Shimasaku}, {Siegmund}, {Smee}, {Smith}, {Snedden}, {Stone}, {Stoughton},
  {Strauss}, {Stubbs}, {SubbaRao}, {Szalay}, {Szapudi}, {Szokoly}, {Thakar},
  {Tremonti}, {Tucker}, {Uomoto}, {Vanden Berk}, {Vogeley}, {Waddell}, {Wang},
  {Watanabe}, {Weinberg}, {Yanny}, {Yasuda}, \& {SDSS Collaboration}}]{york00}
{York} D.~G. {et~al.}, 2000, \aj, 120, 1579

\bibitem[{{Zurita} {et~al}\mbox{.}(2004){Zurita}, {Rela{\~n}o}, {Beckman}, \&
  {Knapen}}]{zurita04}
{Zurita} A., {Rela{\~n}o} M., {Beckman} J.~E., {Knapen} J.~H., 2004, \aap, 413,
  73

\bibitem[{{Zwicky}(1938)}]{zwicky38}
{Zwicky} F., 1938, \pasp, 50, 218

\end{thebibliography}

\end{document}